\documentclass[aps,prd,amsmath,amssymb,superscriptaddress,eqsecnum,notitlepage,nofootinbib,showpacs]{revtex4-1}
\usepackage{color}
\usepackage[pdftex]{graphicx}
\usepackage{dsfont}
\usepackage{yfonts}

\graphicspath{{./Plots/}}

\newcommand{\leaveout}[1]{}
\newcommand{\fixme}[1]{}
\newcommand{\fiximp}[1]{}
\newcommand{\fixmeAO}[1]{}

\newcommand{\Disc}{\delta}
\newcommand{\lam}{\mathcal{\lambda}}

\newcommand{\dt}{t}
\newcommand{\eps}{\ob}
\newcommand{\indmode}{\ell}
\newcommand{\RAM}{\nu}
\newcommand{\fNIA}{\sigma}
\newcommand{\Dg}{\Disc g_{\indmode}}
\newcommand{\g}{g_{\indmode}}
\newcommand{\gm}{g_{\ell-}}

\newcommand{\gp}{g_{\ell+}}
\newcommand{\f}{f_{\indmode}}
\newcommand{\Chi}{\hat f_{\indmode}}

\newcommand{\Xin}[1]{{}_{#1}X^{in}_{\indmode}}
\newcommand{\Xup}[1]{{}_{#1}X^{up}_{\indmode}}
\newcommand{\Xinup}[1]{{}_{#1}X^{in/up}_{\indmode}}

\newcommand{\Xintra}[1]{{}_{#1}X^{in,tra}_{\indmode}} 
\newcommand{\XinincNST}{X^{in,inc}_{s}}
\newcommand{\XinrefNST}{X^{in,ref}_{s}}
\newcommand{\XintraNST}{X^{in,tra}_{s}}
\newcommand{\XuptraNST}{X^{up,tra}_{s}}
\newcommand{\Xinref}[1]{{}_{#1}X^{in,ref}_{\indmode}}
\newcommand{\Xininc}[1]{{}_{#1}X^{in,inc}_{\indmode}}
\newcommand{\Xuptra}[1]{{}_{#1}X^{up,tra}_{\indmode}}

\newcommand{\Xupref}[1]{{}_{#1}X^{up,ref}_{\indmode}}
\newcommand{\Xupinc}[1]{{}_{#1}X^{up,inc}_{\indmode}}

\newcommand{\Rin}[1]{{}_{#1}R^{in}_{\indmode}}
\newcommand{\Rup}[1]{{}_{#1}R^{up}_{\indmode}}
\newcommand{\Rinup}[1]{{}_{#1}R^{in/up}_{\indmode}}

\newcommand{\Rintra}[1]{{}_{#1}R^{in,tra}_{\indmode}}
\newcommand{\RintraNST}{R^{in,tra}_{s}} 
\newcommand{\RuptraNST}{R^{up,tra}_{s}} 
\newcommand{\RinincNST}{R^{in,inc}_{s}} 
\newcommand{\RinrefNST}{R^{in,ref}_{s}} 
\newcommand{\RininctraNST}{R^{in,inc/tra}_{s}} 
\newcommand{\Rinref}[1]{{}_{#1}R^{in,ref}_{\indmode}}
\newcommand{\Rininc}[1]{{}_{#1}R^{in,inc}_{\indmode}}
\newcommand{\Ruptra}[1]{{}_{#1}R^{up,tra}_{\indmode}}

\newcommand{\Rpmu}{R^{\RAM}_{+}}
\newcommand{\Rpmutra}{R^{\RAM,tra}_{+}}
\newcommand{\Apmu}{A^{\RAM}_{+}}
\newcommand{\ApmuRW}{\check{A}^{\RAM}_{+}}
\newcommand{\Ammu}{A^{\RAM}_{-}}
\newcommand{\RupST}{\RTupN}
\newcommand{\RuptraST}{\RuptraNST}
\newcommand{\Rupref}[1]{{}_{#1}R^{up,ref}_{\indmode}}
\newcommand{\Rupinc}[1]{{}_{#1}R^{up,inc}_{\indmode}}

\newcommand{\DRupRuptra}[1]{\Disc \RupN{#1}}
\newcommand{\RinN}[1]{{}_{#1}\hat{R}^{in}_{\indmode}}
\newcommand{\RupN}[1]{{}_{#1}\hat{R}^{up}_{\indmode}}
\newcommand{\RupccN}[1]{{}_{#1}\hat{R}^{up^*}_{\indmode}}
\newcommand{\RinccN}[1]{{}_{#1}\hat{R}^{in^*}_{\indmode}}
\newcommand{\RinprimeN}[1]{{}_{#1}\hat{R}^{in'}_{\indmode}}
\newcommand{\RupprimeN}[1]{{}_{#1}\hat{R}^{up'}_{\indmode}}
\newcommand{\RinincN}[1]{{}_{#1}\hat{R}^{in,inc}_{\indmode}}
\newcommand{\RupmN}[1]{{}_{#1}\hat{R}^{up}_{\indmode -}}
\newcommand{\RuppN}[1]{{}_{#1}\hat{R}^{up}_{\indmode +}}
\newcommand{\RinNfg}[1]{{}_{#1}\tilde{R}^{in}_{\indmode}}
\newcommand{\RupNfg}[1]{{}_{#1}\tilde{R}^{up}_{\indmode}}
\newcommand{\RintraNfg}[1]{{}_{#1}\tilde{R}^{in,tra}_{\indmode}}
\newcommand{\RuptraNfg}[1]{{}_{#1}\tilde{R}^{up,tra}_{\indmode}}
\newcommand{\RinincNfg}[1]{{}_{#1}\tilde{R}^{in,inc}_{\indmode}}
\newcommand{\RW}{\mathcal{X}_s}
\newcommand{\RWsource}{S_s}
\newcommand{\RWind}{{}_s\mathcal{X}_{\indmode m \omega}}
\newcommand{\RWsourceind}{{}_sS_{\indmode m }}
\newcommand{\RRW}[1]{{}_{#1}X_{\indmode}}
\newcommand{\RRWinN}{X^\text{in}_{s}}
\newcommand{\RRWupN}{X^\text{up}_{s}}
\newcommand{\RRWinupN}{X^\text{in/up}_{s}}
\newcommand{\RRWind}{{}_sX_{\ell m}}
\newcommand{\RWtr}{{}_s\mathcal{X}_{\indmode}}
\newcommand{\T}{\Psi_s} 
\newcommand{\RTind}{{}_sR_{\ell m}}
\newcommand{\Tsource}{T_s}
\newcommand{\Tsourceind}{{}_sT_{\indmode m }}
\newcommand{\RT}[1]{{}_{#1}R_{\indmode}} 
\newcommand{\RTinN}{R_s^\text{in}} 
\newcommand{\RTupN}{R_s^\text{up}} 
\newcommand{\RTinupN}{R_s^\text{in/up}} 
\newcommand{\Ttr}{{}_s\Psi_{\indmode}} 
\newcommand{\NC}[1]{N_\text{C}^{#1}}
\newcommand{\RCp}[1]{R_+^{#1}}
\newcommand{\RCm}[1]{R_-^{#1}}
\newcommand{\XCp}[1]{X_+^{#1}}
\newcommand{\XCm}[1]{X_-^{#1}}
\newcommand{\CT}{{}_s\chi_{\indmode m}} 
\newcommand{\CTh}{{}_s\chi_{\indmode}} 
\newcommand{\chiin}{{}_s\chi^{in}_{\indmode}} 
\newcommand{\chiup}{{}_s\chi^{up}_{\indmode}} 
\newcommand{\cz}{c_0}
\newcommand{\co}{c_1}
\newcommand{\ct}{c_2}

\newcommand{\WT}{W^T}

\newcommand{\Gret}{{}_sG_{\text{ret}}}
\newcommand{\GretT}{{}_sG_{\text{ret},T}}
\newcommand{\Gl}{{}_sG_{\indmode}}
\newcommand{\Glret}{{}_sG^\text{ret}_{\indmode}}
\newcommand{\GlT}{{}_sG^{\text{ret},T}_{\indmode}}
\newcommand{\GlwT}{{}_sG^{T}_{\indmode}}
\newcommand{\GlBC}{{}_sG_{\indmode}^{BC}}
\newcommand{\GlBCT}{{}_sG_{\indmode}^{BC,T}}
\newcommand{\DGw}{\Disc {}_sG_{\indmode}}
\newcommand{\DGwT}{\Disc {}_sG^T_{\indmode}}
\newcommand{\SWSH}[1]{{}_{#1}Y_{\ell m}}
\newcommand{\OpRW}{\mathcal{O}}
\newcommand{\OpRWind}{\mathcal{O}_{\omega}}
\newcommand{\OpD}{\mathcal{D}}

\newcommand{\amumT}[1]{a_{#1}^{-\RAM-1,T}}
\newcommand{\anRWnoG}[1]{\check{a}_{#1}}
\newcommand{\anRWnoGNow}[1]{a_{#1}}
\newcommand{\anTnoG}[1]{a_{#1}}
\newcommand{\nb}{\bar\fNIA}
\newcommand{\fb}{\hat f_{\ell}}

\newcommand{\W}[1]{W\left(#1\right)}

\newcommand{\ob}{\bar{\omega}}
\newcommand{\rb}{\bar{r}}

\newcommand{\G}{\Gamma (1-s)}
\newcommand{\RAMmLead}{\RAM_2}
\newcommand{\Kmu}{K_{\RAM}}
\newcommand{\Kmmu}{K_{-\RAM-1}}

\newcommand{\qT}{q^T}
\newcommand{\Aout}{A^{out}_{\indmode}}
\newcommand{\Ain}{A^{in}_{\indmode}}

\newcommand{\ulBC}{u_{\ell}^{BC}}

\newcommand{\iv}{v}

\newcommand{\pRW}{\check{p}}
\newcommand{\pT}{p}
\newcommand{\fRW}{\check{h}}
\newcommand{\fT}{h}
\newcommand{\gRW}{k}

\newcommand{\KT}{K}

\begin{document}


\author{Marc Casals}
\email{mcasals@cbpf.br,marc.casals@ucd.ie}
\affiliation{Centro Brasileiro de Pesquisas F\'isicas (CBPF), Rio de Janeiro, 
Brazil.}
\affiliation{School of Mathematics \& Statistics and Complex \& Adaptive Systems
Laboratory, University College Dublin, Belfield, Dublin 4, Ireland}

\author{Adrian Ottewill}
\email{adrian.ottewill@ucd.ie}
\affiliation{School of Mathematics \& Statistics and Complex \& Adaptive Systems
Laboratory, University College Dublin, Belfield, Dublin 4, Ireland}

\title{
High-order Tail in Schwarzschild Space-time
}

\begin{abstract}
We present an analysis of the behaviour at late-times of linear field perturbations of a Schwarzschild black hole space-time.
In particular, we give explicit analytic expressions for the field perturbations (for a specific $\ell$-multipole) of general spin up to the first four 
  orders
   at late times.
These expressions are valid at arbitrary radius and include, apart from the well-known power-law tail decay
at leading order ($\sim\dt^{-2\ell-3}$), a new logarithmic behaviour
at  third leading order ($\sim\dt^{-2\ell-5}\ln \dt$).
We obtain these late-time results by developing the so-called MST formalism
and by expanding   the various MST Fourier-mode quantities   for small frequency.
While we   give explicit expansions  up to the first four leading orders (for small-frequency for the Fourier modes, for late-time for the field perturbation),
 we give a prescription for obtaining expressions to arbitrary order within a `perturbative regime'.


\end{abstract}

\date{\today}
\maketitle




\section{Introduction}

The study of linear field perturbations of black hole background space-times is important for several purposes.
For example, for the investigation of the (linear) stability of black holes, the effect that black holes have on fields propagating in their neighbourhood
or the binary inspiral of a black hole and another compact, astrophysical object.
In 1955 and subsequent years~\cite{Regge:1957td,Wheeler:1955zz,Price:1971fb,Price:1972pw,PhysRevLett.24.737,PhysRevD.2.2141,ruffini1972electromagnetic}, the equations describing linear field perturbations of a non-rotating (Schwarzschild) black hole space-time were decoupled and separated.
This rendered the equations treatable semi-analytically as the full perturbation may be obtained as a sum of Fourier modes, with the radial part satisfying the so-called
Regge-Wheeler (RW) ordinary differential equation.
It was not until 1972 that a similar feat was achieved by Teukolsky~\cite{teukolsky1972rotating,Teukolsky:1973ha} in the case of a rotating (Kerr) black hole space-time.
In the Schwarzschild limit, the radial part of the  Teukolsky equation reduces  to the so-called Bardeen-Press-Teukolsky (BPT) equation~\cite{bardeen1973radiation},
which was also obtained in 1972.

 The RW and the BPT radial equations are satisfied by different quantities (different combinations of field components and their derivatives) and their
solutions
have been studied thoroughly, both numerically as well as 
with asymptotic analyses.
Of particular interest for this paper, is the result by Price~\cite{Price:1971fb,Price:1972pw} (obtained by studying the field perturbations without Fourier-decomposing)
for the behaviour at late-times  of a RW field perturbation of any spin of a Schwarzschild black hole.
Price found that its radiative $\ell$-multipole
 decays to  leading-order  in the form of a power-law: $\dt^{-2\ell-3}$, where 
  $t$ the Schwarzschild time.
  The analysis in~\cite{hod1998late} of the Teukolsky equation in Kerr shows that the same
   leading-order power-law tail behaviour is satisfied by the radiative multipoles of BPT field perturbations in Schwarzschild.
In this paper we present analytic expressions for the behaviour of
general-spin field perturbations  in Schwarzschild
up to the first {\it four} orders for late-times, revealing a new logarithmic behaviour in the third leading order ($\sim\dt^{-2\ell-5}\ln \dt$).
Furthermore, although most analyses   (though see, e.g.,~\cite{PhysRevD.61.024026} for an exception) that give the
radius-dependent coefficient of the leading-order power-law
have been constrained to large radius, our results for all four orders are valid for {\it arbitrary} radius.
We obtain this late-time behaviour both for the field quantities 
satisying the RW as well as the BPT equation.
We obtain these results by developing a method which is valid not only at late-times, but it may  be used to obtain results valid in principle at any time regime.
We now briefly introduce this method.

In 1986, Leaver~\cite{Leaver:1986a} derived various analytical representations for the solutions of the RW and BPT equations
in terms
of infinite series of special functions.
A series of Japanese researchers
 later `revamped' some of Leaver's series representations 
and derived other new  series representations for these radial solutions~\cite{Mano:Suzuki:Takasugi:1996,Mano:1996mf,Sasaki:2003xr}.
These latter series representations, to which we shall refer as MST expansions,
are  naturally adapted to carrying out small-frequency expansions. Small-frequency expansions yield the late-time behaviour of the full  linear perturbation after integration over frequency.
We note, however, that the MST series in principle converge for any value of the frequency,
although the speed of their
 convergence decreases as the magnitude of the frequency increases.

The MST method is a powerful method which only relatively recently researchers have been starting to use in order to obtain results in black hole perturbation theory.
For example, in the case of a spherically-symmetric background, the MST formalism has found applications in the calculation of the self-force~\cite{Poisson:2011nh} on a point particle~\cite{hikida2004new,hikida2005new,CDOW13}, post-Newtonian coefficients and gauge-invariant quantities~\cite{bini2014high,PhysRevD.90.024039,PhysRevD.89.104047,PhysRevD.89.064042,Shah:2014tka}
and dynamical tidal interaction of compact objects~\cite{delsate2013new}.
The MST method has proven particularly useful for calculating the retarded Green function (GF) of the wave equation satisfied by the field perturbation,
which is a fundamental quantity 
 as it determines the evolution in time of any given initial data.
The Fourier modes of the GF posess poles (the so-called quasinormal modes) and a branch cut (BC) in the complex-frequency plane~\cite{Leaver:1986}.
It is known (e.g.,~\cite{Leaver:1986,Ching:1995tj}) that the small-frequency part of the BC is the contribution to the GF that gives rise to the late-time behaviour;  the non-small-frequency
part of the BC contributes to the behaviour at earlier times (see~\cite{Casals:2012ng,Casals:2011aa,CDOW13,MaassenvandenBrink:2000ru,MaassenvandenBrink:2003as}).

In this paper we derive in detail some of the results on the GF  Fourier modes and black hole perturbations that  we briefly presented in the 
Letter~\cite{PhysRevLett.109.111101}.
Namely, we derive a small-frequency expansion of the MST series in general, and of the BC in particular, which we then use to
obtain
the late-time behaviour of the GF and field perturbations. We obtain the  late-time behaviour up to the first four leading orders at arbitrary radius  in Schwarzschild space-time and
for  general-integer-spin
\footnote{We do not give the expansions explicitly  for polar gravitational (Zerilli) perturbations
 nor for positive BPT spin but these can be derived directly from, respectively,
axial (RW) gravitational perturbations or  negative BPT spin, both of which we do derive explicitly.}
 of the field.  
In~\cite{CDOW13} we used the calculation of the late-time GF derived in~\cite{PhysRevLett.109.111101} in order to find its contribution to the
self-force on a scalar charge in Schwarzschild.
The MST method has also been used to calculate  the quasinormal mode  contribution to the GF
 in Schwarzschild space-time~\cite{CDOW13}  and in Kerr space-time~\cite{zhang2013quasinormal};
 in the former case the GF calculation was applied to obtain the scalar self-force
 and, in the latter case, to obtain the radiation emitted  given a
 specific perturbation source.
The calculations of the quasinormal mode series in~\cite{CDOW13,zhang2013quasinormal} involved evaluating the MST series at frequencies with 
{\it `arbitrarily'} large magnitude.


Apart from an explicit small-frequency/late-time analysis, we shall also generally develop the MST formalism in Schwarzschild space-time.
In particular, we shall present
 MST expressions (valid for general frequency) for the solutions of the RW equation for general integer spin.
To the best of our knowledge, these expressions are new for spin-1 since
the MST formalism has not yet been presented for the RW equation for spin-1 (it only has been for spins-0 and -2).
Furthermore, we
develop, also for the first time in the literature,
 the MST formalism
for calculating the contribution to the GF
of  Fourier modes along the BC.
We also give the relationships between the BPT quantities and the RW quantities via the so-called Chandrasekhar transformation.  
We then 
give  
explicit expressions 
for the main MST quantities
up to the first 
four
leading orders for small 
frequency,
 for general spin and multipole number $\ell$.
We note that alhough small-frequency expansions have been given for some of these quantities already,
that has typically been done within a post-Newtonian framework,
and so for the radial solutions expanded about radial infinity. In here, instead, we give small-frequency expansions 
for the radial solutions which are valid for {\it arbitrary} radius.

For the reader who is not interested in the details and who just wants to use our results in order to obtain the late-time behaviour of the Green function or a field perturbation to high order, 
the main result is in Sec.\ref{sec:late-time GF}.
In particular, in Eq.(\ref{eq:late-time GF}) we give the late-time behaviour of the $\ell$-modes of the GF of the RW equation; above that equation
we indicate how to find the coefficients appearing in the equation; below that equation we indicate how to obtain a similar  late-time expansion in the BPT case. 
Of course, if one wants the late-time behaviour of some given initial data for the field perturbation, one should convolute the obtained small-frequency expansions for the GF with the initial data as in 
Eq.(\ref{eq:perturbation}).

The layout of this paper is as follows.
In Sec.\ref{sec:BC} we present the RW and BPT equations, expressions for their GFs and the BC contributions, as well as the relationships between the RW and BPT quantities.
In Sec.\ref{sec:MST gral s} we develop the MST formalism both for the RW and BPT equations, including our new derivation in the specific case of the RW equation for spin-1.
In Secs.\ref{sec:Tail an and nu} and \ref{sec:Tail} we give explicit expansions of the various MST quantities,
 except for the radial functions, up to the first four leading orders for small frequency.
In Sec.\ref{sec:radial} we  extract the small-frequency expansions  for the radial functions using  a novel method.
In Sec.\ref{sec:plot perturbation} we calculate the late-time behaviour of the GF (Sec.\ref{sec:late-time GF}) and of
a perturbation response (Sec.\ref{sec:pert}) and compare it with  highly-accurate numerical results.
In App.\ref{sec:plots BC} we plot the small-frequency expansions of MST quantities and check that they match with an independent method (presented in~\cite{Casals:2012ng}) which is valid in a `mid-frequency' regime.
In App.\ref{sec:App radial coeffs} we relate the radial coefficients of the solutions of the RW and BPT equations.

In this paper we use geometrized units:
$c=G=1$. 
We shall use a bar over a quantity to indicate that the
quantity has been made dimensionless via an appropriate
factor of `$2M$' (except where otherwise indicated),
where $M$ is the black hole mass; e.g., $\ob\equiv 2M\omega$ indicates a dimensionless frequency $\omega$ and 
$\bar r\equiv r/(2M)$ a dimensionless Schwarzschild radius $r$.

%


\section{Branch cut Green function} \label{sec:BC}

\subsection{Regge-Wheeler equation}

Decoupled and separated equations for linear field perturbations of a Schwarzschild black hole space-time were derived
for axial -- also called `odd' -- gravitational perturbations (spin $s=2$) in~\cite{Regge:1957td},
 for electromagnetic perturbations ($s=1$) in~\cite{Wheeler:1955zz,ruffini1972electromagnetic},
 and for  scalar perturbations ($s=0$) in~\cite{Price:1971fb,Price:1972pw}. 
 All these integer-spin-field perturbation equations can be written compactly as one single partial differential equation, which
 in Schwarzschild coordinates reads:
 \begin{equation}
\label{eq:covRWeqn}
\left[ \nabla^{\alpha}\nabla_{\alpha}
 + s^2\frac{2M}{r^3}
\right]\RW
 =\RWsource,
\end{equation}
where  $\RW=\RW(t,r,\theta,\phi)$ is a scalar function that describes the field perturbation of spin $s$
created by a source $\RWsource$
 and
\begin{equation}
\nabla^{\alpha}\nabla_{\alpha}=-\frac{r^2}{\Delta} \frac{\partial^2 }{\partial t^2} + 
 \frac{1}{r^2} \frac{\partial \ }{\partial r}  \left(\Delta\frac{\partial}{\partial r}\right)+ 
 \frac{1}{r^2 \sin\theta}
   \frac{\partial \ }{\partial \theta}  \left(\sin \theta \frac{\partial }{\partial \theta}\right)  + 
 \frac{1}{r^2 \sin^2\theta}\frac{\partial  }{\partial \phi^2} 
\end{equation}
is the Klein-Gordon operator in Schwarzschild space-time, where  $\Delta\equiv r(r-2M)$ and  $M$ is the mass of the black hole.
Separating variables we may obtain a complete set of solutions of the form
\begin{align} \label{eq: sep vars X}
\RWind (t,r,\theta,\phi) =  e^{-i \omega t} Y_{\ell m}(\theta, \phi) \frac{\RRWind(r,\omega)}{r} ,
\end{align}
 where $\ell$ and $m$ are, respectively, the multipolar and azimuthal numbers and
 $Y_{\ell m}(\theta, \phi)$ are the scalar (spin-weight $0$) spherical harmonics. 
 As in~\cite{Nakano:2000ne}, 
 we treat Eq.(\ref{eq:covRWeqn}) as a {\it scalar} wave equation, although in the electromagnetic and gravitational cases we have to be aware that the non-radiative ($\ell<|s|$)
 modes would drop out when constructing the electromagnetic and gravitational potentials and so these modes would have to be included separately.
The radial functions satisfy the ordinary differential equation
\begin{align} \label{eq:RW}
\left[   \frac{1}{r^2} \frac{\text{d} \ }{\text{d} r}\left( \Delta \frac{\text{d}}{\text{d} r}\right)
-2\frac{\Delta}{r^3} \frac{\text{d}}{\text{d} r}
+\frac{\omega ^2 r^2}{\Delta}- \left(\frac{\ell(\ell+1)}{r^2}+\frac{2 M
   \left(1-s^2\right)}{r^3}\right) \right] \RRWind =\RWsourceind,
 \end{align}
 where $\RWsourceind=\RWsourceind(r,\omega)$ are the corresponding modes of the source $\RWsource$.
We shall refer to Eq.(\ref{eq:RW}) as the (radial) Regge-Wheeler (RW) equation and to Eq.(\ref{eq:covRWeqn}) as the 4-dimensional RW equation (as per~\cite{Nakano:2000ne}).
Introducing the standard `tortoise' coordinate $r_*\equiv r + 2M \ln\bigl(r/(2M)-1\bigr)$, Eq.(\ref{eq:RW}) may also be written as  
\begin{align} \label{eq:RWrstar}
\left[    \frac{\text{d}^2 \ }{\text{d} r_*{}^2}
+\omega^2 - \frac{\Delta}{r^4}\left(\ell(\ell+1) +\frac{2 M
   \left(1-s^2\right)}{r}\right) \right]\RRWind=\frac{\Delta}{r^2}\RWsourceind  .
 \end{align}
The case of polar -- or `even' -- gravitational perturbations was derived in~\cite{PhysRevLett.24.737,PhysRevD.2.2141}; 
the corresponding radial equation is the so-called Zerilli equation. Solutions
of the Zerilli equation can be obtained as  linear combinations of solutions and their radial derivatives of the RW equation for $s=2$~\cite{Chandrasekhar}.

We define the retarded Green function (GF) of the 4-dimensional RW Eq.(\ref{eq:covRWeqn})  as the solution of~\cite{Nakano:2000ne}
\begin{equation}
\left[ \nabla^{\alpha}\nabla_{\alpha}
 + s^2\frac{2M}{r^3}
\right]\Gret(x,x')
 = - \delta(x,x')= - \frac{1}{r\, r'} \delta(t-t') \delta(r-r') \delta(\Omega-\Omega') ,
\end{equation}
that obeys appropriate causal boundary conditions, where $x$ and $x'$ are two space-time points.
Here, $\Omega$ is the solid angle of the 2-sphere.
For notational simplicity, we  use time translation invariance to henceforth take $t'=0$.
We may then use the symmetries of the space-time to write
\begin{align} 
 \label{eq:Greenprelim}
&
\Gret(x,x')= \frac{1}{ r\, r'} \int_{-\infty+ic}^{\infty+ic} \frac{d\omega}{2\pi} \sum_{\ell=0}^{\infty}\sum_{m=-\ell}^{\ell}
Y_{\ell m}(\theta,\phi)Y^*_{\ell m}(\theta',\phi')\ e^{-i\omega \dt} \Gl(r,r';\omega),
\end{align} 
where $c>0$,  and the Fourier modes $ \Gl(r,r';\omega)$ must satisfy
\begin{align} \label{eq:RWrstar GF}
\left\{    \frac{\text{d}^2 \ }{\text{d} r_*{}^2}
+\omega^2 - \frac{\Delta}{r^4}\left(\lambda +\frac{2 M
   \left(1-s^2\right)}{r}\right) \right\}\Gl(r,r';\omega) =-\frac{\Delta}{r^2} \delta(r-r') = - \delta(r_*-r_*'),
 \end{align}
 where\footnote{We note that our definition of $\lambda$ is slightly different from that in~\cite{Sasaki:2003xr}} $\lambda\equiv \ell (\ell+1)$.
Correspondingly,
\begin{align} \label{eq:Greenomega}
\Gl(r,r';\omega)=-\frac{\f(r_<,\omega)\g(r_>,\omega)}{\W{\omega}},
\end{align}
where 
$r_>\equiv \max(r,r')$, $r_<\equiv \min(r,r')$,
and  the Wronskian is given by
\begin{align} 
\W{\omega}\equiv 
W\left[\f(r,\omega),\g(r,\omega)\right]=
\f\frac{d\g}{dr_*}-\g\frac{d\f}{dr_*}.
\end{align}
Here, the  `ingoing' $\f=\f(r,\omega)$ and `upgoing' $\g=\g(r,\omega)$ radial functions are solutions of the
homogeneous version of the RW Eq.(\ref{eq:RWrstar}) which, for general  spin, behave asymptotically as 
\begin{align}
\label{eq: bc f}
\f\sim 
\begin{cases}
e^{-i\omega r_*},& \bar r_*\to -\infty, \\ 
\Ain e^{-i\omega r_*}+\Aout e^{+i\omega r_*},& \bar r_*\to +\infty,
\end{cases} 
\end{align}
where 
$\bar r_*\equiv r_*/(2M)$,
$\Ain$ and $\Aout$ are complex-valued constant coefficients (we give higher-order terms of Eq.(\ref{eq: bc f}) in App.\ref{sec:App radial coeffs}),
and 
\begin{align}
\label{eq: bc g}
\g\sim 
e^{+i\omega r_*}, \quad \bar r_*\to +\infty.
\end{align}
It is then easy to see that 
\footnote{Note that due to a typographical error the right hand side of Eq.(2.3) of Ref.~\cite{Casals:2012ng} should read
 $\f\g'-\f'\g$ instead of  $\g\f'-\f\g'$ so that it is indeed $W=2i\omega \Ain$ as claimed below Eq.2.7~\cite{Casals:2012ng}.}
 \begin{equation}\label{eq:Wronsk}
 W=2i\omega \Ain.
  \end{equation}
 
 It will be convenient for the next section to define similar ingoing and upgoing solutions of Eq.(\ref{eq:RWrstar}) without choosing a specific overall normalization:
\begin{align}
\label{eq: bc Xin}
\Xin{s}\sim
\begin{cases}
 \Xintra{s}e^{-i\omega r_*},&\!\!\! \bar r_*\to -\infty,\\
\Xininc{s} e^{-i\omega r_*}+\Xinref{s} e^{+i\omega r_*},&\!\!\! \bar r_*\to +\infty,
\end{cases} 
\end{align}
and
 \begin{align}
\label{eq: bc Xup}
\Xup{s}\sim
\begin{cases}
\Xupinc{s} e^{+i\omega r_*}+ \Xupref{s} e^{-i\omega r_*},&\!\!\! \bar r_*\to -\infty,\\
\Xuptra{s} e^{+i\omega r_*},&\!\!\! \bar r_*\to +\infty,
\end{cases} 
\end{align}
 where ${}_sX^{inc,in/ref/tra}_{\ell}$ and ${}_sX^{up,inc/ref/tra}_{\ell}$ are ingoing/reflection/transmission coefficients.
 It is clear that
 \begin{equation} \label{eq:rln X and f,g}
 \Xin{s}= \Xintra{s}\f, \quad  \Xup{s}= \Xuptra{s}\g.
 \end{equation}
 
Using the standard spherical harmonic addition theorem, we may now rewrite  Eq.~(\ref{eq:Greenprelim}) as
\begin{align} \label{eq:Green}
&
\Gret(x,x')=
\frac{1}{4 \pi r\, r'}
\sum_{\ell=0}^{\infty}(2\ell+1)P_{\ell}(\cos\gamma){}_sG^\text{ret}_{\ell}(r,r';\dt),
\end{align}
where
\begin{align}
\Glret(r,r';\dt)\equiv
\frac{1}{2\pi}
\int_{-\infty+ic}^{\infty+ic} d\omega\ {}_sG_{\ell}(r,r';\omega)e^{-i\omega \dt},
\end{align}
and
$\gamma$ is the angle between the spacetime points $x$ and $x'$.

 The upgoing radial solution $\g$, unlike $\f$, possesses a branch cut (BC) in the complex-frequency plane starting at the origin $\omega=0$ and extending down the negative imaginary 
 axis~\cite{Leaver:1986a,Leaver:1986,PhysRevLett.109.111101}.
This BC is inherited by $\Ain$ and by the Fourier modes $\Gl(r,r';\omega)$ of the GF.
 This BC in $\g$, however, only occurs as a change of sign in its imaginary part 
 as the frequency crosses the negative imaginary 
 axis, therefore
 $|\g|$ and $|\Ain|$ do not possess a BC.
We will use a new `frequency variable' $\fNIA>0$, so that when $\omega=-i\fNIA$ 
then it lies on the negative imaginary axis and when
 $\omega=+i\fNIA$ then it lies on the positive imaginary axis\footnote{We note that in~\cite{PhysRevLett.109.111101,Casals:2012ng,Casals:2011aa} (and in the BC literature references therein) we used a different symbol for
the frequency $\fNIA=i\omega$.
The symbol used there coincided with the symbol for the `renormalized angular momentum' parameter introduced later on that is used throughout the MST literature and
which we also use in this paper; hence the reason for the change of symbol to $\fNIA$.
}.
We define $\Disc A(\nb)\equiv  A_+(-i\nb)-A_-(-i\nb)$ for any function $A=A(\ob)$ possessing a BC along the NIA,
where $A_{\pm}(-i\nb)\equiv \lim_{ \epsilon\to 0^+}A(\pm \epsilon-i\nb)$, with $\nb>0$, where $\nb\equiv 2M\fNIA$ and $\ob\equiv 2M\omega$.

The contribution from the BC to the
$\ell$-mode ${}_sG^{ret}_{\ell}$ 
  is  given by
\begin{equation} \label{eq: G^BC integral}
\GlBC(r,r';\dt)\equiv
\frac{1}{2\pi i}
\int\limits_{0}^{\infty } d\fNIA\ 
\DGw(r,r';\fNIA)
e^{-\fNIA \dt}.
\end{equation}

Using the obvious symmetry $\g(r,-\omega^*)=\g^*(r,\omega)$ together with the boundary condition $\g(r,+i\fNIA)\sim e^{-\fNIA r_*}$ as $\bar r_*\to \infty$ and the fact that
all three functions $\gp(r,-i\fNIA)$, $\gm(r,-i\fNIA)$ and $\g(r,+i\fNIA)$  satisfy the same homogenous linear second-order differential equation  (namely, the RW equation) with real-valued coefficients along 
both the negative and the positive imaginary axes
(since $\omega=\pm i\fNIA$ with $\fNIA>0$ there),
it follows that 
\begin{equation}\label{eq:def q}
\Dg=i q(\fNIA)\g(r,+i\fNIA),
\end{equation}
 for some real-valued function $q(\fNIA)$, and that $W[\gm(r,-i\fNIA),\gp(r,-i\fNIA)]=-2i \fNIA q(\fNIA)$.
 We refer to  $q(\fNIA)$ as the `BC strength'.
The symmetry  $\f(r,-\omega^*)=\f^*(r,\omega)$ together with the fact that $\f$ has no BC means that $\f(r,-i\fNIA)$, with $\fNIA>0$, is a real-valued function.
Putting all these results together, we find that the discontinuity along the BC of the $\ell$-modes of the GF is given by

\begin{align} 
\label{eq:DeltaG in terms of Deltag}
&
\DGw(r,r';\fNIA)
\equiv
-2i\fNIA \frac{q(\fNIA)}{\left|W\right|^2}\f(r,-i\fNIA)\f(r',-i\fNIA),
\end{align}
which is a purely-imaginary quantity.
The reader may refer to~\cite{Leung:2003ix,Casals:2011aa} for details.

\subsection{Bardeen-Press-Teukolsky equation}

The Newman-Penrose formalism  offers an alternative
 way of describing spin-field perturbations of a Schwarzschild black hole background space-time 
to that provided by the RW formalism (one key advantage of the Newman-Penrose formalism, however, is that it generalizes to Kerr space-time).
The scaled Newman-Penrose scalars $\T=\T(t,r,\theta,\phi) $
obey the following equation~\cite{bardeen1973radiation,teukolsky1972rotating,Teukolsky:1973ha}:
\begin{equation} \label{eq:BPT}
\left\{ \nabla^{\alpha}\nabla_{\alpha}
 + \frac{2 s}{r^2}
 \left[\left(-r + \frac{Mr^2}{\Delta}\right)\frac{\partial}{\partial t} + (r - M) \frac{\partial  }{\partial r} 
+ i \frac{\cos\theta}{\sin^2\theta} \frac{\partial }{\partial \phi} + \frac{1 - s \cot^2\theta}{2} \right]
\right\}\T
 =\Tsource,
\end{equation}
where $\Tsource$ is the matter source term.
The scalings of the Newman-Penrose scalars here are given by
\begin{align*}
\Psi_s = \begin{cases} \psi_0 \ \text{or}\ r^{4}\psi_4 , & s=+2 \ \text{or} \ s=-2,\\
\phi_0\ \text{or}\ r^{2}\phi_2 , & s=+1 \ \text{or} \ s=-1,\\
\varphi , & s=0,
\end{cases}
\end{align*}
where $\psi_{0/4}$ are the radiative Weyl scalars, $\phi_{0/2}$ are the radiative Maxwell scalars and $\varphi$ is a massless scalar field.
We note that in the scalar case ($s=0$), Eq.(\ref{eq:BPT}) is the same as the 4-dimensional RW Eq.(\ref{eq:covRWeqn}).

As for the RW equation Eq.(\ref{eq: sep vars X}),
we may obtain a complete set of solutions of  Eq.(\ref{eq:BPT})  in the form
\begin{align}\label{eq:BPT sln}
\T(t,r,\theta,\phi) = 
e^{-i \omega t}  {}_{s}Y_{\ell m}(\theta, \phi) \RTind(r,\omega),
\end{align}
where ${}_{s}Y_{\ell m}(\theta, \phi)$ are the spin-weighted spherical harmonics~\cite{goldberg1967spin,newman1966note}.
  These are given explicitly by
${}_{0}Y_{\ell m}(\theta, \phi)=Y_{\ell m}(\theta, \phi)$,
\begin{align}
{}_{\pm1}Y_{\ell m}(\theta, \phi)&=\frac{1}{\sqrt{\ell (\ell+1)}}\left[\mp \partial_\theta - \dfrac{i}{\sin\theta} \partial_\phi \right]Y_{\ell m}(\theta, \phi),  & \ell\geq 1,\nonumber\\
{}_{\pm2}Y_{\ell m}(\theta, \phi)&=\frac{1}{\sqrt{(\ell-1)\ell (\ell+1)(\ell+2)}}\left[\partial_\theta^2-\cot\theta\partial_\theta \pm  \dfrac{2i}{\sin\theta} \partial_\theta \partial_\phi \mp  \dfrac{2i \cos\theta}{\sin^2\theta} \partial_\theta \partial_\phi- \dfrac{1}{\sin^2\theta} \partial_\phi^2 \right]Y_{\ell m}(\theta, \phi),  & \ell\geq 2.
\label{eq:scalar harms-SWSH}
\end{align}
The spin-weighted spherical harmonics satisfy the following relations:
\begin{itemize}
\item the conjugation relation ${}_{s}Y_{\ell m}^*(\theta, \phi)=(-1)^{m+s}{}_{-s}Y_{l(-m)}(\theta, \phi)$;
\item the orthonormality relation
\begin{align*}
\int \text{d}^2\Omega \>{}_{s}Y_{\ell m}^*(\theta, \phi) {}_{s}Y_{\ell' m'}(\theta, \phi) = \delta_{\ell\ell'}\delta_{mm'};
\end{align*}
\item the completeness relation
\begin{align*}
\sum\limits_{\ell=|s|}^\infty \sum\limits_{m=-\ell}^{\ell} {}_{s}Y_{\ell m}^*(\theta, \phi) {}_{s}Y_{\ell m}(\theta', \phi') = \delta(\cos\theta-\cos\theta')\delta(\phi-\phi');
\end{align*}
\item the parity relation ${}_{s}Y_{\ell m}(\pi-\theta, \pi+\phi)=(-1)^{\ell}{}_{-s}Y_{\ell(-m)}(\theta, \phi)$;
\item and the generalised addition relation
\begin{equation} 
\sum_m  \SWSH{s_1}(\theta,\varphi) \SWSH{s_2}^*(\theta',\varphi') = \sqrt{\frac{2l+1}{4\pi}} {}_{s_1}Y_{\ell (-s_2)}(\gamma,\alpha)e^{-i s_1 \beta}
\end{equation}
where
\begin{align*}
\alpha&=\tan ^{-1}\left(\cos \theta \sin \theta'-\sin\theta \cos \theta' \cos (\phi-\phi'),
-\sin \theta \sin (\phi - \phi')\right),\\
\beta&=\tan ^{-1}\left(\sin \theta \cos \theta'-\cos\theta \sin \theta' \cos (\phi-\phi'),
-\sin \theta' \sin (\phi - \phi')\right) .
\end{align*}
In particular, when $s_1=s_2=s$,
\begin{equation} 
\sum_m  \SWSH{s}(\theta,\varphi) \SWSH{s}^*(\theta',\varphi') = \frac{2l+1}{4\pi} (-1)^s\left(\frac{1+\cos\gamma}{2}\right)^{2s} P_{\ell-s}^{(0,2s)}(\cos\gamma)e^{-i s (\alpha+\beta)} ,
\end{equation}
where $P_{k}^{(a,b)}(x)$ are the Jacobi polynomials.
\end{itemize}

The radial part of the functions in Eq.(\ref{eq:BPT sln}) satisfies the ordinary differential equation
\begin{align} \label{eq:radial BPT}
\left[\Delta\frac{\text{d}^2 }{\text{d} r^2}+2(r-M)(s+1) \frac{\text{d} }{\text{d} r}
+ \left(\frac{\left(r^2\omega-2is(r-M)\right)r^2\omega}{\Delta}+4is\omega r+s(s+1)-\lambda
\right) \right]\RTind=r^2\,\Tsourceind, 
 \end{align}
 where  $\Tsourceind=\Tsourceind(r,\omega)$  are  the corresponding modes of the source $ \Tsource$.
We shall refer to Eq.(\ref{eq:radial BPT}) as the (radial) Bardeen-Press-Teukolsky (BPT) equation
and to Eq.(\ref{eq:BPT}) as the 4-dimensional BPT equation.
We may write the former in  self-adjoint form as
\begin{align} \label{eq:radial BPTv1}
\left[\frac{\text{d} }{\text{d} r}\Delta^{s+1} \frac{\text{d} }{\text{d} r}
+ \Delta^{s}\left(\frac{\left(r^2\omega-2is(r-M)\right)r^2\omega}{\Delta}+4is\omega r+s(s+1)-\lambda
\right) \right]\RTind=r^2\Delta^{s}\,\Tsourceind, 
 \end{align} 
or, writing $\RTind(r,\omega)=(\Delta^{-s/2} / r) \CT(r,\omega)$,
\begin{align} 
\label{eq:radial BPTv2}
\left[
\frac{\text{d}^2 \ }{\text{d} r_*^{\,2}}
+\omega ^2 +\frac{2 i s(r-3M)  \omega}{r^2} -\lambda\frac{\Delta }{r^4}
-\frac{2  M\Delta}{r^5} -\frac{M^2 s^2}{r^4}
\right]\CT=\frac{\Delta^{s/2+1}}{r}\Tsourceind.
 \end{align} 

Let us denote by $\CTh$ a general homogeneous solution of  Eq.~(\ref{eq:radial BPTv2}).
Asymptotically, the homogeneous version of Eq.~(\ref{eq:radial BPTv2}) takes the form 
\begin{align*} 
&\left[\frac{\text{d}^2 \ }{\text{d} r_*^{\,2}}
+\left(\omega- i s \kappa \right)^2 
\right]\CTh \sim 0,& \bar r_*\to-\infty, \\
&\left[\frac{\text{d}^2 \ }{\text{d} r_*^{\,2}}
+\omega ^2 +\frac{2 i s \omega}{r} 
\right]\CTh\sim 0,&\bar r_*\to+\infty, \\
\end{align*} 
where $\kappa\equiv 1/(4M)$ and correspondingly the solutions behave (omitting dimensionful constant  factors) as linear combinations of
\begin{align*} 
\CTh&\sim 
e^{\pm i\left(\omega-i s \kappa \right)r_*} \sim e^{\pm s/2} \left(\frac{\Delta}{(2M)^2}\right)^{\pm s/2} e^{\pm i \omega r_*},&\bar r_*\to-\infty, \\
\CTh&\sim 
r^{\mp s}
e^{\pm i\omega r_*} ,& \bar r_*\to+\infty.
\end{align*} 
We may define `ingoing' solutions of the homogeneous version of Eq.~(\ref{eq:radial BPTv2}) by their asymptotic behaviour as 
\begin{align}
\label{eq: bc chiin}
\chiin\sim
\begin{cases}
\Rintra{s} \left(2M\right)^{1-s}\left(\dfrac{r}{2M}-1\right)^{-s/2}e^{-i\omega r_*},&\!\!\! \bar r_*\to -\infty,\\
\Rininc{s} r^{s}e^{-i\omega r_*}+\Rinref{s} r^{-s}e^{+i\omega r_*},&\!\!\! \bar r_*\to +\infty,
\end{cases} 
\end{align}
and the `upgoing' solutions as
\begin{align}
\label{eq: bc chiup}
\chiup\sim
\begin{cases}
\Rupinc{s}  \left(2M\right)^{1+s} \left(\dfrac{r}{2M}-1\right)^{s/2}e^{+i\omega r_*}+ \Rupref{s} \left(2M\right)^{1-s} \left(\dfrac{r}{2M}-1\right)^{-s/2}e^{-i\omega r_*},&\!\!\! \bar r_*\to -\infty,\\
\Ruptra{s} r^{-s}e^{+i\omega r_*},&\!\!\! \bar r_*\to +\infty,
\end{cases} 
\end{align}
for general BPT spin.
The corresponding `ingoing' and `upgoing'  solutions
\footnote{We keep a spin $s$ subindex in the solutions $\Rin{s}$ and $\Rup{s}$ of the BPT equation while we did not for the solutions $\f$ and $\g$ of the RW equation because of the explicit 
spin-dependence in the asymptotic Eqs.(\ref{eq: bc Rin}) and (\ref{eq: bc Rup}) for the former set of solutions (as opposed to Eqs.(\ref{eq: bc f}) and (\ref{eq: bc g}) for the latter set); this
leads to the explicit spin-dependence in Eq.(\ref{eq:def qT}).}
 of the homogeneous version of the  BPT Eq.(\ref{eq:radial BPT}) respectively behave asymptotically as 
\begin{align}
\label{eq: bc Rin}
\Rin{s}\sim
\begin{cases}
 \Rintra{s} \Delta^{-s}e^{-i\omega r_*},&\!\!\! \bar r_*\to -\infty,\\
\Rininc{s} r^{-1}e^{-i\omega r_*}+\Rinref{s} r^{-1-2s}e^{+i\omega r_*},&\!\!\! \bar r_*\to +\infty,
\end{cases} 
\end{align}
and 
\begin{align}
\label{eq: bc Rup}
\Rup{s}\sim
\begin{cases}
\Rupinc{s} e^{+i\omega r_*}+ \Rupref{s} \Delta^{-s}e^{-i\omega r_*},&\!\!\! \bar r_*\to -\infty,\\
\Ruptra{s} r^{-1-2s}e^{+i\omega r_*},&\!\!\! \bar r_*\to +\infty,
\end{cases} 
\end{align}
where
${}_sR_{\ell}^{in,inc/ref/tra}$ are the incidence/reflection/transmission
 coefficients of the ingoing radial BPT solution; similarly for 
 ${}_sR_{\ell}^{up,inc/ref/tra}$
  for the upgoing solution.
It is convenient to define the following  `ingoing' and  `upgoing'  solutions and coefficients with a hat on, which are normalized with respect to the corresponding transmission coefficients:
$\RinN{s}\equiv \Rin{s}/\Rintra{s}$, $\RupN{s}\equiv\Rup{s}/\Ruptra{s}$, $\RinincN{s}\equiv  \Rininc{s}/\Rintra{s}$.
Then  it is easy to see that 
\begin{equation} \label{eq:WT}
\WT(\omega) \equiv 
\WT\left[
\RinN{s},\RupN{s}
\right]\equiv\Delta^{s+1}
 \left(\RinN{s} \RupprimeN{s}-\RupN{s}\RinprimeN{s}\right)
=2i\omega
 \RinincN{s},
\end{equation}
where primes denote differentiation wrt $r$.
The quantity $\WT$ is a `generalized Wronskian' in the sense that it is independent of $r$.
Similarly to the RW case, the solution $\RupN{s}$ has a BC along the negative imaginary axis of the complex frequency plane, which is 
inherited by $\WT$, whereas $\RinN{s}$ has no BCs~\cite{Leaver:1986a}.

The  GF of the $4$-dimensional BPT equation is the solution of Eq.(\ref{eq:BPT}) with the source $\Tsource$ replaced by the distribution
$\delta(r-r')\delta(\cos\theta-\cos\theta')\delta(\phi-\phi')/\sqrt{-g}$.
It may be expressed as
\begin{align}
\label{eq:GF Teuk s=2}
&\GretT(x,x')=
\sum_{\ell,m}  \Delta^{s}(r')\GlT(r,r';\dt)\SWSH{s}(\theta,\varphi)\SWSH{s}^*(\theta',\varphi'),
\end{align}
where
\begin{align}
 &
\GlT(r,r';\dt)\equiv
\int_{-\infty+ic}^{\infty+ic} d\omega\ \GlwT(r,r';\omega)e^{-i\omega \dt},
\quad
\GlwT(r,r';\omega)\equiv 
- \frac{\RinN{s}(r_<,\omega)\RupN{s}(r_>,\omega)}{\WT(\omega)}.
\nonumber
\end{align}

In order to obtain an expression for the BC contribution $\GlBCT$  to the
$\ell$-mode $\GlT$ of the   GF of the $4$-dimensional BPT equation we proceed similarly to the previous subsection for the RW equation.
This contribution can be expressed as in Eq.(\ref{eq:DeltaG in terms of Deltag}) for the RW case, but with $\GlBC$  and $\DGw$ replaced by $\GlBCT$ and $\DGwT$, respectively.
We then note that
$\RupN{s}(r,-\omega^*)=\RupccN{s}(r,\omega)$ 
and that 
$\Delta^{-s}\RupN{-s}(r,+i\fNIA)$ goes like $e^{-\fNIA r_*}/r$   as $\rb\to \infty$ (neglecting a constant factor).
It can be shown that all three functions
$\RuppN{s}(r,-i\fNIA)$, $\RupmN{s}(r,-i\fNIA)$ and $\Delta^{-s}\RupN{-s}(r,+i\fNIA)$
satisfy the 
 same homogenous linear second-order differential equation (namely, the BPT equation) with real-valued coefficients along the imaginary-$\omega$ axis (since $\omega=\pm i\fNIA$ with $\fNIA>0$ there).
 The two former solutions have the same large-$\rb$ behaviour, which is linearly independent from that of the latter.
 Therefore, it follows that
 \begin{equation} \label{eq:def qT}
 \DRupRuptra{s}=i\qT(\fNIA)\Delta^{-s}\RupN{-s}(r,+i\fNIA),
 \end{equation}
  for some real-valued function $\qT(\fNIA)$, and
that 
$\WT\left[\RupmN{s},\RuppN{s}\right]=-2i \fNIA \qT(\fNIA)$. 
We emphasize that the `BC strength' function $ \qT$ is calculated using the `upgoing' radial function  along the positive imaginary axis with the {\it opposite} spin-sign, as 
opposed to the same spin-sign for calculating $q$ in the  RW case (see Eq.(\ref{eq:def q})).
The property 
$\RinN{s}(r,-\omega^*)=\RinccN{s}(r,\omega)$ 
together with the fact that 
$\RinN{s}$
has no BC 
means that 
$\RinN{s}(r,-i\fNIA)$, 
with $\fNIA>0$, is a real-valued function.
Putting all these results and properties together, we find that the discontinuity  along the BC of the $\ell$-modes of the $4$-dimensional BPT GF is given by
\begin{align} 
\label{eq:DeltaG in terms of Deltag,Teuk}
&
\DGwT(r,r';\fNIA)
=
-2i\fNIA \frac{\qT(\fNIA)}{\left|\WT\right|^2}
\RinN{s}(r,-i\fNIA)\RinN{s}(r',-i\fNIA). 
\end{align}
In the next subsection we relate the BPT quantities to the RW quantities and, in particular,
we present an alternative way (namely via the RW `BC strength' $q(\fNIA)$) of calculating the BPT discontinuity $\DGwT$.


\subsection{Relationship between RW and BPT quantities} \label{sec:transf RW-BPT}

The so-called Chandrasekhar transformation relates, in the homogeneous case, solutions of the RW equation
to solutions of the BPT equation.
We note that while the RW Eqs.(\ref{eq:covRWeqn}) and (\ref{eq:RW}) are symmetric under $s \leftrightarrow -s$,
the BPT Eqs.(\ref{eq:BPT}) and (\ref{eq:radial BPT}) are not.
We here write the Chandrasekhar transformation compactly for spin $s=0,-1,-2$ (see~\cite{chandrasekhar1975equations} for spin-2 and, e.g.~\cite{jensen1991renormalized}
for spin-1; there are similar
transformations in the case of positive spin -- which of course only changes the BPT equation, not the RW equation -- but we do not deal with these in this paper).

Let us generically denote by $\RWtr(t,r)$ a homogeneous solution of the 4-D RW Eq.(\ref{eq:covRWeqn}) after factorizing out the angle-dependence via scalar spherical harmonics;
similarly, we generically denote by $\Ttr(t,r)$ a homogeneous solution of the  4-D BPT Eq.(\ref{eq:BPT}) after factorizing out the angle-dependence via spin-weighted spherical harmonics.
The corresponding Chandrasekhar transformation is then~\cite{Nakano:2000ne}:
\begin{align} \label{eq:RW toTeuk}
\Ttr(t,r)&=
\OpRW(t,r) \RWtr(t,r),\quad 
\OpRW(t,r)\equiv \left(\frac{\Delta}{r}\right)^{|s|}\left(\partial_r-\frac{r^2}{\Delta}\partial_t\right)^{|s|}r^{|s|-1},
\end{align}
up to a normalization constant.
We already gave in Eq.(\ref{eq:scalar harms-SWSH})   the angular counterpart of the above transformation, i.e., the transformation from the angular 
factor in the 4-D RW $\ell$-modes
(namely, the scalar spherical harmonics) to the angular factor in the 4-D BPT $\ell$-modes (namely, the spin-weighted  spherical harmonics).
We note that if Eq.(\ref{eq:scalar harms-SWSH}) were naively applied to the modes $\ell<|s|$ it would yield the zero function.

It is useful to write  the Chandrasekhar transformation in the frequency domain explicitly for each spin $s=0$, $-1$  and $-2$ separately.
We now generically denote by $\RRW{s}$ and $\RT{s}$ homogeneous solutions to the (radial) RW Eq.(\ref{eq:RW})  and (radial) BPT Eq.(\ref{eq:radial BPT}), respectively.
Introducing the operators
\begin{align*}
\OpD \equiv \frac{\text{d} \  }{\text{d} r} + \frac{i \omega}{f(r)} , \qquad
\OpD^\dagger =\frac{\text{d} \  }{\text{d} r}- \frac{i \omega}{f(r)} ,
\end{align*}
where $f(r)\equiv \Delta/r^2$,
we may express the  BPT solutions in terms of the RW solutions as
\begin{align}\label{eq:Chandr RW to BPT}
\RT{0}
 &=\frac{\cz}{r}
 \RRW{0},
 \\ 
\RT{-1}
 &=\co (r-2M) \OpD
  \RRW{-1}
 = \co r \left[f(r)  
  \RRW{-1}'
  + i \omega 
   \RRW{-1}
   \right], \nonumber\\ 
\RT{-2}
 &=\ct  (r - 2 M)^2 \OpD{}^2\bigl(r 
  \RRW{-2}
 \bigr) \nonumber\\
& = \ct  
\left[ 
 2 ( i r^2 \omega +r -3 M ) r (f(r) 
  \RRW{-2}'
 + i \omega
  \RRW{-2}
 )+ ( \ell (\ell + 1) r-6 M ) f(r) 
  \RRW{-2}
  \right],\nonumber
\end{align}
where $\cz$, $\co$ and $\ct$ are constants of proportionality
and primes denote differentiation with respect to $r$.
Conversely, we have
\begin{align}\label{eq:Chandr BPT to RW}
 \RRW{0}
 &=\frac{1}{\cz} r \RT{0},
 \\ 
 \RRW{-1}
 &=\frac{1}{\lambda \co}r^2 \OpD^\dagger \left(\frac{\RT{-1}
 }{r} \right)  =\frac{1}{\lambda \co} \left[ r \RT{-1}'
  - \left(1 + \frac{i \omega r}{f(r)} \right)\RT{-1} 
  \right],\nonumber\\ 
 \RRW{-2}
 &=\frac{1}{\bigl((\ell-1)\ell(\ell+1)(\ell+2)-12 i M\omega\bigr) \ct}r^3 \OpD^{\dagger\,2} \left(\frac{\RT{-2}
 }{r^2} \right)\nonumber \\
&=\frac{1}{\bigl((\ell-1)\ell(\ell+1)(\ell+2)-12 i M\omega\bigr) \ct}\left[\frac{2 (3 M -r  - i \omega r^2 )}{r-2 M} \RT{-2}' 
+\right.\nonumber\\
&\qquad \qquad  \left. \frac{r^2 \left(\ell(\ell+1)-2 r^2
   \omega ^2+8 i \omega r +4\right) -2 M r \left(\ell(\ell+1) +9 i \omega r +10\right)+24 M^2}{r (r-2 M)^2} \RT{-2} 
   \right].\nonumber
\end{align}

We will now use the Chandrasekhar transformation in order to relate the BPT and RW Wronskians as well as the BPT and RW `BC strengths'.
The specific normalizations
(\ref{eq: bc f})  and (\ref{eq: bc g}) 
of the RW radial functions $\f$ and $\g$ yield specific normalizations for the corresponding
 BPT radial functions via the transformation Eq(\ref{eq:RW toTeuk}).
We will denote the BPT radial functions and coefficients with these specific normalizations with a tilde superscript, i.e.,
$\RinNfg{s}\equiv \OpRWind(r) \f $, $\RupNfg{s} \equiv \OpRWind(r) \g $ and $\RintraNfg{s}$, $\RuptraNfg{s}$, \dots are, respectively, the coefficients
$\Rintra{s}$, $\Ruptra{s}$, \dots\  of $\RinNfg{s}$ and $\RupNfg{s}$.
In App.\ref{sec:App radial coeffs} we use the Chandrasekhar transformation in order to find these BPT radial coefficients and Wronskian in terms of the 
RW radial coefficients and Wronskian.
We are here using the obvious notation of $\OpRWind(r)$ for $\OpRW(t,r)$ after  replacing `$\partial_t$' by `$-i\omega$' in it.

Let us now try to find an alternative expression to Eq.(\ref{eq:DeltaG in terms of Deltag,Teuk}) for the BPT $\DGwT$ in terms of the RW $q(\fNIA)$.
We first re-express the BPT modes $\GlwT$ of Eq.(\ref{eq:GF Teuk s=2}) in terms of the RW solutions $\f$ and $\g$ by using the 
Chandrasekhar transformation Eq.(\ref{eq:RW toTeuk}):
\begin{equation}
\GlwT(r,r';\fNIA)=-\OpRWind(r_<)\OpRWind(r_>)\frac{\f(r_<,\omega)\g(r_>,\omega)}{\RintraNfg{s}\RuptraNfg{s}\WT}.
\end{equation}
Using this expression we can find the discontinuity of the BPT $\ell$-modes as
\begin{equation}\label{eq: DGwT via q}
\DGwT(r,r';\fNIA)=\frac{W}{\RintraNfg{s}\RuptraNfg{s}\WT}\OpRWind(r)\OpRWind(r')\DGw(r,r';\fNIA)=
-\frac{2i\fNIA q(\fNIA)\RintraNfg{s}}{\WT \cdot W^*\RuptraNfg{s}}
\RinN{s}(r,\omega)\RinN{s}(r',\omega). 
\end{equation} 
In deriving Eq.(\ref{eq: DGwT via q})
we have used the fact that $W/(\RintraNfg{s}\cdot \RuptraNfg{s}\WT)$ does not have a BC (as can be seen from 
the expressions in App.\ref{sec:App radial coeffs})
 and we have made
use of Eq.(\ref{eq:DeltaG in terms of Deltag}).

Comparing Eqs.(\ref{eq:DeltaG in terms of Deltag,Teuk}) and (\ref{eq: DGwT via q}) we immediately obtain a relationship between the RW and BPT `BC strengths':
\begin{equation} \label{eq:qT vs q}
\qT(\fNIA)=
\left(\frac{\WT}{W}\right)^*\cdot
\frac{\RintraNfg{s}}{\RuptraNfg{s}}q(\fNIA).
\end{equation}

Using the results in App.\ref{sec:App radial coeffs}, we have that
\begin{equation} \label{eq:qT vs q final}
\qT(\fNIA)=
q(\fNIA)
\begin{cases}
1 & s=0,\\
\displaystyle
-\frac{\alpha^*_{\infty}\alpha_+}{2\fNIA\alpha_+^*},
& s=-1, \\
\displaystyle
\frac{\beta_{\infty}^*(\alpha_++2M^2\beta_+)}{2\fNIA^2(\alpha^*_++2M^2\beta^*_+)},
 & s=-2,
\end{cases} 
\end{equation}
and
\begin{equation} \label{eq:qT/WT^2 vs q/W^2}
\frac{\qT}{|\WT|^2}=
\frac{\RintraNfg{s}}{\RuptraNfg{s}}\frac{W}{\WT}\frac{q}{|W|^2}
=
\frac{q}{|W|^2}
\begin{cases}
\frac{1}{4M^2} & s=0,\\
\displaystyle-\frac{\alpha_{+}^2}{2\fNIA\alpha_{\infty}}, & s=-1, \\
\displaystyle\frac{(\alpha_++2M^2\beta_+)^2}{2M^2\fNIA^2\beta_{\infty}},  & s=-2.
\end{cases} 
\end{equation}

Inverting the relationship and using  the results in App.\ref{sec:App radial coeffs}, we explicitly obtain:
\begin{equation} \label{eq:q/W^2 vs qT/WT^2}
\frac{q}{|W|^2}
=
\mathcal{C}\frac{\qT}{|\WT|^2}
\end{equation}
with 
\begin{align}\label{eq:C}
\mathcal{C}&=4M^2,  & s=0,\\
\mathcal{C}&=
-\frac{16 \lambda  M^4}{\left(\lambda -s^2+1\right)^2}\left(1-4 \nb+4 \nb^2\right), & s=-1,\nonumber\\
\mathcal{C}&=
\frac{256 M^6 (1-2 \nb)^2 (1-\nb)^2}{\lambda ^2-2 \lambda -6 \nb}, & s=-2.\nonumber
\end{align}
The small-frequency behaviour of $\mathcal{C}$ for $s=0$ and $-1$ is already manifest in the exact result above; for $s=-2$ we expand it as
\begin{equation}\label{eq:C spin-2}
\mathcal{C}=
\frac{256 M^6}{(\lambda -2) \lambda }
\left(1-\frac{6 \left(\lambda ^2-2 \lambda -1\right) \nb}{(\lambda -2) \lambda }+\frac{\left(13 \lambda ^4-52 \lambda ^3+16 \lambda ^2+72
   \lambda +36\right) \nb^2}{(\lambda -2)^2 \lambda ^2}\right)+O\left(\nb^3\right), \quad s=-2.
\end{equation}

From 
Eqs.(\ref{eq:DeltaG in terms of Deltag,Teuk}), (\ref{eq:q/W^2 vs qT/WT^2})
and the fact that the radial functions $\f$ and $\RinN{s}$ are generically of the same leading order (order zero) in $\nb$ as $\nb\to 0$ (see
Secs.\ref{sec:MST gral s}\ref{sec:radial}), it  follows that the RW and the BPT BC modes $\DGw$ and $\DGwT$ are of the same leading order in $\nb$ as $\nb\to 0$. 
As a consequence, the RW and BPT GFs are of the same leading order in $\bar t$ as $\bar t\to \infty$, as we explicitly see in Sec.\ref{sec:plot perturbation}.
That is, the RW and the BPT quantities describing black hole perturbations decay at the same rate at late times.


 \section{MST formalism for the RW and BPT equations for general spin } \label{sec:MST gral s}


 The MST method for the Teukolsky equation in Kerr (and, therefore, for the BPT equation in Schwarzschild)  was given in ST for general  spin,
 and earlier in \cite{Mano:Suzuki:Takasugi:1996} (henceforth MSTa) just for spin-2.
To the best of our knowledge, however, the MST method for the RW equation has only been given for spin-2, which was done in~\cite{Mano:1996mf} (henceforth MSTb).
Therefore, the MST method for the RW equation still has not been developed for spin-1 (obviously, the RW spin-0 case is essentially just the same as in ST with $s=0$).
In this section, we 
develop the MST method for the RW equation for general spin:
for spin-2 we recover MSTb,
for spin-0 we  essentially recover ST with $s=0$ and,
for spin-1, to the best of our knowledge, the results are new.
We also develop the MST method for the BPT equation which, although already existing in the literature, will allow us to emphasise the connections 
between the MST formalism for the RW and BPT equations for general spin $s=0$, $-1$ and $-2$.
In particular, we write the expansions for the  radial solutions of  both the RW and BPT equations in terms of just one set of `universal' coefficients (namely, $\anTnoG{n}$).
When referring to the literature, we shall use the generic term MST to refer to all MSTa, MSTb and ST.


We shall we use the notation of 
$\RRWinupN$ and $\RTinupN$
(i.e., with a slight change in the subindices 
with respect to the homogeneous RW solutions   $\Xinup{s}$ and the  BPT solutions  $\Rinup{s}$, 
  respectively)
 for the ingoing/upgoing solutions of the RW and BPT equations when using the specific normalization as in MST (i.e., the one in Eqs.(\ref{eq:coeff tra norm MST}) 
 and (\ref{eq:MST uptra}) below).
A similar change in the subindices notation applies to their  incidence, reflection and transmission coefficients.

\subsection{Series of hypergeometric functions }


In this section we shall assume that $s\leq 0$. For the RW functions we write
\begin{align}\label{eq:RW Xin}
\RRWinN = e^{i\ob}\rb^{s+1} e^{-i \omega r_*} \pRW_s^\text{in}(x)= \bar{r}^{s+1} e^{-i \eps (\bar{r}-1)} \left( \bar{r} -1 \right)^{-i \eps} \pRW_s^\text{in}(x)= (1-x)^{s+1} e^{i \eps x} \left( -x \right)^{-i \eps} \pRW_s^\text{in}(x),
\end{align}
where
$x \equiv1 -\bar r$.
This leads to the ordinary differential equation
\begin{align}\label{eq:eq for RW p}
&x(1-x) 
   \pRW_s^\text{in}{}''(x)+\bigl(1-2 i \eps(1-x)^2 -2 (1+s) x \bigr) \pRW_s^\text{in}{}'(x)\\
&\qquad\qquad + \bigl((\ell-s)(\ell+s+1) +2 i \eps(1+s)(1-x) \bigr)\pRW_s^\text{in}(x) =0 .
\nonumber
\end{align}

Correspondingly, for the BPT functions we write (as in ST)
\begin{align*}
\RTinN =  
e^{i\ob}\left( \bar{r}  -1 \right)^{-s} e^{-i \omega r_* }  \pT_s^\text{in}(x)=  e^{-i \eps ( \bar{r}-1) } \left( \bar{r} -1 \right)^{-s- \eps i} \pT_s^\text{in}(x) =  e^{i \eps x} \left( -x \right)^{-s- \eps i} \pT_s^\text{in}(x),
\end{align*}
which leads to the equation
\begin{align}\label{eq:eq for BPT p}
&x(1-x) 
   \pT_s^\text{in}{}''(x)+\bigl(4 i M \omega x (1-x) + 2 (2 i M \omega -1) x +
  (1-s-4 i M \omega) \bigr) \pT_s^\text{in}{}'(x)\\
&\qquad\qquad + \bigl(\ell(\ell+1)+4 i M \omega(1-s)(1-x)\bigr)\pT_s^\text{in}(x) =0 .
\nonumber
\end{align}

In terms of $\pRW_s^\text{in}(x)$ and $\pT_s^\text{in}(x)$ the Chandrasekhar transformations take the form 
\begin{align*}
\pT_0^\text{in}(x) &=\cz \pRW_0^\text{in}(x),\\ 
\pT_{-1}^\text{in}(x)&=\co \pRW_{-1}^\text{in}{}'(x),\\ 
\pT_{-2}^\text{in}(x) &=\ct  \pRW_{-2}^\text{in}{}''(x),
\end{align*}
and conversely
\begin{align*}
\pRW_0^\text{in}(x) &=\frac{1}{\cz} \pT_0^\text{in}(x),\\ 
\pRW_{-1}^\text{in}(x)&=\frac{1}{\ell(\ell+1)\co} \left[ x(1-x)  \pT_{-1}^\text{in}{}'(x) + \left(1 -2 i \eps (1-x)^2 \right)\pT_{-1}^\text{in}(x) \right],\\ 
\pRW_{-2}^\text{in}(x)&=\frac{1}{\bigl((\ell-1)\ell(\ell+1)(\ell+2)-6 i \eps\bigr) \ct}\left[x(1-x)  \left(2 x+1-2 i \eps (1-x)^2 \right) \pT_{-2}^\text{in}{}'(x)\right.\\
&\qquad \left.-\left(\ell(\ell+1) x (1-x)+4 \eps^2 (1-x)^4 +6 i \eps (x+1) (1-x)^2 
    -2 (2 x+1)\right)\pT_{-2}^\text{in}(x)\right].
\end{align*}

We now follow  MST and introduce the expansion in terms of hypergeometric functions,
\begin{subequations}
\begin{align}
\label{eq:MSTRWin}
\pRW_s^\text{in}(x) &=
N_s^\text{in}
 \sum_{n=-\infty}^{\infty} \anRWnoG{n} \frac{\Gamma (-n-\RAM+s-i \eps ) \Gamma (n+\RAM +s+1- i
   \eps )}{\Gamma (1-2 i \eps )} \times\nonumber \\
&\qquad\qquad\qquad\qquad  \, {}_2F_1(-n-\RAM +s- i  \eps ,n+\RAM +s+1-
   i \eps ;1-2 i \eps ;x),\\
\label{eq:MSTTin}
\pT_s^\text{in}(x) &= 
N_s^\text{in}
\sum_{n=-\infty}^{\infty} \anTnoG{n} \frac{\Gamma (-n-\RAM - i \eps ) \Gamma (n+\RAM+1 - i 
   \eps)}{\Gamma (1-s-2 i \eps)} 
\times\nonumber  \\
&\qquad\qquad\qquad\qquad  \, _2F_1(-n-\RAM - i \eps , n+\RAM +1-i \eps;1-s-2 i \eps;x),
\end{align}
\end{subequations}
where $N_s^\text{in}$ is a normalization constant which we specify later on.
Here, the parameter $\RAM$ is referred to as the renormalized angular momentum and is determined by the 
requirement that these series converge both as $n\to\infty $ and $n\to -\infty$, it has the property that 
either $\RAM=\ell+O(\ob)$ or $\RAM=-\ell-1+O(\ob)$; see 
MSTa or ST
for a full discussion.
These series representations in terms of hypergeometric functions converge $\forall r\in [2M,\infty)$.

In these terms, the Chandrasekhar transformations follow from the standard identity (Eq.15.5.1~\cite{NIST:DLMF})
\begin{align*}
\frac{\text{d}\ }{\text{d}x} \,
   _2F_1(a,b;c;x)=\frac{ab  }{c}\,
   _2F_1(a+1,b+1;c+1;x),
\end{align*}
for some parameters $a$, $b$ and $c$.
Inserting into their respective differential equations and using the standard hypergeometric function identities
\begin{align*}
&x \,
   _2F_1(a,b;c;x) =\frac{a (b-c) }{(a-b) (a-b+1)}\, _2F_1(a+1,b-1;c;x)+\\
&\qquad\qquad\qquad\frac{c(a+b-1) -2ab }{(a-b-1) (a-b+1)}\, _2F_1(a,b;c;x)+\frac{b  (a-c) }{(a-b-1) (a-b)}\, _2F_1(a-1,b+1;c;x),\\
&x(1-x) \frac{\text{d}\ }{\text{d}x} \,
   _2F_1(a,b;c;x)=\frac{ab (c-b) }{(a-b) (a-b+1)}\,
   _2F_1(a+1,b-1;c;x)+\\
&\qquad\qquad\qquad \frac{ab (2c - a-b-1) }{(a-b+1)(a-b-1)}\, _2F_1(a,b;c;x)+\frac{ab (a-c) }{(a-b) (a-b-1)}\, _2F_1(a-1,b+1;c;x),
\end{align*}
we find that $\anRWnoG{n}$ and $\anTnoG{n}$ must satisfy the same three-term recurrence relation:
\begin{align}
\label{eq:recursion}
\alpha_n \anTnoG{n+1} + \beta_n \anTnoG{n} + \gamma_n\anTnoG{n-1} =0, 
\end{align}
where
\begin{align*}
\alpha_n&=-\frac{ i \eps (n+\RAM +1- i \eps ) (n+\RAM +1+s- i \eps)(n+\RAM +1+s + i  \eps)(n+\RAM)}{ (2
   n+2\RAM+3) },\\
\beta_n&=-\lam(n+\RAM)(n+\RAM+1) + \bigl((n+\RAM)(n+\RAM+1)+\eps^2 \bigr)^2+s^2 \eps^2,\\
\gamma_n&=\frac{ i \eps  (n+\RAM+ i \eps) (n+\RAM-s+i \eps )
  (n+\RAM-s- i \eps)(n+\RAM +1)} {(2 n+2\RAM -1) } .
\end{align*}
The overall normalisation of the coefficients in the homogeneous Eq.(\ref{eq:recursion}) is, of course, irrelevant for the value of the $a_n$ 
but the above form has the advantage that all denominators are bounded away from 0 in the perturbative (small-frequency) regime.
By `perturbative regime' we essentially mean the frequency regime where $\RAM$ is real -- see the end of Sec.\ref{sec:Tail an and nu} for further details.
We choose the normalization $\anRWnoG{0}=\anTnoG{0}=1$, so then we have  $\anRWnoG{n}=\anTnoG{n}$, $\forall n\in \mathbb{Z}$, and from now on we will write down
all series using the `universal' set of coefficients $\anTnoG{n}$.
These coefficients $\anTnoG{n}$ are equal, for  $s=-2$, to the $a_n^{\nu}$ in MSTb as long
as  the same  normalization  is chosen for the two sets.

We could alternatively choose to include the $\Gamma$-functions in the coefficients, that is, write
\begin{align}
a_n^\text{RW} &\propto\anRWnoGNow{n} \frac{\Gamma (-n-\RAM+s-i \eps ) \Gamma (n+\RAM +s-i  \eps +1)}{\Gamma (1-2 i \eps )} , \\
a_n^\text{T} &\propto \anTnoG{n} \frac{\Gamma (-n-\RAM -i \eps ) \Gamma (n+\RAM -i \eps +1)}{\Gamma (1-s-2 i \eps)} ,
\end{align}
where the constants of proportionality are independent of $n$ and so just reflect the normalisation of the series.
A particularly convenient choice is
\begin{align}
\label{eq:aRW}
a_n^\text{RW} &\equiv\anRWnoGNow{n} (-\RAM+s-i \eps )_{-n}  (\RAM +s- i \eps +1)_n, \\
\label{eq:aT}
a_n^\text{T} &\equiv 
\anTnoG{n}  (-\RAM -i \eps )_{-n}  (\RAM -i \eps +1)_n
=\anTnoG{n}  \frac{(\RAM -i \eps +1)_n}{(\RAM +i \eps +1)_n  }(-1)^n .
\end{align}
As we choose the normalization $\anTnoG{0}=1$, we also have $a_0^\text{T} =a_0^\text{RW}=1$.
Using this convention the corresponding three-term recurrence relations have coefficients
\begin{alignat}{4}
\label{eq:coefRW}
\alpha_n^\text{RW} &{}=  -\frac{n+\RAM+1-s+i \eps}{n+\RAM+1+s- i \eps}\alpha_n,&\qquad   \beta_n^\text{RW} &{}=\beta_n,&\qquad \gamma_n^\text{RW} &{}=-\frac{n+\RAM+s-i \eps}{n+\RAM-s+i \eps}\gamma_n, \\
\label{eq:coefT}
\alpha_n^\text{T} &{}=  -\frac{n+\RAM+1+i \eps}{n+\RAM+1-i \eps}\alpha_n, &\qquad   \beta_n^\text{T} &{}=\beta_n,&\qquad \gamma_n^\text{T} &{}=-\frac{n+\RAM-i \eps}{n+\RAM+i \eps}\gamma_n.
\end{alignat}
Note that, in the perturbative regime, where $\nu$ is real, Eq.~(\ref{eq:coefT}) may be reexpressed as 
\begin{align*}
\alpha_n^\text{T} &{}=   \alpha_n^*,&\qquad   \beta_n^\text{T} &{}=\beta_n^*,&\qquad \gamma_n^\text{T} &{}=\gamma^*_n  .
\end{align*}
Up to irrelevant overall normalisation, the coefficients $\alpha_n^\text{T}$, $\beta_n^\text{T} $ and $\gamma_n^\text{T} $ and corresponding $a_n^T$ are the same as the corresponding quantities in Eq.123 ST.

 As the corresponding coefficients differ by a scaling  that tends to 1 for large $|n|$, $\RAM$ is the same for RW and for BPT.
In particular, $\RAM$ depends only on $|s|$ which can be seen directly since under the transformation $s\to -s$, $n+\RAM\to -n-\RAM-1$, $\beta_n$ is invariant while $\alpha_n$ and $\gamma_n$ are simply interchanged. 

As the event horizon of the Schwarzschild black hole is approached, we have
\begin{align*}
\RRWinN \sim
 e^{i\ob}e^{-i \omega r_*} \pRW_s^\text{in}(0)\qquad \text{and} \qquad  \RTinN \sim e^{i\ob} \left( \bar r -1 \right)^{-s} e^{-i \omega r_* } \pT_s^\text{in}(0),\quad  r\to 2M.
\end{align*}
That is, our solutions (\ref{eq:MSTRWin}) and (\ref{eq:MSTTin}) are normalised according to (see Eqs.(\ref{eq: bc Xin}) and (\ref{eq: bc Rin}))
\begin{subequations}
\label{eq:coeff tra norm MST}
\begin{align}
\XintraNST &= 
e^{i\ob}\pRW_s^\text{in}(0) = 
N_s^\text{in}e^{i\ob}
\sum_{n=-\infty}^{\infty}\anRWnoGNow{n} \frac{\Gamma (-n-\RAM+s-i \eps ) \Gamma (n+\RAM +s+1- i
   \eps )}{\Gamma (1-2 i \eps )}, \\
\RintraNST &=
e^{i\ob}(2M)^{2s}\pT_s^\text{in}(0) =
N_s^\text{in}
e^{i\ob}(2M)^{2s} \sum_{n=-\infty}^{\infty} \anTnoG{n} \frac{\Gamma (-n-\RAM - i \eps ) \Gamma (n+\RAM+1 - i 
   \eps)}{\Gamma (1-s-2 i \eps)}.
\end{align}
\end{subequations}
We note that the particular normalization choice,
\begin{equation}
N_s^\text{in}=\frac{\Gamma(1-s-2i\ob)}{\Gamma(-\RAM-i\ob)\Gamma(1+\RAM-i\ob)},
\end{equation}
yields the specific normalization used in ST for the ingoing BPT solutions, and so that is our choice henceforth.


\subsection{Series of Coulomb wave functions}

An alternative expansion is useful for the construction of the `up' solutions. In terms of the variable $z\equiv \omega r = \eps \bar{r}$, the RW equation may be written as 
\begin{align*}
\RRW{s}''(z) + \left(\frac{1}{z-\eps}-\frac{1}{z}\right)\RRW{s}'(z) +
 \left(1+ \frac{2\eps}{z-\eps}+\frac{\eps^2}{(z-\eps)^2}-\frac{\ell(\ell+1)}{z(z-\eps)}-\frac{(1-s^2)\eps}{z^2(z-\eps)}\right)\RRW{s}(z)=0.
\end{align*}
Writing $\RRW{s}(z)=\left(1-\frac{\eps}{z}\right)^{-i\eps}\fRW_s(z)$
 this becomes
\begin{align}
\label{eq:confluentRW}
z^2\fRW_s''+\left[z^2+2\eps z-\ell (\ell+1)\right]\fRW_s=
\eps\left[z(\fRW_s''+\fRW_s)-(1-2i\eps)\fRW_s'-\left(s^2-(1-i\eps)^2\right)\frac{1}{z}\fRW_s-\eps \fRW_s\right] .
\end{align}
The left hand side is the operator defining the Coulomb wave equation (Eq.33.14.1~\cite{NIST:DLMF}) with solution satisfying `up' boundary conditions at infinity given by
\begin{align*}
\hat{H}^+_\ell(-\eps,z)  = W_{i\eps,\ell+\frac{1}{2}}(-2 i z) = e^{i z} (-2i z)^{\ell+1} U(\ell+1-i\eps,2\ell+2,-2iz) ,
\end{align*}
where $\hat{H}^+_\ell(-\eps,z)$ denotes the  (unnormalised) irregular Coulomb function (Sec.33.2(iii)~\cite{NIST:DLMF}) and $W$ denotes the Whittaker function (Sec.13.14~\cite{NIST:DLMF}),  
We use a `hat' to denote that in writing the above we have dropped the conventional normalisation prefactor $e^{-i\pi \ell/2+i\sigma_\ell(\-\eps)} e^{-\pi\eps/2}$,
where $\sigma_\ell(-\eps)$ is the Coulomb phase shift, which is irrelevant to our current discussion. $U$ denotes the 
irregular confluent hypergeometric
function  (Sec.13.2~\cite{NIST:DLMF}, $\Psi$ in the notation of~\cite{GradRyz}).
Again, following Leaver~\cite{Leaver:1986a} and MST, this suggests that we introduce the expansion for the upgoing solution
\begin{align}
\fRW_s^\text{up}(z)&=
N^\text{up}_{s}\sum_{n=-\infty}^{\infty}\frac{\Gamma(n+\RAM+1+s-i\eps)\Gamma(n+\RAM+1-i\eps)}{\Gamma(n+\RAM+1-s+i\eps)\Gamma(n+\RAM+1+i\eps)} \anRWnoGNow{n}
\hat{H}^+_{n+\nu}(-\eps,z)\nonumber\\
&= N^\text{up}_{s}e^{iz}(-2 i z)^{\RAM+1}\!\!
\sum_{n=-\infty}^{\infty}\frac{\Gamma(n+\RAM+1+s-i\eps)\Gamma(n+\RAM+1-i\eps)}{\Gamma(n+\RAM+1-s+i\eps)\Gamma(n+\RAM+1+i\eps)}\anRWnoGNow{n}\times\nonumber\\
&\qquad\qquad\qquad\qquad\qquad\qquad(-2i z)^nU\left(n+\RAM+1-i\eps,2n+2\RAM+2,-2iz\right) ,
\end{align}
where $N^\text{up}_{s}$ is a normalisation constant that we will specify later so that our normalisation agrees with ST.
Inserting into Eq.(\ref{eq:confluentRW}), eliminating second derivatives using the differential equation satisfied by $\hat{H}^+_L(-\eta,z)$,   and noting the following identities which follow from standard properties of the 
 functions $U$:
\begin{align*}
&\frac{1}{z} \,
  \hat{H}^+_L(-\eta,z) =-i\frac{(L+1-i \eta) }{(L+1) (2L+1)}i 
   \hat{H}^+_{L+1}(-\eta,z)+\frac{\eta }{L(L+1)} \hat{H}^+_{L}(-\eta,z)+i\frac{(L+i \eta) }{L (2L+1)}
   i \hat{H}^+_{L-1}(-\eta,z),\\
& \frac{\text{d}\ }{\text{d}z} \,
   \hat{H}^+_L(-\eta,z) = i\frac{L(L+1-i \eta) }{(L+1) (2L+1)}i 
   \hat{H}^+_{L+1}(-\eta,z)+\frac{\eta }{L(L+1)} \hat{H}^+_{L}(-\eta,z)+\frac{(L+1)(L+i \eta) }{L (2L+1)}
   i \hat{H}^+_{L-1}(-\eta,z),
\end{align*}
we find that $\anRWnoGNow{n}$  must satisfy the same three-term recurrence relation Eq.(\ref{eq:recursion}).

Similarly, the BPT equation becomes 
\begin{align*}
\RT{s}''(z) + (1+s)\left(\frac{1}{z-\eps}+\frac{1}{z}\right)\RT{s}'(z) +
 \left(1+ \frac{2(\eps+is)}{z-\eps}+\frac{\eps(\eps-is) }{(z-\eps)^2}-\frac{\ell(\ell+1)-s(s+1)}{z(z-\eps)}\right)\RT{s}(z)=0.
\end{align*}
Writing $\RT{s}(z)=z^{-1-s}\left(1-\frac{\eps}{z}\right)^{-s-i\eps}\fT_s(z)$ this becomes
\begin{align}
\label{eq:confluent}
z^2\fT_s''+\left[z^2+2(\eps+is) z-\ell (\ell+1)\right]\fT_s=
\eps\left[z(\fT_s''+\fT_s)-(1-s-2i\eps)\fT_s'-(1-i\eps)(s-1+i\eps)\frac{1}{z}\fT_s+i(s+i\eps) \fT_s\right] .
\end{align}
The appropriate Coulomb function is now $\hat{H}^+_l(-\eps-is ,z)$ and the corresponding expansion for the upgoing solution is
\begin{align*}
\fT^\text{up}_{s}&=(-1)^sN^\text{up}_{s}
\sum_{n=-\infty}^{\infty}\frac{\Gamma(n+\RAM+1+s-i\eps)\Gamma(n+\RAM+1-i\eps)}{\Gamma(n+\RAM+1-s+i\eps)\Gamma(n+\RAM+1+i\eps)}a_{n}
\hat{H}^+_{n+\nu}(-\eps-is,z)\\
&= (-1)^sN^\text{up}_{s} e^{iz}(-2 i z)^{\RAM+1}\!\!
\sum_{n=-\infty}^{\infty}\frac{\Gamma(n+\RAM+1+s-i\eps)\Gamma(n+\RAM+1-i\eps)}{\Gamma(n+\RAM+1-s+i\eps)\Gamma(n+\RAM+1+i\eps)}a_{n}\times\nonumber\\
&\qquad\qquad\qquad\qquad\qquad\qquad
(-2i z)^nU\left(n+\RAM+1+s-i\eps,2n+2\RAM+2,-2iz\right).
\end{align*}

The Chandrasekhar transformations in this case follow term by term from the Whittaker function identities:
\begin{align*}
z \left(1-\frac{\eps}{z}\right) \bar{\mathcal{D}}_0 \left(\left(1-\frac{\eps}{z}\right)^{-i\eps}W_{i\eps,L+\frac{1}{2}}(-2 i z) \right) &=-\left(1-\frac{\eps}{z}\right)^{1-i\eps}W_{1+i\eps,L+\frac{1}{2}}(-2 i z), \\
z^2 \left(1-\frac{\eps}{z}\right)^2 \bar{\mathcal{D}}_0^2 \left(z \left(1-\frac{\eps}{z}\right)^{-i\eps}W_{i\eps,L+\frac{1}{2}}(-2 i z) \right)&=z\left(1-\frac{\eps}{z}\right)^{2-i\eps}W_{2+i\eps,L+\frac{1}{2}}(-2 i z),
\end{align*}
or equivalently
\begin{align*}
z \left(1-\frac{\eps}{z}\right) \bar{\mathcal{D}}_0 \left(\left(1-\frac{\eps}{z}\right)^{-i\eps}\hat{H}^+_{n+\nu}(-\eps,z)\right) &=-\left(1-\frac{\eps}{z}\right)^{1-i\eps}\hat{H}^+_{n+\nu}(-\eps+i,z), \\
z^2 \left(1-\frac{\eps}{z}\right)^2 \bar{\mathcal{D}}_0^2 \left(z \left(1-\frac{\eps}{z}\right)^{-i\eps}\hat{H}^+_{n+\nu}(-\eps,z)\right)&=z\left(1-\frac{\eps}{z}\right)^{2-i\eps}\hat{H}^+_{n+\nu}(-\eps+2i,z),
\end{align*}
where
\begin{align*}
\bar{ \mathcal{D}}_0 =  \eps \left(\frac{\text{d}}{\text{d}z} + i \left(1-\frac{\eps}{z}\right)^{-1}\right).
\end{align*}

Using the relationship between $a_n$ and $a^T_n$ we may write our solution in the alternate form
\begin{align}
\label{eq:STcompareup}
\fT^\text{up}_{s}&=(-1)^s \tfrac{1}{2} e^{-\pi\eps}e^{iz}(-2 z)^{\RAM+1}
\sum_{n=-\infty}^{\infty}\frac{(\RAM+1+s-i\eps)_n}{(\RAM+1-s+i\eps)_n}a^T_{n}
(2i z)^nU\left(n+\RAM+1+s-i\eps,2n+2\RAM+2,-2iz\right),
\end{align}
where we have made the choice 
\begin{align*}
N^\text{up}_{s}&=\tfrac{1}{2} e^{-\pi\eps}\frac{\Gamma(\RAM+1-s+i\eps)\Gamma(\RAM+1+i\eps)} {\Gamma(\RAM+1+s-i \eps)\Gamma(\RAM+1-i\eps) }e^{-i\frac{\pi}{2}(\RAM+1)} .
\end{align*}
so as to agree with the normalisation of ST. However our original form serves to highlight the boundary conditions and the link to the RW solution.

The corresponding upgoing BPT solution is given by
 \begin{equation}\label{eq:upgoing MST BPT}
\RTupN=z^{-1-s}\left(1-\frac{\eps}{z}\right)^{-s-i\eps}\fT^\text{up}_{s}(z).
 \end{equation}
 \fiximp{You had commented out this equation (which I need to refer to later on) - is it wrong?}

This series representation in terms of  irregular confluent hypergeometric functions
converges $\forall r> 2M$. Since $U(a,c,x) \sim x^{-a}$ as $|x|\to \infty$, it is straightforward to write down the asymptotic 
forms
\begin{align*}
\RRWupN&\sim (2i)^s \Ammu e^{i(z+\eps \ln z)},\\
\RTupN &\sim \Ammu z^{-1-2s} e^{i(z+\eps \ln z)},
\end{align*}
where
\begin{align}
\label{eq:A_-}
\Ammu\equiv  (2i)^{-s} e^{-\frac{\pi}{2}\eps} e^{-i\frac{\pi}{2}(\RAM+1)}  2^{-1+i\eps}
\sum_{n=-\infty}^{\infty}\frac{(\RAM+1+s-i\eps)_n(\RAM+1-i\eps)_n}{(\RAM+1-s+i\eps)_n(\RAM+1+i\eps)_n}
\anRWnoGNow{n}.
\end{align}
Clearly, then, from Eqs.(\ref{eq: bc Rup}),
\begin{align} \label{eq:MST uptra}
\XuptraNST &=(2i)^s \Ammu e^{i\ob \ln\ob}
, \\
\RuptraNST &=\Ammu \omega^{-1-2s}   e^{i\ob \ln\ob}.
\nonumber
\end{align}

To conclude this subsection, we note that we could have used the ansatz $\RRW{s}(z)=\left(1-\frac{\eps}{z}\right)^{i\eps}\gRW_s(z)$ 
for the RW equation, giving
\begin{align}
\label{eq:confluent}
z^2\gRW_s''+\left[z^2+2\eps z-\ell (\ell+1)\right]\gRW_s=
\eps\left[z(\gRW_s''+\gRW_s)-(1+2i\eps)\gRW_s'-\left(s^2-(1+i\eps)^2\right)\frac{1}{z}\gRW_s-\eps \gRW_s\right],
\end{align}
and the upgoing solution
\begin{align*}
\gRW_s^\text{up}(z)&=
\sum_{n=-\infty}^{\infty}\frac{\Gamma(n+\RAM-s+1-i\eps)\Gamma(n+\RAM+1+i\eps)}{\Gamma(n+\RAM+s+1+i\eps)\Gamma(n+\RAM+1-i\eps)}\anRWnoGNow{n}
\hat{H}^+_{n+\nu}(-\eps,z)\\
&= e^{iz}(-2 i z)^{\RAM+1}\!\!
\sum_{n=-\infty}^{\infty}\frac{\Gamma(n+\RAM-s+1-i\eps)\Gamma(n+\RAM+1+i\eps)}{\Gamma(n+\RAM+s+1+i\eps)\Gamma(n+\RAM+1-i\eps)}\anRWnoGNow{n}
(-2i z)^nU\left(n+\RAM+1-i\eps,2n+2\RAM+2,-2iz\right) .
\end{align*}
While this expansion seems to more naturally capture the boundary conditions of the RW equation, the Chandrasekhar transformation
is less natural in terms of it, so we will discuss it no further.

\subsection{Relation between the two solutions}

Using the standard relation Eq.15.8.3~\cite{NIST:DLMF}, we may reexpress $\RRWinN$ in terms that are better suited for discussing its behaviour at radial infinity:
\begin{align*}
\RRWinN &=N_s^\text{in} (1-x)^{\RAM +1+i \eps } e^{i \eps x} \left( -x \right)^{-i \eps} \sum_{n=-\infty}^{\infty}  \frac{\Gamma (-n-\RAM+s-i \eps ) \Gamma (2n+2\RAM +1 )}{\Gamma (n+\RAM+1-s-i \eps )}\anRWnoGNow{n}\times\\
&\qquad\qquad\qquad\qquad  \, (1-x)^n{}_2F_1\left(-n-\RAM +s- i  \eps ,-n-\RAM -s-
   i \eps ;-2n-2\RAM  ;\frac{1}{1-x}\right) + \\
&\qquad\qquad+  N_s^\text{in}(1-x)^{-\RAM +i \eps } e^{i \eps x} \left( -x \right)^{-i \eps} \sum_{n=-\infty}^{\infty}  \frac{\Gamma (n+\RAM+1+s-i \eps ) \Gamma (-2n-2\RAM -1 )}{\Gamma (-n-\RAM-s-i \eps )}\anRWnoGNow{n} \times\\
&\qquad\qquad\qquad\qquad  \, (1-x)^{-n}{}_2F_1\left(n+\RAM+1 +s- i  \eps ,n+\RAM+1 -s-
   i \eps ;2n+2\RAM +2;\frac{1}{1-x}\right) .
\end{align*}
The second term can be obtained from the first by the substitution $n\to-n$, $\nu\to-\nu-1$ and correspondingly the terms are denoted by $X_0^\nu$ and 
$X_0^{-\nu-1}$ respectively, so 
\begin{equation}\label{eq:Xin X0nu}
\RRWinN =X_0^\nu+X_0^{-\nu-1}.
\end{equation}
 (Note that $N_s^\text{in}$ is invariant under this transformation.) It is easily checked that each of these terms is independently a solution of the RW equation and moreover are linearly independent (MSTb).
In terms of $z=\eps\bar{r}=\eps(1-x)$,
\begin{align}
X_0^\RAM(z) &=  N_s^\text{in} e^{i\ob} e^{-i z} \left(\frac{z}{\eps}-1 \right)^{-i \eps} \left(\frac{z}{\eps}\right)^{\RAM +i \eps }
\sum_{n=-\infty}^{\infty} \frac{\Gamma (-n-\RAM+s-i \eps ) \Gamma (2n+2\RAM +1 )}{\Gamma (n+\RAM+1-s-i \eps )}a_n  \times\nonumber\\
&\qquad\qquad\qquad\qquad  \, \left(\frac{z}{\eps}\right)^n{}_2F_1\left(-n-\RAM - i  \eps ,-n-\RAM -s-
   i \eps ;-2n-2\RAM  ;\frac{\eps}{z}\right).
\end{align}
In an identical fashion we can write 
\begin{equation}\label{eq:Rin R0nu}
\RTinN =R_0^\nu+R_0^{-\nu-1}
\end{equation}
   with 
\begin{align}\label{eq:R0}
R_0^\nu &= 
N_s^\text{in} e^{i\ob}
 e^{-i z} \left(\frac{z}{\eps}-1 \right)^{-s-i \eps}  \left(\frac{z}{\eps} \right)^{\RAM+i \eps}  \sum_{n=-\infty}^{\infty} \anTnoG{n} \frac{\Gamma (-n-\RAM - i \eps ) \Gamma (2n+2\RAM+1)}{\Gamma (n+\RAM+1-s- i \eps)} 
\times\nonumber\\
&\qquad\qquad\qquad\qquad  \,\left(\frac{z}{\eps}\right)^n {}_2F_1\left(-n-\RAM - i \eps , -n-\RAM -s-i \eps;-2n - 2\RAM;\frac{\eps}{z}\right) .
\end{align}
This series representation for $\RTinN$ converges $\forall r> 2M$.

To obtain solutions in terms of confluent hypergeometric functions suitable for discussing the behaviour at the horizon, we introduce the auxiliary solutions involving the
regular Coulomb wave function
\begin{align*}
\hat{F}_\ell(-\eps,z)  &= \frac{\Gamma(\ell+1+i\eps)}{\Gamma(2\ell+2)}M_{i\eps,\ell+\frac{1}{2}}(-2 i z) = \frac{\Gamma(\ell+1+i\eps)}{\Gamma(2\ell+2)}e^{i z} (-2i z)^{\ell+1} M(\ell+1-i\eps,2\ell+2,-2iz)\\
&= \frac{\Gamma(\ell+1+i\eps)}{\Gamma(2\ell+2)}e^{-i z} (-2i z)^{\ell+1} M(\ell+1+i\eps,2\ell+2,2iz) ,
\end{align*}
where we include the prefactor for reasons that will become clear when we consider the Chandrasekhar transformation below.
The function $\hat{F}_\ell(-\eta,z)$ then satisfies the following identities
\begin{align*}
&\frac{1}{z} \,
  \hat{F}_L(-\eta,z) =i\frac{(L+1-i \eta) }{(L+1) (2L+1)}
   \hat{F}_{L+1}(-\eta,z)+\frac{\eta }{L(L+1)} \hat{F}_{L}(-\eta,z)-i\frac{(L+i \eta) }{L (2L+1)}
   \hat{F}_{L-1}(-\eta,z),\\
& \frac{\text{d}\ }{\text{d}z} \,
   \hat{F}_L(-\eta,z) = -i\frac{L(L+1-i \eta) }{(L+1) (2L+1)}
   \hat{F}_{L+1}(-\eta,z)+\frac{\eta }{L(L+1)} \hat{F}_{L}(-\eta,z)-i\frac{(L+1)(L+i \eta) }{L (2L+1)}
   \hat{F}_{L-1}(-\eta,z).
\end{align*}

Proceeding as before we construct the corresponding solutions:
\begin{align} \label{eq:XCnu}
X_\text{C}^{\RAM}(z)
&= \NC{\RAM}e^{-iz}\left(1-\frac{\eps}{z}\right)^{-i\eps} (-2 i z)^{\RAM+1}\!\!
\sum_{n=-\infty}^{\infty}\anTnoG{n}\frac{\Gamma(n+\RAM+1+s-i\eps)\Gamma(n+\RAM+1-i\eps)}{\Gamma(n+\RAM+1-s+i\eps)\Gamma(2n+2\RAM+2)}\times\nonumber\\
&\qquad\qquad
(2i z)^nM\left(n+\RAM+1+i\eps,2n+2\RAM+2,2iz\right) .
\end{align}
and
\begin{align} \label{eq:RCnu}
R_\text{C}^{\RAM}
(z)
&=\NC{\RAM} e^{-iz}\left(1-\frac{\eps}{z}\right)^{-s-i\eps} z^{-1-s} (-2 i z)^{\RAM+1}\!\!
\sum_{n=-\infty}^{\infty}\anTnoG{n}\frac{\Gamma(n+\RAM+1+s-i\eps)\Gamma(n+\RAM+1-i\eps)}{\Gamma(n+\RAM+1+i\eps)\Gamma(2n+2\RAM+2)}\times\nonumber\\
&\qquad\qquad(2i z)^nM\left(n+\RAM+1-s+i\eps,2n+2\RAM+2,2iz\right) ,
\end{align}
where 
\begin{align}\label{eq:NsC}
\NC{\RAM} = \frac{\Gamma(\RAM+1-s+i \eps)\Gamma(\RAM+1+i \eps)}{2\Gamma(\RAM+1+s-i \eps)\Gamma(\RAM+1-i \eps)}e^{i\frac{\pi}{2}(\nu+1)},
\end{align}
for agreement with Eq.139 ST when $z>0$ \fixme{For $z<0$ it only agrees with Eq.139 ST if we take the branch $\arg z \in (-\pi,\pi]$. ST say that their results are only valid for $\omega>0$ but we used them for QNM and BC calculations? is it that their results are actually valid for $Re(\omega)>0$?}.
Our normalization for $X_\text{C}^{\RAM}$ follows from that of $R_\text{C}^{\RAM}$ via the  Chandrasekhar transformation\fixme{Check}, and so for $s=-2$
it does not  coincide with the normalization choice in MSTb.
Specifically, for $s=-2$, our $X_\text{C}^{\RAM}$ is equal to that in MSTb times
\begin{equation*}
\frac{\Gamma(\RAM+1-s+i\ob)\Gamma(\RAM+1+i\ob)}{\Gamma(\RAM-1-i\ob)\Gamma(\RAM+1-i\ob)}.
\end{equation*}
The Chandrasekhar transformation
when using these series representations
follows term by term from the Whittaker function identities:
\begin{align*}
z \left(1-\frac{\eps}{z}\right) \bar{\mathcal{D}}_0 \left(\left(1-\frac{\eps}{z}\right)^{-i\eps}M_{i\eps,L+\frac{1}{2}}(-2 i z) \right) &=(L+1+i\eps)\left(1-\frac{\eps}{z}\right)^{1-i\eps}M_{1+i\eps,L+\frac{1}{2}}(-2 i z), \\
z^2 \left(1-\frac{\eps}{z}\right)^2 \bar{\mathcal{D}}_0^2 \left(z \left(1-\frac{\eps}{z}\right)^{-i\eps}M_{i\eps,L+\frac{1}{2}}(-2 i z) \right)&=(L+1+i\eps)(L+2+i\eps)z\left(1-\frac{\eps}{z}\right)^{2-i\eps}M_{2+i\eps,L+\frac{1}{2}}(-2 i z),
\end{align*}
or equivalently
\begin{align*}
z \left(1-\frac{\eps}{z}\right) \bar{\mathcal{D}}_0 \left(\left(1-\frac{\eps}{z}\right)^{-i\eps}\hat{F}^+_{n+\nu}(-\eps,z)\right) &=\left(1-\frac{\eps}{z}\right)^{1-i\eps}\hat{F}^+_{n+\nu}(-\eps+i,z), \\
z^2 \left(1-\frac{\eps}{z}\right)^2 \bar{\mathcal{D}}_0^2 \left(z \left(1-\frac{\eps}{z}\right)^{-i\eps}\hat{F}^+_{n+\nu}(-\eps,z)\right)&=z\left(1-\frac{\eps}{z}\right)^{2-i\eps}\hat{F}^+_{n+\nu}(-\eps+2i,z),
\end{align*}
where
\begin{align*}
\bar{ \mathcal{D}}_0 =  \eps \left(\frac{\text{d}}{\text{d}z} + i \left(1-\frac{\eps}{z}\right)^{-1}\right).
\end{align*}
These solutions can be related to $\RRWupN$ and $\RTupN$ using the identity 
in Eq.6.7(7) Vol.1~\cite{Erdelyi:1953}
(valid for $b\notin\mathbb{Z}$):
\begin{align*}
M(a,b,2iz) =& 
\begin{cases} 
\dfrac{\Gamma(b)}{\Gamma(b-a)}e^{a \pi i} U(a,b,2iz) +  \dfrac{\Gamma(b)}{\Gamma(a)}e^{(a-b) \pi i}e^{2iz}U(b-a,b,-2iz),& \text{Re}(z)>0,\\
\dfrac{\Gamma(b)}{\Gamma(b-a)}e^{-a \pi i} U(a,b,2iz) +  \dfrac{\Gamma(b)}{\Gamma(a)}e^{-(a-b) \pi i}e^{2iz}U(b-a,b,-2iz),&\text{Re}(z)<0,
\end{cases}
\end{align*}
the first and second terms on each line yielding the incoming and outgoing wave solutions at infinity, respectively. Thus, $X_\text{C}^{\RAM}(z)=\XCp{\RAM}(z)+\XCm{\RAM}(z)$, where
\begin{align} \label{eq:XCnu+-}
\XCp{\RAM}(z)
&= \NC{\RAM} e^{\mp\pi \eps}e^{-iz}\left(1-\frac{\eps}{z}\right)^{-i\eps} (2 i z)^{\RAM+1}\!\!
\sum_{n=-\infty}^{\infty}\anTnoG{n}\frac{\Gamma(n+\RAM+1+s-i\eps)}{\Gamma(n+\RAM+1-s+i\eps)}\times\nonumber\\
&\qquad\qquad
(-2i z)^nU\left(n+\RAM+1+i\eps,2n+2\RAM+2,2iz\right) \\
\XCm{\RAM}(z)
&= \NC{\RAM}e^{\mp\pi \eps} e^{\mp 2(\RAM+1) \pi i} e^{iz}\left(1-\frac{\eps}{z}\right)^{-i\eps} (2 i z)^{\RAM+1}\!\!
\sum_{n=-\infty}^{\infty}\anTnoG{n}\frac{\Gamma(n+\RAM+1+s-i\eps)\Gamma(n+\RAM+1-i\eps)}{\Gamma(n+\RAM+1-s+i\eps)\Gamma(n+\RAM+1+i\eps)}\times\nonumber\\
&\qquad\qquad
(-2i z)^nU\left(n+\RAM+1-i\eps,2n+2\RAM+2,-2iz\right),
\nonumber
\end{align}
\fixme{A quick check indicates that $\XCp{\RAM}$ differs from $X_{C-}^{\RAM}$ in Eq.3.10 MSTb by a factor which differs in 
$e^{\pi i (\RAM+1)/2} e^{-i\pi(\RAM+1)}$ from the proportionality factor given above between our $X_\text{C}^{\RAM}$ and the $X_\text{C}^{\RAM}$ in MSTb.
Another quick check indicates that our (and ST's) $\KT_{\RAM}$ differs from that in MSTb by a factor different from the inverse of $e^{\pi i (\RAM+1)/2} e^{-i\pi(\RAM+1)}$?}
where the signs correspond to ${Re(z)>0}\atop{Re(z)<0}$.
Similarly, $R_\text{C}^{\RAM}(z)=\RCp{\RAM}(z)+\RCm{\RAM}(z)$, where
\begin{align} \label{eq:RCnu+-}
\RCp{\RAM}
(z)
&=(-1)^s \NC{\RAM} e^{\mp\pi \eps}  e^{-iz}\left(1-\frac{\eps}{z}\right)^{-s-i\eps} z^{-1-s} (2 i z)^{\RAM+1}\!\!
\sum_{n=-\infty}^{\infty}\anTnoG{n}\frac{\Gamma(n+\RAM+1-i\eps)}{\Gamma(n+\RAM+1+i\eps)}\times\nonumber\\
&\qquad\qquad(-2i z)^nU\left(n+\RAM+1-s+i\eps,2n+2\RAM+2,2iz\right) ,\\
\RCm{\RAM}
(z)
&=(-1)^s \NC{\RAM} e^{\mp\pi \eps}  e^{\mp 2(\RAM+1) \pi i} e^{iz}\left(1-\frac{\eps}{z}\right)^{-s-i\eps} z^{-1-s} (2 i z)^{\RAM+1}\!\!
\sum_{n=-\infty}^{\infty}\anTnoG{n}\frac{\Gamma(n+\RAM+1+s-i\eps)\Gamma(n+\RAM+1-i\eps)}{\Gamma(n+\RAM+1-s+i\eps)\Gamma(n+\RAM+1+i\eps)}\times\nonumber\\
&\qquad\qquad(-2i z)^nU\left(n+\RAM+1+s-i\eps,2n+2\RAM+2,-2iz\right).
\nonumber
\end{align}
Note that our naming convention here follows ST and is opposite in sign to MSTb.
In particular, the minus solutions are just multiples of the corresponding `up' solutions.


Critically as noted by MST,  
The functions $X_0^\RAM$ and $X_\text{C}^{\RAM}$ solve the same differential equation and have the same analytical behaviour as functions of $z$;
similarly for $R_0^\RAM$ and $R_\text{C}^{\RAM}$.
Therefore, they must be proportional:
\begin{align}\label{eq:sR0nu}
X_0^\RAM=\KT_{\RAM}X_\text{C}^{\RAM}, \qquad \text{and} \qquad
R_0^\RAM=\KT_{\RAM}R_\text{C}^{\RAM},
\end{align}
where $\KT_{\RAM}$ is the constant of proportionality.
Equating the corresponding Laurent series we can obtain explicit expressions for  $\KT_{\RAM}$
in terms of $\nu$ and our $a_n$ coefficients.  The results are given for general spin by ST (based on their Teukolsky equation analysis),  we repeat them here for completeness specialised to Schwarzschild  space-time:
\begin{align}\label{eq:Knu}
&\KT_{\RAM} =
\frac{e^{i\ob}(2\ob  )^{s-\nu-r}2^{-s}i^{r}
\Gamma(1-s-2i\ob)\Gamma(r+2\nu+2)}
{\Gamma(r+\nu+1-s+i\ob)
\Gamma(r+\nu+1+i\ob)\Gamma(r+\nu+1+s+i\ob)}
\nonumber\\
&\times \left ( \sum_{n=r}^{\infty}
\frac{\Gamma(n+r+2\nu+1)}{(n-r)!}
\frac{\Gamma(n+\nu+1+s+i\ob)}{\Gamma(n+\nu+1-s-i\ob)}
\frac{\Gamma(\nu+1+i\ob)}{\Gamma(\nu+1-i\ob)}
\,\anTnoG{n}\right)
\nonumber\\
&\times \left(\sum_{n=-\infty}^{r}
\frac{1}{(r-n)!
(r+2\nu+2)_n}\frac{(\nu+1+s-i\ob)_n}{(\nu+1-s+i\ob)_n}
\frac{(\nu+1-i\ob)_n}{(\nu+1+i\ob)_n}
\anTnoG{n}\right)^{-1}.
\end{align}
The parameter $r$ here is an arbitrary integer number.

The above expressions lay out the foundations for taking the $\bar r\to\infty$ limit of the ingoing solutions and thus finding their incidence and reflection coefficients.
First we relate the quantities at $-\RAM-1$ to those at $\RAM$. 
From Eq.(\ref{eq:NsC}) we have
\begin{equation}
\NC{-\RAM-1}= \frac{\sin\left(\pi(\nu+i\eps)\right)\sin\left(\pi(\nu+i\eps-s)\right)}{\sin\left(\pi(\nu-i\eps)\right)\sin\left(\pi(\nu-i\eps+s)\right)}             e^{-i \pi \nu}         e^{-i\frac{\pi}{2}} \NC{\RAM}.
\end{equation}
Then, from Eqs.(\ref{eq:XCnu+-}) it follows that
\begin{equation}
\XCp{-\RAM-1}= \frac{\sin\left(\pi(\nu+i\eps)\right)}{\sin\left(\pi(\nu-i\eps)\right)}             e^{-i \pi \nu}         e^{-i\frac{\pi}{2}} \XCp{\RAM}
\end{equation}
and
\begin{equation}
\XCm{-\RAM-1}= 
e^{\pm 2 i \pi \nu}e^{- i \pi \nu}
e^{i\frac{\pi}{2}} 
\XCm{\RAM},
\end{equation}
where the upper/lower sign corresponds to $\text{Re}(z)$ positive/negative.
\fixme{The above agree with  Eqs.3.8 and 4.3 MSTb for $Re(z)>0$ but the result for $\XCm{-\RAM-1}$ does not agree with Eqs.3.8 and 4.3 MSTb for
$Re(z)<0$? ST say that their results are only valid for $\omega>0$ but MSTb do not seem to say that}
\fiximp{In fact, it seems that the above should not agree with  Eqs.3.8 and 4.3 MSTb, because the proportionality factor indicated in fixme above between
our $\XCp{\RAM}$ and MSTb's $X_{C-}^{\RAM}$ is not invariant under $\RAM \to -\RAM-1$?}

From Eqs.(\ref{eq:Xin X0nu}) and (\ref{eq:XCnu+-}) and 13.7.3~\cite{NIST:DLMF} we can take the $z\to \infty$ limit and obtain  the incidence and reflection coefficients of the ingoing RW solution:
\begin{equation}\label{eq:XinincNST}
\XinincNST=
\left(\KT_{\RAM} -ie^{-i\pi \RAM}\frac{\sin\left(\pi(\RAM+i\ob)\right)}{\sin\left(\pi(\RAM-i\ob)\right)}\KT_{-\RAM-1}\right)\ApmuRW e^{-i\ob\ln\ob}
\end{equation}
and
\begin{equation}\label{eq:XinrefNST}
\XinrefNST=
\left(\KT_{\RAM} +
ie^{i\pi \RAM}
 \KT_{-\RAM-1}\right)(2i)^s\Ammu e^{i\ob\ln\ob}
\end{equation}
\fiximp{In the above I've assumed $z>0$}
where
\begin{align}
\label{eq:A_+}
\ApmuRW\equiv 
e^{-\pi \ob/2}e^{i\frac{\pi}{2}(\nu+1)}\ 2^{-1-i\ob}  \frac{\Gamma(\RAM+1-s+i \eps)\Gamma(\RAM+i \eps+1)}{\Gamma(\RAM+1+s-i \eps)\Gamma(\RAM-i \eps+1)} 
 \sum_{n=-\infty}^{\infty}(-1)^n \frac{\Gamma(n+\RAM -i \eps +1+s)}{\Gamma(n+\RAM +i \eps +1-s) } \anTnoG{n} .
\end{align}

We can proceed similarly for the BPT solution.
By taking $z\to \infty$ in Eq.(\ref{eq:Rin R0nu})
we
obtain the incidence and reflection coefficients of the ingoing BPT solution:
\begin{equation}\label{eq:RinincNST}
\RinincNST=
\omega^{-1}\left(\KT_{\RAM} -ie^{-i\pi \RAM}\frac{\sin\left(\pi(\RAM-s+i\ob)\right)}{\sin\left(\pi(\RAM+s-i\ob)\right)}\KT_{-\RAM-1}\right)\Apmu e^{-i\ob\ln\ob}
\end{equation}
and
\begin{equation}\label{eq:RinrefNST}
\RinrefNST=
\omega^{-1-2s}\left(\KT_{\RAM} +ie^{i\pi \RAM}\KT_{-\RAM-1}\right)\Ammu e^{i\ob\ln\ob},
\end{equation}
where
\begin{align}
\label{eq:A_+}
\Apmu\equiv   e^{-\pi \ob/2} e^{i\frac{\pi}{2}(\RAM+1-s)}  2^{-1+s-i\eps}
 \frac{\Gamma(\RAM+1-s+i\ob)\Gamma(\RAM +i \eps +1)}{\Gamma(\RAM+1+s-i\ob)\Gamma(\RAM -i \eps +1)}\sum_{n=-\infty}^{\infty}(-1)^n \frac{\Gamma(n+\RAM -i \eps +1)}{\Gamma(n+\RAM +i \eps +1)}
 \anTnoG{n} .
\end{align}
We note that one can obtain the radial incidence, reflection and transmission coefficients of the RW solution from those of the BPT solution (or viceversa)
via the Chandrasekhar transformation Eq.(\ref{eq:Chandr BPT to RW}) (or Eq.(\ref{eq:Chandr RW to BPT})).
In doing so, the leading order for large-$r$ would be annihilated and one would require a higher order term.

\section{Low-frequency Expansion of the Coefficients and of $\RAM$}\label{sec:Tail an and nu}

In this Section our goal is to provide the low-frequency behaviour of the MST series renormalised angular momentum  $\nu$
and series coefficients $a_n$. 
We will provide the expansions explicitly up to the first five leading orders.
The behaviour of the coefficients $a_n^\text{RW}$ and $a_n^\text{T} $ may  be deduced immediately from Eqs.(\ref{eq:aRW}) and~(\ref{eq:aT}).

We start by noting that Eqs.(\ref{eq:eq for RW p}) and (\ref{eq:eq for BPT p}) reduce to the  hypergeometric equation when $\eps=0$, indeed it was precisely for this reason that Leaver~\cite{Leaver:1986a} and MST wrote them in this way.
For the `in' RW and BPT solutions we want the regular solutions corresponding to, respectively,
\begin{align*}
 {}_2F_{1}(\ell+1+s,-\ell+s,1;x)={}_2F_{1}(-\ell+s,\ell+1+s,1;x)
\end{align*}
and
\begin{align*}
 {}_2F_{1}(\ell+1,-\ell,1-s;x)={}_2F_{1}(-\ell,\ell+1,1-s;x).
\end{align*}
The left hand sides of these expressions correspond to $\RAM=\ell$ and the right hand sides to $\RAM=-\ell-1$ when $\eps=0$. 
In fact, under $\RAM \to -\RAM-1$, $\alpha_n$ equals $\gamma_{-n}$ and $\beta_n$ equals $\beta_{-n}$
and therefore, $a_n$ satisfies the same recurrence relation as $a_{-n}$ under $\RAM \to -\RAM-1$; the equivalent  symmetries hold
for the RW counterparts ($a_n^\text{RW} $, $\alpha_n^\text{RW} $, $\beta_n^\text{RW} $ and $\gamma_n^\text{RW}$) and
 for the BPT counterparts ($a_n^\text{T} $, $\alpha_n^\text{T} $, $\beta_n^\text{T} $ and $\gamma_n^\text{T} $).
This symmetry stems from the fact that the renormalized angular momentum $\RAM$ was introduced into the Ordinary Differential Equation Eq.(119) ST 
\fixme{would it not be good to write out explicitly Eq.119ST?} in the form
$\RAM(\RAM+1)$, which is invariant under  $\RAM \to -\RAM-1$.
  In addition, we may
determine the expansion of $\RAM$ about $\ell$ from that about `$-\ell-1$'.




With the natural ansatz that $a_n=O(\eps^{|n|})$, the 3-term recurrence relation Eq.(\ref{eq:recursion}) can be solved directly yielding:
\begin{equation} \label{eq:small-omega nu}
\nu+\tfrac{1}{2}=\left(\ell+\tfrac{1}{2}\right)\Bigl[1-\frac{\lambda  (15 \lambda -11)+3 s^4+6 (\lambda -1) s^2}{\lambda  (4 \lambda -3) (4 \lambda +1)}\eps ^2+\frac{P_{0,4}^{(8,4)}\bigl(\lambda,s^2\bigr)}{4 (\lambda -2) \lambda ^3 (4 \lambda -15) (4 \lambda -3)^3 (4 \lambda +1)^2} \eps ^4	 +O(\eps ^6)\Bigr]
\end{equation}
for the renormalized angular momentum.
Note that the expansion of $\nu$ must be even in $s$ in Schwarzschild space-time since it also arises through the expansion of the RW
equation which manifestly has this property\fixme{Not sure this justification is sufficient (same justification could apply to other RW quantities which don't have this symmetry)?}. 
As it will be needed later on, let us define $\RAMmLead$ as minus the coefficient of $\ob^2$ in the above expansion for $\nu$, ie,
\begin{align}\label{eq:2nd order nu}
\RAM= \ell - \RAMmLead \ob^2 + O(\ob^3).
\end{align}

For the series coefficients themselves, the above procedure yields
\begin{align} 
\label{eq:gen}
&a_4=\frac{ (\ell-s+1)^2 (\ell-s+2)^2 (\ell-s+3)^2 (\ell-s+4)^2}{96 (\ell+1) (\ell+2) (2 \ell+1) (2 \ell+3)^2 (2 \ell+5)^2 (2 \ell+7)}\eps ^4+O(\eps ^5),\\
&a_3=\frac{ (\ell-s+1)^2
   (\ell-s+2)^2 (\ell-s+3)^2}{24 (\ell+1) (\ell+2) (2 \ell+1) (2 \ell+3)^2 (2 \ell+5)}i\eps ^3-\frac{\left(3 \ell^2+12 \ell+11\right)  (\ell-s+1)^2 (\ell-s+2)^2 (\ell-s+3)^2}{24 (\ell+1)^2 (\ell+2)^2 (\ell+3) (2 \ell+1) (2 \ell+3)^2 (2 \ell+5)}\eps ^4+O(\eps ^5)\nonumber,\\
&a_2=-\frac{ (\ell-s+1)^2 (\ell-s+2)^2}{4 (\ell+1) (2 \ell+1) (2 \ell+3)^2}\eps ^2- \frac{ (\ell-s+1)^2 (\ell-s+2)^2}{4 (\ell+1)^2 (\ell+2) (2 \ell+1) (2 \ell+3)}i \eps ^3+\nonumber,\\
&\qquad\qquad\qquad\frac{
  P_{2,4}^{(15,8)}(\ell,s)}{48 \ell (\ell+1)^3 (\ell+2) (\ell+3) (2 \ell-1) (2 \ell+1)^3 (2 \ell+3)^4 (2 \ell+7)}\eps ^4+O(\eps ^5)\nonumber,\\
&a_1=-\frac{
   (\ell-s+1)^2  }{2 (\ell+1) (2 \ell+1)}i\eps+\frac{(\ell-s+1)^2 }{2 (\ell+1)^2 (2 \ell+1)}\eps ^2+\frac{P_{1,3}^{(12,6)}(\ell,s) }{8 \ell (\ell+1)^3 (\ell+2) (2 \ell-1) (2 \ell+1)^3 (2 \ell+3)^2 (2 \ell+5)}i\eps ^3\nonumber,\\
&\qquad\qquad\qquad+\frac{ P_{1,4}^{(12,6)}(\ell,s)}{8 \ell (\ell+1)^4
   (\ell+2) (2 \ell-1) (2 \ell+1)^3 (2 \ell+3)^2 (2 \ell+5)}\eps ^4+O(\eps ^5)\nonumber,\\
&a_{-1}=-\frac{(\ell+s)^2}{2 \ell (2 \ell+1)}i\eps -\frac{ (\ell+s)^2}{2 \ell^2 (2 \ell+1)}\eps ^2+\frac{P^{(12,6)}_{-1,3}(\ell,s)}{8 (\ell-1)  \ell^3 (\ell+1) (2\ell-3) (2 \ell-1)^2(2 \ell+1)^3 ( 2\ell+3)}i \eps ^3 \nonumber,\\
&\qquad\qquad\qquad+\frac{ P_{-1,4}^{(12,6)}(\ell,s)}{8 (\ell-1)  \ell^4 (\ell+1) (2\ell-3) (2 \ell-1)^2(2 \ell+1)^3 ( 2\ell+3)}\eps ^4+O(\eps ^5)\nonumber,\\
&a_{-2}=-\frac{ (\ell+s-1)^2 (\ell+s)^2}{4  \ell (2 \ell-1)^2(2 \ell+1)}\eps ^2+\frac{  (\ell+s-1)^2 (\ell+s)^2}{4 (\ell-1) \ell^2 (2 \ell-1) (2 \ell+1)}i \eps ^3+\nonumber,\\
&\qquad\qquad\qquad\frac{ P_{-2,4}^{(15,8)}(\ell,s)}{48 (\ell-2) (\ell-1) \ell^3 (2 \ell-5)(2 \ell-1)^4 (2 \ell+1)^3(2\ell+3)}\eps ^4+O(\eps ^5)\nonumber,\\
&a_{-3}=\frac{ (\ell+s-2)^2 (\ell+s-1)^2
   (\ell+s)^2}{24 (\ell-1) \ell (2 \ell-3)  (2 \ell-1)^2 (2 \ell+1)}i \eps ^3+\frac{ P_{-3,4}{(15,8)}(\ell,s)}{24 (\ell-2) (\ell-1)^2 \ell^2 (2 \ell-3)  (2 \ell-1)^2(2 \ell+1)}\eps ^4+O(\eps ^5)\nonumber,\\
&a_{-4}=\frac{ (\ell+s-3)^2 (\ell+s-2)^2 (\ell+s-1)^2 (\ell+s)^2}{96 (\ell-1) \ell (2 \ell-5) (2\ell-3)^2(2\ell-1)^2(2 \ell+1) }\eps ^4+O(\eps ^5),\nonumber
\end{align}
where $P^{(m,n)}(x,y)$ denotes a real polynomial with integer coefficients of degree $m$ in $x$ and $n$ in $y$ and recall $\lambda=l(l+1)$.  The precise form of the polynomials is easily determined to high order but is too long to be useful in printed form for general $s$.  For completeness, we give, as the most important example, the term in the renormalised angular momentum for $s=0,-1,-2$:
\begin{align*} 
P_{0,4}^{(8,4)}\bigl(\lambda,0\bigr)&=
-\lambda^2(\lambda-2)  (3240 + 8733 \lambda - 
   82625 \lambda^2 + 155295 \lambda^3 - 105000 \lambda^4 + 
   18480 \lambda^5), \\
P_{0,4}^{(8,4)}\bigl(\lambda,1\bigr)&=-(\lambda-2) (405 + 1917 \lambda - 3240 \lambda^2 - 
   16305 \lambda^3 + 19435 \lambda^4 + 54495 \lambda^5 - 
   84840 \lambda^6 + 18480 \lambda^7),\\
P_{0,4}^{(8,4)}\bigl(\lambda,4\bigr)&=51840 + 102816 \lambda - 953424 \lambda^2 + 51522 \lambda^3 + 
 123233 \lambda^4 + 85775 \lambda^5 - 2415 \lambda^6 + 
 61320 \lambda^7 - 18480 \lambda^8.
\end{align*}
  
The naive pattern evident in the leading behaviour  under the ansatz $a_n=O(\eps^{|n|})$ is
\begin{equation}
\label{eq:adiag}
a_{n}\overset{\text{naive}}{=}\begin{cases}
\dfrac{
\left((\ell+1-s)_n\right)^2
(\ell+1)_n 2^n}
{(2\ell+2)_n (2\ell+1)_{2n} n!}(-i \eps)^{|n|}
+O\left(\ob^{n+1}\right)&n>0,\\
\\
\dfrac{
\left((-\ell-s)_{|n|}\right)^2
(-\ell)_{|n|} 2^{|n|}}
{(-2\ell)_{|n|} (-2\ell-1)_{2|n|}|n|!}(-i \eps)^{|n|}
+O\left(\ob^{n+1}\right)&n<0,
\end{cases}
\end{equation}
\fixme{
Eq.(\ref{eq:adiag}) for $a_{n}$ for $n<0$ does not seem to agree with my
\begin{equation}\label{eq:a_n lead ST}
\amumT{n}=\frac{
e^{-3\pi i n/2} 2^{1-n}\Gamma\left(2\ell-n+1\right)\Gamma\left(\frac{3}{2}+\ell-n\right)}{\Gamma\left(\ell+\frac{1}{2}\right)\Gamma(2\ell+2)\Gamma(n+1)}
\left\{\begin{array}{lr} 
\displaystyle \frac{1}{\Gamma^2(\ell+s-n+1)}, & n-\ell\leq s \\
\Gamma^2(n-\ell-s)\ob^2, & s<n-\ell\leq 0
\end{array}
\right\}
\amumT{0}\ob^n
+\left(\ob^{n+1}\right)
\end{equation}
(Eq.(\ref{eq:adiag})  is for $a_n$ whereas Eq.(\ref{eq:a_n lead ST}) is for $a_n^T$, but the transformation is merely via a $(-1)^n$ according to Eq.5.5MSTa)?
}
where $(x)_n$ denotes the Pochhammer symbol $(x)_n=\Gamma(x+n)/\Gamma(x)= x(x+1)\dots(x+n-1)$.
Note that reflecting the symmetry noted at the beginning of the section, we have  $a_n \to a_{-n}$  under $\ell \to -\ell-1$.
We have also derived Eq.(\ref{eq:adiag}) via an alternative method: by imposing that 
Eq.(\ref{eq:Knu})
for $K_{\RAM}$ is independent of the parameter $r$.

Critically,  Eq.~(\ref{eq:adiag}) reveals that the ansatz $a_n=O(\eps^{|n|})$ is flawed since the denominator 
vanishes whenever $|n|\geq 2\ell+1$ while, in addition, the numerator vanishes when $|n|\geq \ell+s+1$.  A detailed analysis reveals that
the correct ansatz for $\RAM=\ell+O(\eps^2)$ and $s= 0$ is\fiximp{Could this be phrased better particularly in how Eq.(\ref{eq:an exc s=0}) relates to/corrects Eq.(\ref{eq:adiag})?}:
\begin{equation}
\label{eq:an exc s=0}
a_{n}=\begin{cases}
O(\eps^{|n|})&n\geq-\ell,\\
O(\eps^{|n|+1})&-(2\ell+1)< n < -\ell,\\
O(\eps^{|n|-1})&n\leq -(2\ell+1).\\
\end{cases}
\end{equation}
The same rules apply to $s\neq 0$ with the overrides\fiximp{Are these strictly the only overrides wrt Eq.(\ref{eq:adiag})?}:
\begin{align}
\label{eq:an exc s<>0}
s&=-2:\quad  a_{n}= O(\eps^{|n|+2}),&&n= -\ell+1\ \text{and}\ n=-\ell,\\
s&=-1:\quad  a_{n}= O(\eps^{|n|+2}),&& n=-\ell,\\
s&=+1:\quad  a_{n}= O(\eps^{|n|-1}),&& n=-\ell-1,\\
s&=+2:\quad  a_{n}= O(\eps^{|n|-1}),&&n= -\ell-1\  \text{and}\  n=-\ell-2 .
\end{align}

\fiximp{Could this paragraph be phrased better?}Inserting this revised ansatz into Eq.(\ref{eq:recursion}), together with an expansion for $\nu$, yields equations that can be solved recursively along
rising and falling diagonals (treating $\nu$ as if it were $a_0$) to very high order.
Through the diagonal nature of this procedure, it becomes clear in all cases that the general terms given above work for 
$a_n$ through the following orders while beyond this they must be supplemented by expansions based on the corrected ansatz:
\begin{align*}
\text{General expressions of Eq.~(\ref{eq:gen})  valid for\ } s\leq 0 \text{\ to order} \ 
\begin{cases} 
O(\eps^{n+2\ell+1}),&n\geq-\ell,\\
O(\eps^{-n}),&-(2\ell+1)<n<-\ell,\\
O(\eps^{-n-2}),&n\leq-(2\ell+1).
\end{cases}
\end{align*}
We may usefully turn this around, if we wish to work uniformly to order  $\eps^N$, the most anomalous behaviour occurs at either $n=-\ell/-\ell-1$ or
at $n=-(2\ell+1)$ and  we may use the general expansions except when $\min(\ell+1,2\ell-1)<N$. For example, the expansions to order $\eps^4$ of Eq.~(\ref{eq:gen}) are valid for any $\ell>2$ while we need to calculate the expansions for $\ell\leq2$ using the revised, correct ansatz to order $\eps^4$:
\begingroup
\allowdisplaybreaks
\begin{align}\label{eq:specific an's}
s&=0:\  \ell=0&&
a_4=\frac{1}{525}\eps ^4, \quad 
a_3=\frac{1}{60}i\eps^3-\frac{11 }{360}\eps ^4, \quad 
a_2=-\frac{1}{9}\eps^2-\frac{1}{6}i\eps^3-\frac{943}{4536} \eps ^4,\\*
&&&
a_1=-\frac{1}{2}i\eps+\frac{1}{2}\eps^2-\frac{331}{360}i\eps^3+\frac{541}{360}\eps^4,\quad 
\nu=-\frac{7}{6}\eps^2  -\frac{9449 }{7560}\eps ^4,\nonumber\\*
&&&
a_{-1}=-\frac{2 }{9}+\frac{7}{27}i\eps-\frac{1591}{2835}\eps^2+\frac{31723}{34020}i \eps ^3-\frac{73956143 }{50009400}\eps ^4,\quad 
a_{-2}=\frac{1 }{9}i \eps +\frac{1}{54}\eps^2+\frac{853}{3780}i\eps^3+\frac{9679 }{34020}\eps ^4,\nonumber\\*
&&&
a_{-3}=\frac{2 }{81}\eps ^2+\frac{2 }{243}i \eps ^3+\frac{1243 \eps ^4}{34020},\quad 
a_{-4}=-\frac{1}{270}i \eps ^3+\frac{1 }{405}\eps ^4,\quad a_{-5}=-\frac{2 }{4725}\eps ^4.\nonumber\\
\phantom{s}&\phantom{=0:}\quad \ell=1&& 
a_4=\frac{1}{1323}\eps ^4, \quad 
a_3=\frac{4}{525}i\eps^3-\frac{13 }{1575}\eps ^4, \quad 
a_2=-\frac{3}{50}\eps^2-\frac{1}{20}i\eps^3-\frac{1741}{67500} \eps ^4,\nonumber\\*
&&&
a_1=-\frac{1}{3}i\eps+\frac{1}{6}\eps^2-\frac{2447}{18900}i\eps^3+\frac{2221}{18900}\eps^4,\quad 
\nu=1-\frac{19}{30}\eps^2  -\frac{1325203}{3591000}\eps ^4,\nonumber\\*
&&&
a_{-1}=-\frac{1}{6}i\eps-\frac{1}{6}\eps^2-\frac{169}{20520}i \eps ^3-\frac{467}{4104}\eps ^4,\quad 
a_{-2}=-\frac{545}{4332}i\eps^3-\frac{5341}{25992}\eps ^4,\nonumber\\*
&&&
a_{-3}=-\frac{25}{722}\eps ^2+\frac{5}{228}i \eps ^3-\frac{5070221 }{32840892}\eps ^4,\quad 
a_{-4}=\frac{25}{2166}i \eps ^3+\frac{5}{3249}\eps ^4,\quad a_{-5}=\frac{3 }{1444}\eps ^4.\nonumber\\
\phantom{s}&\phantom{=0:}\quad \ell=2&& 
a_4=\frac{5}{9702}\eps ^4, \quad 
a_3=\frac{5}{882}i\eps^3-\frac{47}{10584}\eps ^4, \quad 
a_2=-\frac{12}{245}\eps^2-\frac{1}{35}i\eps^3-\frac{87826}{9904125} \eps ^4,\nonumber\\*
&&&
a_1=-\frac{3}{10}i\eps+\frac{1}{10}\eps^2-\frac{21793}{441000}i\eps^3+\frac{38383}{1323000}\eps^4,\quad 
\nu=2-\frac{79}{210}\eps^2  -\frac{708247 }{9261000}\eps ^4,\nonumber\\*
&&&
a_{-1}=-\frac{1}{5}i\eps-\frac{1}{10}\eps^2-\frac{1613}{31500}i \eps ^3-\frac{1399}{31500}\eps ^4,\quad 
a_{-2}=-\frac{1}{90}\eps^2+\frac{1}{60}i\eps^3-\frac{5104}{5599125}\eps ^4,\nonumber\\*
&&&
a_{-3}=-\frac{7}{1422}\eps ^4,\quad 
a_{-4}=0,\quad a_{-5}=-\frac{49}{112338}\eps ^4.\nonumber\\
\phantom{s}&\phantom{=0:}\quad \ell=3&& 
a_4=\frac{35}{84942}\eps ^4, \quad 
a_3=\frac{10}{2079}i\eps^3-\frac{37}{12474}\eps ^4, \quad 
a_2=-\frac{25}{567}\eps^2-\frac{5 }{252}i\eps^3-\frac{2323091}{468087984} \eps ^4,\nonumber\\*
&&&
a_1=-\frac{2}{7}i\eps+\frac{1}{14}\eps^2-\frac{351521}{12224520}i\eps^3+\frac{117151}{9779616}\eps^4,\quad 
\nu=3-\frac{169}{630}\eps^2  -\frac{74380421}{2750517000}\eps ^4,\nonumber\\*
&&&
a_{-1}=-\frac{3}{14}i\eps-\frac{1}{14}\eps^2-\frac{6019}{205800}i \eps ^3-\frac{29887}{1852200}\eps ^4,\quad 
a_{-2}=-\frac{3}{175}\eps^2+\frac{1}{70}i\eps^3-\frac{15979}{5145000}\eps ^4,\nonumber\\*
&&&
a_{-3}=\frac{1}{2100}i \eps ^3+\frac{11}{12600}\eps ^4,\quad 
a_{-4}=0,\quad a_{-5}=0.\nonumber
\end{align}
\begin{align*}
s&=-1:\  \ell=1&&a_4=\frac{1}{147}\eps ^4, \quad a_3=\frac{1}{21}i\eps^3-\frac{13 }{252}\eps ^4, \quad a_2=-\frac{6}{25}\eps^2-\frac{1}{5}i\eps^3-\frac{463 }{2500}\eps ^4,\\*
&&&a_1=-\frac{3}{4}i\eps+\frac{3}{8}\eps^2-\frac{14977}{33600}i\eps^3+\frac{24847}{67200}\eps^4,\quad 
\nu=1-\frac{47}{60}\eps^2  -\frac{43908007}{71064000} \eps ^4,\\*
&&&a_{-1}=-\frac{17 }{282}i \eps ^3-\frac{17 }{282}\eps ^4,\quad a_{-2}=-\frac{5 }{47}i \eps ^3-\frac{107 }{564}\eps ^4,\\*
&&&a_{-3}=-\frac{200 }{2209}\eps ^2+\frac{10 }{141}i \eps ^3-\frac{152936921 }{307419903}\eps ^4,\quad 
a_{-4}=\frac{150 }{2209}i \eps ^3+\frac{85 }{4418}\eps ^4,\quad a_{-5}=\frac{48 }{2209}\eps ^4.\\
\phantom{s}&\phantom{=-1:}\quad \ell=2&& 
a_4=\frac{5}{1782}\eps ^4, \quad a_3=\frac{10}{411}i\eps^3-\frac{47}{2646}\eps ^4, \quad 
a_2=-\frac{20}{147}\eps^2-\frac{5}{63}i\eps^3-\frac{440941}{14261940}\eps ^4,\\*
&&&a_1=-\frac{8}{15}i\eps+\frac{8}{45}\eps^2-\frac{32987}{330750}i\eps^3+\frac{56647}{992250}\eps^4,\quad 
\nu=2-\frac{169}{420}\eps^2  -\frac{6832249}{74088000} \eps ^4,\\*
&&&a_{-1}=-\frac{1}{20}i\eps-\frac{1}{40}\eps^2-\frac{5707}{168000}i \eps ^3-\frac{7397}{336000}\eps ^4,\quad 
a_{-2}=-\frac{11}{6760}\eps ^4,\\*
&&&a_{-3}=-\frac{7}{6084}\eps ^4,\quad 
a_{-4}=0,\quad a_{-5}=-\frac{49}{57122}\eps ^4.
\end{align*}
\begin{align*}
s&=-2:\  \ell=2&&a_4=\frac{10}{891}\eps ^4, \quad 
a_3=\frac{5}{72}i\eps^3-\frac{47}{864}\eps ^4, \quad 
a_2=-\frac{15}{49}\eps^2-\frac{5}{28}i\eps^3-\frac{730781}{6338640}\eps ^4,\\*
&&&a_1=-\frac{5}{6}i\eps+\frac{5}{18}\eps^2-\frac{12029}{52920}i\eps^3+\frac{19519}{158760}\eps^4,\quad 
\nu=2-\frac{107}{210}\eps^2  -\frac{1695233 }{9261000} \eps ^4,\\*
&&&a_{-1}=-\frac{1 }{20}i \eps ^3-\frac{1}{40}\eps ^4,\quad a_{-2}=\frac{11 }{12840}\eps ^4,\\*
&&&a_{-3}=-\frac{7}{1926}\eps ^4,\quad 
a_{-4}=0,\quad a_{-5}=-\frac{98}{11449}\eps ^4.
\end{align*}%

In order to evaluate $ \qT(\fNIA)$ via Eqs.(\ref{eq:qT ST}) and (\ref{eq:ratio coeffs in q})
we also need the sum of the $a_n$. Using Eq.(\ref{eq:gen}) we obtain, to order $\eps^4$,
\begin{align} 
\label{eq:suman}
&
\sum_{n=-\infty}^{\infty}a_n=1+
\frac{  \left(\lambda+s^2\right)}{2 \lambda}i \ob
\nonumber \\ 
&\quad-\frac{\lambda ^2 \left(4 \lambda ^2-7 \lambda +4\right)+\lambda  (4 \lambda +9) s^4+6 \lambda  (4 \lambda
   -3) s^3+3 (2 \lambda -1) \left(4 \lambda ^2+\lambda -6\right) s^2+2 \lambda  (4 \lambda -3) (7
   \lambda -6) s}{4 \lambda ^2 (4 \lambda -3)^2}\ob ^2\nonumber\\
   &\qquad+\frac{P_3}{24 (3-4 \lambda )^2 (\lambda -2) \lambda ^3}i\ob^3+\frac{P_4}{48 (\lambda -2)^2 \lambda ^4 (4 \lambda -15)^2 (4 \lambda -3)^4}\ob^4+O(\ob^5),
   \end{align}
where
\begin{align} \label{eq:explicit an's}
P_3&= (\lambda -2) \lambda ^2 \left(4 \lambda ^3+213 \lambda ^2-309 \lambda +108\right)+i \left(4 \lambda
   ^3+69 \lambda ^2-81 \lambda +54\right) s^6\nonumber\\*
&-18 i (\lambda -2) \lambda  (4 \lambda -3) s^5-i \lambda 
   \left(36 \lambda ^3-175 \lambda ^2+432 \lambda -252\right) s^4-12 i (\lambda -2) \lambda  (4
   \lambda -3) (5 \lambda -3) s^3\nonumber\\*
&-i \lambda  \left(36 \lambda ^4-279 \lambda ^3+686 \lambda ^2-738
   \lambda +288\right) s^2-6 i (\lambda -2) \lambda ^2 (4 \lambda -3) (7 \lambda -6) s,\\
P_4&=-(\lambda -2)^2 \lambda ^2 (448 \lambda ^8+13584 \lambda ^7-68220 \lambda ^6-372693 \lambda
   ^5+2636097 \lambda ^4-5229081 \lambda ^3+4967622 \lambda ^2-2347380 \lambda +437400)\nonumber\\*
&-(\lambda
   -2) \lambda  (448 \lambda ^6+6288 \lambda ^5+19620 \lambda ^4-334989 \lambda ^3+958311 \lambda
   ^2-841995 \lambda +109350) s^8\nonumber\\*
&+18 (\lambda -2)^2 \lambda  (4 \lambda -15) (4 \lambda -3)^2
   (4 \lambda ^2-102 \lambda +45) s^7\nonumber\\*
&+(-256 \lambda ^9-24896 \lambda ^8+45680 \lambda
   ^7+1948740 \lambda ^6-12101346 \lambda ^5+28846314 \lambda ^4\nonumber\\*
&-34305849 \lambda ^3+21625542 \lambda
   ^2-7333740 \lambda +1312200) s^6\nonumber\\*
&+6 (\lambda -2) \lambda  (4 \lambda -15) (4 \lambda -3)^2
   (100 \lambda ^4-1194 \lambda ^3+4425 \lambda ^2-3744 \lambda +540) s^5\nonumber\\*
&+(9600
   \lambda ^{10}-142432 \lambda ^9+532648 \lambda ^8+2117966 \lambda ^7-23736786 \lambda ^6+79670331
   \lambda ^5\nonumber\\*
&-130327029 \lambda ^4+113471766 \lambda ^3-54566136 \lambda ^2+14346720 \lambda
   -1749600) s^4\nonumber\\*
&+2 (\lambda -2) \lambda ^2 (4 \lambda -15) (4 \lambda -3)^2 (540 \lambda
   ^4-6334 \lambda ^3+25641 \lambda ^2-40896 \lambda +15660) s^3\nonumber\\*
&-\lambda  (256 \lambda
   ^{10}+57152 \lambda ^9-505072 \lambda ^8+7324 \lambda ^7+14370666 \lambda ^6-68251182 \lambda
   ^5\nonumber\\*
&+151062669 \lambda ^4-183278700 \lambda ^3+123738192 \lambda ^2-43759440 \lambda +6415200)
   s^2\nonumber\\*
&+2 (\lambda -2) \lambda ^2 (4 \lambda -15) (4 \lambda -3)^2 (84 \lambda ^5-3342 \lambda
   ^4+21295 \lambda ^3-47598 \lambda ^2+42048 \lambda -11880) s,
 \end{align}
\endgroup
with special cases
\begin{align*}
s&=0:\  \ell=0&&\sum_{n=-\infty}^{\infty}a_n=\frac{7}{9}-\frac{7  }{54}i \eps-\frac{122 }{945}\eps ^2+\frac{6343 }{68040}i \eps ^3+\frac{276257 }{2500470}\eps ^4+O(\eps ^5),\\
\phantom{s}&\phantom{=0:}\  \ell=1&&\sum_{n=-\infty}^{\infty}a_n=1-\frac{1 }{2}i \eps-\frac{854 }{9025}\eps ^2-\frac{19669}{72200} i \eps ^3-\frac{3079726441}{7982161250} \eps ^4+O(\eps ^5),\\
\phantom{s}&\phantom{=0:}\  \ell=2&&\sum_{n=-\infty}^{\infty}a_n=1-\frac{1 }{2}i \eps-\frac{53 }{882}\eps ^2-\frac{377}{3528} i \eps ^3-\frac{167323193 }{4855023684} \eps ^4+O(\eps ^5),\\
s&=-1:\  \ell=1&&\sum_{n=-\infty}^{\infty}a_n=1-\frac{3 }{4}i \eps +\frac{19643 }{441800}\eps^2-\frac{2212413}{3534400} i \eps^3-\frac{10841072759029 }{19128349520000}\eps^4+O(\eps ^5),\\
\phantom{s}&\phantom{=-1:}\  \ell=2&&\sum_{n=-\infty}^{\infty}a_n=1-\frac{7  }{12}i
   \eps +\frac{59 }{3528}\eps^2-\frac{329 }{1728}i \eps^3-\frac{3421065415 }{236995065216}\eps^4+O(\eps ^5),\\
s&=-2:\  \ell=2&&\sum_{n=-\infty}^{\infty}a_n=1-\frac{5    }{6}i \eps-\frac{25 }{882}\eps ^2-\frac{2045 }{5292}i \eps ^3-\frac{1706608417 }{23750538336}\eps ^4+O(\eps ^5).
\end{align*}

Finally we should note that the   power series for $\RAM$ cannot be valid $\forall \ob\in\mathbb{C}$.
The reason is that, for example for $\ob \in\mathbb{R}$, the exact value of $\RAM$ is known to be real for small values of $\ob$
 but, as $\ob$ increases,
$\RAM$ reaches some half-integer
value and then it suddenly picks up an imaginary part (see, e.g., Table 1 in ST; we have found a similar behaviour for $\ob$ on the negative imaginary axis).
The power series for $\RAM$, however, is purely real for $\ob$ real and so it cannot reproduce this behaviour.
In this paper, however, we are only interested in the small-$\ob$ behaviour, where the power series does converge.
We have been referring to such regime as the 
 `perturbative (small-frequency) regime'.
 \fixme{This paragraph is to be checked} 


\section{Low-frequency Expansion of the Branch Cut Integrand} \label{sec:Tail}

 It is clear from Eqs.(\ref{eq:Green}) and (\ref{eq: G^BC integral}) that the 
 late-time asymptotics of the BC contribution  to the  GF
 is provided by the  small-$\nb$ expansion of the modes $\DGw$.
  The  small-$\nb$ expansion of  $\DGw$, in its turn, is given by the  small-$\nb$   asymptotics of the radius-independent quantity $q/|W|^2$ and the radial solution $\f$.
  In this section we will derive  the  small-$\nb$   asymptotics of $q/|W|^2$.
We will use these asymptotics later in
 Eq.(\ref{eq:GF late-times}) in order to prove that $\GlBC\sim \dt^{-2\ell-3}$, to leading order for large-$\dt$, for general integer spin.

Specifically, in the following subsections we provide explicit expansions for  small-$\ob$  
up to the first three
 leading powers (which actually correspond to the first four leading orders, since the third order has the same power of $\ob$ as the fourth order but, as we shall show, it  contains a logarithm in $\ob$) 
 for the various perturbation quantities (except for the $a_n$ and $\RAM$, which we gave in the previous section, and for the radial functions, which we give in the next section).
 We provide these expansions for general integer-spin $s$ and multipole number $\ell$.
These expansions will yield the first three leading powers (so four leading orders) for late-times of the multipole-$\ell$ GF and field perturbations.
The above comments apply to the RW GF and field perturbations but a similar argument applies to the BPT ones.
In fact, we shall give the small-$\nb$ expressions explicitly  for 
BPT quantities;
 one can then readily find the corresponding expansions for the RW quantities
 via the transformations given in Sec.\ref{sec:transf RW-BPT}.

From now on and for the rest of the paper we shall restrict ourselves to the case that the BPT spin is a negative-integer, $s=0,-1,-2$ (RW spin, of course, is indistinctively positive or negative).
BPT quantities for positive spin can be obtained from those for negative spin via the Teukolsky-Starobinski\u{\i} identities~\cite{teukolsky1974perturbations,Chandrasekhar}.

Also, as mentioned, from now on we will focus on small-frequeny expansions up to  the first three leading powers of the frequency.
It is easy to see from the results in the previous section (e.g., compare the general-$\ell$ and -$s$ expressions in Eq.(\ref{eq:gen}) with the specific-mode expressions in Eq.(\ref{eq:specific an's})) that the expressions that we shall obtain for general multipole-$\ell$ and spin-$s$ in principle are not 
necessarily valid, up to  the first three leading powers of the frequency, for the three specific modes 
$\ell=0$ (and $s=0$) and $\ell=1$ (and $s=0,-1$).
Indeed, some general-$s$ and -$\ell$ expressions that we shall give appear to have singularities at the $s$ and $\ell$ values for these modes.
We have dealt with these three modes separately by carrying out small-frequency expansions after setting the corresponding values of $s$ and $\ell$ right from the start.
Remarkably, we have found that our general-$s$ and -$\ell$ expansions up to  the first three leading powers of the frequency, for $\qT$, $\RinincN{s}$ and all subsequent quantities derived from these
two quantities, actually give the correct result  for  two of these anomalous modes, namely for 
$\ell=1$, with $s=0$  and $-1$
\footnote{Specifically, for, e.g., $s=0$ and $\ell=1$, the general-$s$ and -$\ell$ expression for $\Kmu$ gives the wrong coefficient in the third leading power of $\ob$ and for $\Kmmu$ it gives the wrong leading order.
However, for this mode, the two incorrections in $\Kmu$  and in $\Kmmu$  somehow miraculously cancel each other out to give the right result for $\RinincN{s}$ up to  the first three leading powers of $\ob$.}.
For the mode $s=\ell=0$, the  general-$s$ and -$\ell$ expressions do not give the correct result and we present the results for this mode (as well as for $\ell=-s=1$ and $2$, for completeness)
in Sec.\ref{Sec:-s}.



\subsection{$\Kmu$}

The quantity $\Kmu$ introduced in Eq.(\ref{eq:sR0nu}) is needed
in order to obtain the Wronskian below (in Eq.(\ref{eq:Ain gral s,l})). 
We obtain the following small-$\ob$ expansion of $\Kmu$ from Eq.(\ref{eq:Knu}):
\begin{align}
&
\Kmu=
\frac{2^\ell \Gamma \left(\ell+\frac{1}{2}\right) \Gamma (2 \ell+2)
\G
 }{\sqrt{\pi }
   \Gamma^2 (\ell-s+1)}\ob^{-\ell+s}
   +
\nonumber   \\&
   \frac{i 2^{\ell+1} \Gamma (2 \ell) \Gamma \left(\ell+\frac{3}{2}\right) 
   \G
    \left(2
   (\ell+1) \ell \left(H_{\ell}-2 H_{-s}+\gamma_E \right)+\ell^2+\ell-s^2\right)}{\sqrt{\pi } (\ell+1) \Gamma^2 (\ell-s+1)}\ob^{-\ell+s+1}+
\nonumber     \\&
   \frac{2^{-\ell-3} \Gamma (2 \ell+1) \Gamma (2 \ell+2) \Gamma (1-s)}{\Gamma (\ell+1) \Gamma (\ell-s+1)^2}K_{\nu}^{(3)}\ob^{-\ell+s+2}+   
   O(\ob^{-\ell+s+3}),
   \label{eq:Kmu}
\end{align}
where 
\begin{align}
&
K_{\nu}^{(3)}\equiv 
   \frac{4 \left(6 \left(\ell^2+\ell-1\right) s^2+\ell (\ell+1) \left(15 \ell^2+15 \ell-11\right)+3 s^4\right)
   \left(2 H_{\ell-s}+H_{\ell}-2 H_{2 \ell}-2 H_{2 \ell+1}+\ln (2 \ob )+\gamma_E \right)}{\ell (\ell+1)
   (2 \ell-1) (2 \ell+1) (2 \ell+3)}-
   \nonumber   \\&
   4 \left(-\frac{s^2}{\ell (\ell+1)}+2 \gamma_E +1\right) \left(H_{\ell}-2
   H_{-s}+\gamma_E \right)-4 \left(H_{\ell}-2 H_{-s}\right){}^2-4 \left(2
   H_{\ell-s}^{(2)}+H_{\ell}^{(2)}-4 H_{-s}^{(2)}+\frac{\pi ^2}{6}-\gamma_E ^2\right)-
      \nonumber   \\&
   \frac{1}{\ell^2 (\ell+1)^2 (2 \ell-1)^2 (2 \ell+3)^2}
   \Bigg\{
   2   \left((\ell+1) \left(8 \ell^3-16 \ell^2-24 \ell+9\right) s^4+\ell^2 (\ell+1)^2 \left(8 \ell^4+16 \ell^3-4
   \ell^2-13 \ell+4\right)-
\right.    \nonumber   \\& \left.
   \left(16 \ell^6+104 \ell^5+164 \ell^4-16 \ell^3-131 \ell^2-3 \ell+18\right)
   s^2\right)
   \Bigg\}.
\end{align}
Doing similarly for $\Kmmu$, we obtain
\begin{align}
&
\Kmmu=
\frac{(-1)^{\ell}i2^{\ell-1}\G\Gamma(\ell+1)\Gamma^2(\ell+s+1)}{\Gamma(2\ell+1)\Gamma(2\ell+2)\RAMmLead^2}\ob^{s+\ell}+O(\ob^{\ell+s-1}).
     \label{eq:Kmmu}
\end{align}
\fixme{Check the latter with Mathematica}
In obtaining Eq.(\ref{eq:Kmmu}), we have used Eq.(\ref{eq:adiag}) together with the fact that $\Gamma(-n+\ob)\sim\frac{(-1)^n}{\Gamma(n+1)\ob}$ as $\ob\to 0$
when $n$ is a nonnegative integer.


We note that the asymptotics of $\Kmu$ in Eq.(\ref{eq:Kmu}) and those of $\Kmmu$ in Eq.(\ref{eq:Kmmu})
imply, via Eq.168 ST (and since $\ell\ge |s|$), that  $\Kmmu$ is not necessary for obtaining
$\RinincN{s}$  to the first four leading orders ($\Kmmu$  starts playing a part only in the next order),
except in the cases $\ell=0$ and $\ell=1$. 
As mentioned at the 
start of this section,
though, Eqs.(\ref{eq:Kmu}) and (\ref{eq:Kmmu}) are in principle not valid for $\ell=0$ and $1$,
although  the general-$s$ and -$\ell$ expressions that we give for quantities from now on are, somewhat surprisingly,  also valid for $\ell=1$  (with $s=0,1$).
The case $s=\ell=0$ we treat separately in Sec.\ref{Sec:s=l=0}.

\subsection{Radial coefficients and Wronskian}

We now turn to the coefficients in Eq.(\ref{eq: bc Rin}) of the ingoing radial (BPT) solution.
We have obtained the following
expansion for the
radial coefficient $\RinincN{s}$
by carrying out small-$\nb$ asymptotics of 
$\RinincNST/\RintraNST$, where $\RininctraNST$ are the incidence/transmission coefficients of $\RTinN$ (i.e., with the specific normalization used in 
Sec.\ref{sec:MST gral s}, which is the same normalization as in MST).
The expression for the coefficient $\RintraNST$ is given in Eq.(\ref{eq:coeff tra norm MST}) whereas for $\RinincNST$ we used Eq.168 ST (which requires
$\Kmu$  and $\Kmmu$).
We obtain:
\begin{align} \label{eq:Ain gral s,l}
&
\RinincN{s}
=
M^{1-2s}\frac{2^{-\ell-s} e^{\frac{1}{2} i \pi  (\ell-s+1)} \Gamma (2 \ell+1) \Gamma (2 \ell+2) \Gamma (1-s)}{\Gamma (\ell+1) \Gamma
   (\ell-s+1) \Gamma (\ell+s+1)}
   \frac{1}{\ob^{\ell-s+1}}
   \Bigg\{1
   +
\nonumber   \\ &
\left[i \left(H_{\ell-s}+H_{\ell+s}+H_{\ell}-2 H_{-s}-\ln (2 \ob )+\frac{i \pi }{2}-\gamma_E \right)-\frac{i \left(\ell (\ell+1)+s^2\right)}{2 \ell (\ell+1)}\right]\ob+
\nonumber \\ &
\left[
\frac{\left(6 \left(\ell^2+\ell-1\right) s^2+\ell (\ell+1) \left(15 \ell^2+15 \ell-11\right)+3 s^4\right) \left(H_{\ell-s}+H_{\ell+s}+H_{\ell}-2 H_{2 \ell}-2 H_{2 \ell+1}+\ln (2\ob)-\frac{i \pi }{2}+\gamma_E\right)}{2 \ell (\ell+1) (2 \ell-1) (2 \ell+1) (2 \ell+3)}-
\right. \\ & \left.
\frac{1}{2} \left(\frac{s^2}{\ell (\ell+1)}+1\right) \left(-H_{\ell-s}-H_{\ell+s}-H_{\ell}+2 H_{-s}+\ln (2\ob )-\frac{i \pi }{2}+\gamma_E\right)-
   \right. \nonumber\\ & \left.
   \frac{1}{2} \left(-H_{\ell-s}-H_{\ell+s}-H_{\ell}+2 H_{-s}+\ln(2\ob )-\frac{i \pi }{2}+\gamma_E\right){}^2+\frac{1}{2} \left(-H_{\ell-s}^{(2)}-H_{\ell+s}^{(2)}-H_{\ell}^{(2)}+4
   H_{-s}^{(2)}-\frac{\pi ^2}{6}\right)-
   \right.\nonumber \\ & \left.
   \frac{\left(8 \ell^3-16 \ell^2-24 \ell+9\right) s^4}{4 \ell^2 (\ell+1) (2 \ell-1)^2 (2 \ell+3)^2}-\frac{8 \ell^4+16 \ell^3-4 \ell^2-13 \ell+4}{4 (2 \ell-1)^2 (2 \ell+3)^2}-\frac{\left(16
   \ell^6-8 \ell^5-116 \ell^4-48 \ell^3+101 \ell^2+21 \ell-18\right) s^2}{4 \ell^2 (\ell+1)^2 (2 \ell-1)^2 (2 \ell+3)^2}\right]\ob^{2}
\Bigg\}
\nonumber\\ &
+o\left(\frac{1}{\ob^{\ell-s-1}}\right).
\nonumber
\end{align}

We note that the leading order of Eq.(\ref{eq:Ain gral s,l}) agrees with 
 Eq.3.6.13~\cite{th:CasalsPhD}, which is obtained via an independent method based on Page's~\cite{ar:PageI'76}.
As a token example of the more simplified form that adopts the expansion  for a particular value of $s$, 
we give  Eq.(\ref{eq:Ain gral s,l}) specifically for $s=0$:
\begin{align} \label{eq:Ain s=0 gral l}
&
\frac{1}{M}\RinincN{s}
=   \frac{2^{-\ell} e^{\frac{1}{2} i \pi  (\ell+1)} \Gamma (2 \ell+1) \Gamma (2 \ell+2) }{\Gamma^3 (\ell+1)}\frac{1}{\ob ^{\ell+1}}+
\nonumber   \\ &
   \frac{i
   2^{-\ell-1} e^{\frac{1}{2} i \pi  (\ell+1)} \Gamma (2 \ell+1) \Gamma (2 \ell+2)(6 \psi(\ell+1)-2 \ln (2 \ob
   )+i \pi +4 \gamma_E -1)}{\Gamma^3 (\ell+1)}\frac{1}{ \ob ^{\ell} }
   \nonumber \\&
\frac{2^{3 \ell} e^{\frac{1}{2} i \pi  (\ell+1)} \Gamma \left(\ell+\frac{1}{2}\right) \Gamma \left(\ell+\frac{3}{2}\right)}{\pi  \Gamma
   (\ell+1)}
   \left((\ln(2i\ob)-i \pi -2 \gamma_E ) \left(\frac{15 \ell^2+15 \ell-11}{(2 \ell-1) (2 \ell+1) (2 \ell+3)}+6 \psi(\ell+1)-1\right)+
   \right. \nonumber \\ &\left.
   \frac{\left(15 \ell^2+15 \ell-11\right) (3 \psi(\ell+1)-2 \psi(2 \ell+1)-2 \psi(2 \ell+2)+2
   \gamma_E )}{(2 \ell-1) (2 \ell+1) (2 \ell+3)}+\frac{2 \ell+1}{4 (2 \ell-1)^2 (2 \ell+3)^2}-
      \right. \nonumber \\ &\left.
   9 \psi(\ell+1)^2+3 \psi(\ell+1)+3 \psi
   ^{(1)}(\ell+1)-(\ln(2i\ob)-i \pi -2 \gamma_E )^2-\frac{2 \pi ^2}{3}-\frac{1}{4}\right)\frac{1}{ \ob ^{\ell-1}}
   +\left(\frac{1}{ \ob ^{\ell-2}}\right),\quad s=0.
   \end{align}
   Here, the function $\psi(z)\equiv \Gamma'(z)/\Gamma(z)$ is the digamma function~\cite{NIST:DLMF}
   and  $\psi^{(n)}(z)$ its $n$th-derivative.
   We note that the digamma function may be expressed as $\psi(\ell)=H_{\ell-1}-\gamma_E$, in terms of the harmonic numbers used above.

\subsection{BC strength $\qT$}

In this section we will provide a small-$\nb$ for the BC strength  $\qT$.
The solution $\RCp{\RAM}$ of the BPT equation has the following asymptotics, from Eqs.(\ref{eq:RinincNST}) and (\ref{eq:RCnu+-}) (taking the upper sign)\fiximp{But we should find this generally?},
\begin{equation} \label{eq:Rpmu large-r}
\frac{\Rpmu}{\Rpmutra}\sim  \frac{1}{r}e^{-i\omega r_*}, \quad \bar r\to \infty,
\end{equation} 
where
\begin{equation}
 \Rpmutra(\omega)\equiv \Apmu\omega^{-1}e^{-i\ob \ln \ob},
\end{equation} 
and $\Apmu$ is given in Eq.(\ref{eq:A_+}).
Comparing with Eq.(\ref{eq: bc Rup}), it follows that
\begin{equation}
\frac{\Rpmu(r,\omega)}{\Rpmutra(\omega)}=
\RupN{-s}(r,-\omega)\Delta^{-s}.
\end{equation}

Now, from 
Eq.(\ref{eq:upgoing MST BPT})
together with the analytic continuation of the irregular hypergeometric $U$-functions on the complex-$\omega$ plane (Eq.13.2.41~\cite{bk:onlineAS}),
we find:
\begin{equation}
\RupST(r,\omega e^{2\pi i})=e^{-2\pi \ob}\RupST(r,\omega)+\left[e^{-2\pi\ob}-e^{-2\pi i\RAM}\right]\Rpmu(r,\omega).
\end{equation}
Using the definition Eq.(\ref{eq:def qT}) of the BC strength, it then follows that
 \begin{equation} \label{eq:qT ST}
 \qT(\fNIA)=i\left[1-e^{2\pi (\ob-i\RAM)}\right]\frac{\Rpmutra(-i\fNIA)}
{ \RuptraST(-i\fNIA)}.
 \end{equation}
We can therefore calculate the ratio of coefficients on the right hand side via:
 \begin{equation}\label{eq:ratio coeffs in q}
\frac{\Rpmutra(\omega)}
{ \RuptraST(\omega)}
=\frac{\Apmu}{\Ammu}\omega^{2s}\ob^{-2i\ob},
 \end{equation}
 where we have made used of Eqs.(\ref{eq:Rpmu large-r}) and (\ref{eq:MST uptra}).
From Eqs.(\ref{eq:qT ST}) and (\ref{eq:ratio coeffs in q})
and expanding
  Eq.(\ref{eq:A_+}) for $\Apmu$ and Eq.(\ref{eq:A_-}) for $\Ammu$,
we find the first five leading orders for the BC strength:
\begin{align} \label{eq:q gral l}
& \qT(\fNIA)=\frac{\nb^{2s+1}}{M^{2s}}
\frac{2 \pi  (-1)^{\ell+s} \Gamma (\ell-s+1)}{\Gamma (\ell+s+1)}
\Bigg\{
1
+
\nonumber \\ &
\Bigg[
H_{\ell-s}+H_{\ell+s}+\frac{3 \left(\ell^2+\ell-1\right) s^2}{\ell (\ell+1) (2 \ell-1) (2 \ell+1) (2 \ell+3)}+\frac{16 \ell^3+39 \ell^2+11 \ell-17}{2 (2 \ell-1) (2 \ell+1) (2 \ell+3)}+\frac{3 s^4}{2 \ell (\ell+1) (2 \ell-1) (2 \ell+1) (2
   \ell+3)}-
  \nonumber \\ &
   2 \ln (2 \nb )-2 \gamma_E
   \Bigg]
   \nb
+
\nonumber \\ &
\left[
\left(-2 H_{\ell-s}-2 H_{\ell+s}-\frac{6 \left(\ell^2+\ell-1\right) s^2+\ell (\ell+1) \left(16 \ell^3+39 \ell^2+11 \ell-17\right)+3 s^4}{\ell (\ell+1) (2 \ell-1) (2 \ell+1) (2 \ell+3)}+4 \gamma_E \right) \left(\gamma_E-H_{\ell-s}+\ln (2\nb) -\frac{1}{2}\right)-
\right. \nonumber \\ &\left.
2 \left(\gamma_E-H_{\ell-s} -\frac{1}{2}\right){}^2+\frac{1}{2} \left(H_{\ell+s}-H_{\ell-s}\right){}^2+\frac{1}{2}
   \left(H_{\ell+s}^{(2)}-H_{\ell-s}^{(2)}\right)+\frac{s \left(7 \ell^2+7 \ell+3 s^2-6\right)}{2 \ell (\ell+1) (2 \ell-1) (2 \ell+3)}+
   2\ln^2(2\nb)
   -\frac{\pi ^2}{6}
\right]
\nb^{2}
\Bigg\}
+
\nonumber \\ & 
o(\nb)^{2s+3}.
\end{align}
Using Eq.(\ref{eq:qT vs q final}) to relate BPT's $\qT$ with RW's $q$, it is easy to check that the leading order in Eq.(\ref{eq:q gral l}) agrees with Eq.41~\cite{Leaver:1986} (the `$iq$' here being `$K$' in~\cite{Leaver:1986})
for all spins $s=0,-1,-2$ (the leading order of RW's $q$ is independent of the spin).

Again, as a token example of the more simplified form  for a particular value of $s$, 
we give  Eq.(\ref{eq:q gral l}) specifically for $s=0$:
\begin{align} \label{eq:q s=0 gral l}
&
 \qT(\fNIA)=q(\fNIA)=
 i \pi  (-1)^{\ell}\ob \left\{2
+ \left(\frac{i \left(16 \ell^3+39 \ell^2+11 \ell-17\right)}{8 \ell^3+12 \ell^2-2 \ell-3}+4 i \psi (\ell+1)-4 i
   \ln (2 \nb ) \right)\ob-
\right.\\& \left.
 \left(\frac{2 \left(16 \ell^3+39 \ell^2+11 \ell-17\right) (\psi (\ell+1)-\ln (2 \nb))}{(2 \ell-1) (2 \ell+1) (2 \ell+3)}+\frac{8 \ell^3+27 \ell^2+13 \ell-14}{(2 \ell-1) (2 \ell+1) (2 \ell+3)}+4 (\psi(\ell+1)-\ln   (2 \nb))^2-\frac{\pi ^2}{3}\right)\ob^2
   \right\}
   +
\nonumber  \\ &  o(\ob^3), \quad s=0.
\nonumber 
\end{align}


\subsection{$\qT/|\WT|^2$} \label{sec:q/W^2}

We are finally in a position to give an expansion for the main target of this section: the radius-independent quantity in the BC integrand Eq.(\ref{eq: G^BC integral}).
From Eqs.(\ref{eq:Ain gral s,l}) and
(\ref{eq:q gral l}) it follows that
\begin{align}
&
\frac{\qT(\fNIA)}{\left|\WT\right|^2}=M^{2s}
\frac{(-1)^{\ell+s}\pi ^2 2^{-2 \ell+2 s+1} \Gamma^3 (\ell-s+1) \Gamma (\ell+s+1)}{\Gamma^2   \left(\ell+\frac{1}{2}\right) \Gamma^2 (2 \ell+2) \Gamma^2 (1-s)}
 \left[\nb^{2\ell+1} -
\right. \nonumber \\ &\left.
\frac{\nb^{2\ell+2}}{2}
 \left(8 H_{-s}-2 H_{\ell-s}-2 H_{\ell+s}-4 H_{\ell}+
 \frac{2 (\ell+1) (\ell (8 \ell+7)-6) s^2+\ell (\ell+1) (\ell (\ell (32 \ell+63)+7)-23)+3 s^4}{\ell (\ell+1) (2 \ell-1) (2 \ell+1) (2 \ell+3)}+\right.\right.\nonumber\\&\left.\left.
 \frac{2 (l+1) \left(8 l^2+7 l-6\right) s^2+l (l+1) \left(32 l^3+63 l^2+7 l-23\right)+3 s^4}{l (l+1) (2 l-1) (2 l+1) (2 l+3)}
 \right)
-Q\nb^{2\ell+3}
\right] 
+o(\nb^{2\ell+3}),
\label{q/W^2 gral s,l}   
\end{align}
where 
\begin{align} \label{eq:Q}
&
Q\equiv
\frac{\left(-6 \left(\ell^2+\ell-1\right) s^2+\ell (\ell+1) (11-15 \ell (\ell+1))-3 s^4\right) \left(H_{\ell-s}-4 H_{2 \ell}+2 H_{-s}+\ln (2\nb )+\gamma_E \right)}{\ell (\ell+1) (2 \ell-1) (2
   \ell+1) (2 \ell+3)}-
  \nonumber \\ &
   4 H_{-s} \left(2 H_{-s}+\frac{s^2}{\ell (\ell+1)}+2\right)+\left(-2 H_{\ell}+4 H_{-s}+\frac{s^2}{\ell (\ell+1)}+2\right) \left(H_{\ell-s}+H_{\ell+s}+2 H_{\ell}\right)-\frac{1}{2}
   \left(H_{\ell-s}+H_{\ell+s}\right){}^2+
  \nonumber    \\ &
   2 \left(H_{\ell}\right){}^2+\frac{3 H_{\ell-s}^{(2)}}{2}+\frac{H_{\ell+s}^{(2)}}{2}+H_{\ell}^{(2)}-4 H_{-s}^{(2)}-\frac{3 s^6}{2 \ell^2 (\ell+1)^2 (2 \ell-1)
   (2 \ell+1) (2 \ell+3)}-\frac{\left(7 \ell^2+7 \ell-6\right) s}{2 \ell (\ell+1) (2 \ell-1) (2 \ell+3)}-
 \nonumber     \\ &
   \frac{\left(128 \ell^6+768 \ell^5+1088 \ell^4+120 \ell^3-364 \ell^2+18 \ell+45\right) s^4}{4 \ell^2 (\ell+1)^2
   (2 \ell-1)^2 (2 \ell+1)^2 (2 \ell+3)^2}+\frac{-512 \ell^6-2016 \ell^5-1616 \ell^4+1472 \ell^3+1128 \ell^2-722 \ell+59}{4 (2 \ell-1)^2 (2 \ell+1)^2 (2 \ell+3)^2}-
  \nonumber    \\ &
   \frac{\left(256 \ell^8+1464 \ell^7+2660
   \ell^6+1082 \ell^5-1361 \ell^4-926 \ell^3+122 \ell^2+93 \ell+18\right) s^2}{2 \ell^2 (\ell+1)^2 (2 \ell-1)^2 (2 \ell+1)^2 (2 \ell+3)^2}-\frac{3 s^3}{2 \ell (\ell+1) (2 \ell-1) (2 \ell+3)}+\frac{\pi ^2}{3}.
\end{align}

Here we give the value of $q/|W|^2$ specifically for the case  $s=0$.
From Eqs.(\ref{eq:Ain s=0 gral l}) and (\ref{eq:q s=0 gral l}), or equivalently from Eq.(\ref{q/W^2 gral s,l}) with $s=0$ (or, equivalently, from~\cite{PhysRevLett.109.111101}\footnote{We note a typo of an extra overall `-1' in Eq.8~\cite{PhysRevLett.109.111101}.}), 
we have
\begin{widetext}
\begin{align}\label{q/W^2 s=0 gral l}
&
4\frac{\qT(\fNIA)}{\left|\WT\right|^2}=
\frac{q(\fNIA)}{|W|^2}= \frac{ (-1)^{\ell}\pi}{2^{2\ell-3}}\left(\frac{(2\ell+1) \ell!}{\left( \left(2\ell+1\right)!!\right)^2}\right)^2 \left[\nb^{2\ell+1} -
\nb^{2\ell+2}\left(\frac{-32 \ell^3-63 \ell^2-7 \ell+23}{2(2\ell+3)(2\ell+1)(2\ell-1)}+4 H_{\ell}\ \right)\right] \nonumber\\
&
+\frac{ (-1)^{\ell} \pi} { 2^{2 \ell-1}} \left(\frac{(2\ell+1)\ell!}{((2 \ell+1)!!)^2}\right)^2 \nb^{2\ell+3}
    \left[\frac{4(15\ell^2+15\ell-11)}{(2 \ell-1)(2\ell+1) (2 \ell+3)} \left(\ln (2 \nb )+H_{\ell}-4 H_{2 \ell}+
\gamma_E
   \right)\right.   \\ &\left.
-4  \left(-8 H_{\ell}{}^2+8 H_{\ell}+3 H_{\ell}^{(2)}+
   2H_{\infty}^{(2)}
    \right)+
  \frac{512\ell^6+2016 \ell^5 +1616 \ell^4-1472 \ell^3- 1128 \ell^2+722 \ell-59}{(2 \ell-1)^2  (2 \ell+1)^2 (2 \ell+3)^2}\right] +o(\nb^{2\ell+3}),
  \quad s=0,
   \nonumber
\end{align}
\end{widetext}
where $H^{(r)}_{\ell}$ is the $\ell$-th harmonic number of order $r$.

We note the appearance in (\ref{q/W^2 gral s,l}) of a logarithmic behaviour in $\nb$ at order $\nb^{2\ell+3}$  for small-frequency.
It is worth pointing out the following `curious' fact.
Even though the logarithmic behaviour appears already at second leading order both in the BC strength $\qT$  (see Eq.(\ref{eq:q gral l}))
and in the radial coefficient $\RinincN{s}$ (see Eq.(\ref{eq:Ain gral s,l})), there is a delicate cancellation between the terms which leads to the logarithmic behaviour for  $\qT/\left|\WT\right|^2$
appearing {\it not} at second order but at third leading order instead.
As shown later in Sec.\ref{sec:plot perturbation}, this implies that the logarithmic behaviour of the Green function or a field perturbation will also appear at third -- as opposed to second, as one might have expected -- leading order for late times.


\subsection{Cases $\ell=-s$} \label{Sec:-s}

As pointed out at the
start of this section,
the expression in Eq.(\ref{q/W^2 s=0 gral l}) might not be valid for the case $s=\ell=0$.
In this section we present the results obtained by carrying out a specific calculation for this case by setting $s=\ell=0$ right at the start of the calculation.
Although the BPT results for $\ell=-s=1,2$ can be obtained directly by putting in these values into the general $s$ and $\ell (> 0)$ expressions found above, 
we also include these cases for completeness and because they require an extra step in order to obtain the RW results from the BPT results.


\subsubsection{Case $s=\ell=0$} \label{Sec:s=l=0}

We have  carried out a specific calculation for the case $s=\ell=0$  by setting these values right from the start and we have obtained:
\begin{align}\label{Ain s=0=l}
&
\frac{1}{M}\RinincN{s}=
-\frac{1}{\nb }+\ln (2 \nb )-i \pi +\gamma_E
   +\frac{1}{2} +
    \nonumber  \\&
  \frac{1}{36} \nb  \left(15 \pi ^2-6 \ln (2 \nb ) \left(2\ln \left(2 \nb \right)-6 i \pi +6 \gamma_E -8\right)+36 i \gamma_E  \pi -48 i \pi +6 (8-3 \gamma_E ) \gamma_E -82\right)+o(\nb).
   \end{align}
 We note that the first two leading orders agree with Eq.(\ref{eq:Ain s=0 gral l}) with $\ell=0$ but the third leading orders differ slightly
 in the term
 which does not contain any `$ \ln (2 \nb )$'.

From Eq.(\ref{Ain s=0=l}) we readily obtain (assuming evaluation on the NIA, i.e., $\nu>0$)
\begin{align}\label{W^2 s=0=l}
&
\left|\WT\right|^2=4\fNIA^2|\RinincN{s}|^2=
1-(2\left(\ln (2 \nb )+\gamma_E\right)+1) \nb +
\nonumber  \\ &
\frac{1}{36} \left(72 \left(\ln (2 \nb )+\gamma_E\right)^2-60\left(\ln (2 \nb )+\gamma_E\right)+6 \pi ^2+173\right) \nb ^2
+
   o\left(\nb^2\right).
\end{align}

For the `BC strength' we obtain

\begin{align}\label{q s=0=l}
&
\qT(\fNIA)=q(\fNIA)=
   2 \pi  \nb
   +\frac{\nb ^2}{3} \left( 17 \pi -12 \pi  \left(  \ln (2 \nb )+ \gamma_E \right) \right)
   +  \nonumber \\ &
   \frac{ \pi  \nb ^3}{3} \left(12 \ln ^2(2 \nb )+2(12 \gamma_E -17) \ln (2 \nb )+
  2 \gamma_E (6\gamma_E -17 )
   +20-\pi ^2\right)+
    o\left(\nb^3\right).
\end{align}
We note that if instead of calculating the BC strength directly for $s=\ell=0$ we had used Eq.(\ref{eq:q gral l}) (or Eq.(\ref{eq:q s=0 gral l})) with $s=\ell=0$ we would have obtained
the result in Eq.(\ref{q s=0=l}) 
but with the number $14$ in the place of the number  $20$  in the 3rd leading order.

Finally, from Eqs.(\ref{W^2 s=0=l}) and (\ref{q s=0=l}) it follows that
\begin{align}\label{q/W^2 s=0=l}
&
\frac{\qT(\fNIA)}{\left|\WT\right|^2}=
2 \pi  \nb
+\frac{23 \pi  \nb   ^2}{3} -
\frac{\pi  \nb ^3}{18}  \left(12 \pi ^2-132 \left( \ln (2 \nb )+ \gamma_E \right)-85\right)+
o\left(\nb^4\right),\quad \ell=s=0.
\end{align}
 The third  leading order differs from Eq.(\ref{q/W^2 s=0 gral l}) in that the `-85'
is a `+59' in Eq.(\ref{q/W^2 s=0 gral l}); this is a consequence of the corresponding discrepancies in both $q(\fNIA)$ and $\Ain$.


\subsubsection{Cases $\ell=-s=1,2$} \label{Sec:l=-s=1}

By inserting the values $\ell=-s=1$ and $\ell=-s=2$ into Eq.(\ref{q/W^2 gral s,l}) we respectively obtain
\begin{align}\label{q/W^2 -s=1=l}
4M^2\frac{\qT(\fNIA)}{\left|\WT\right|^2}=\frac{4 \pi }{9}\nb^3+\frac{227 \pi }{135}\nb^4+
\frac{\pi  \left(-1200 \pi ^2+5640\left( \ln (2\nb )+ \gamma_E\right) +10457\right)}{8100}\nb^5
+o\left(\nb^5\right), \quad \ell=-s=1,
\end{align}
and
\begin{align}\label{q/W^2 -s=2=l}
4M^4\frac{\qT(\fNIA)}{\left|\WT\right|^2}=\frac{\pi }{75}\nb^5+\frac{191 \pi }{3500}\nb^{6}-
\frac{\pi  \left(19600 \pi ^2-59920 \left(\ln (2\nb )+ \gamma_E\right) -348727\right)}{4410000}\nb^{7}
+o\left(\nb^7\right),\quad \ell=-s=2.
\end{align}


%
%


\section{Radial solutions} \label{sec:radial}

In this section we obtain the small-frequency behaviour of the upgoing and ingoing radial solutions  valid for {\it arbitrary} radius $r$.
For this purpose, we use a special trick based on the Barnes integral representation of the hypergeometric function.
As shown in Sec.\ref{sec:BC}, the radial-dependence of the BC contribution to the GF only
comes in through the ingoing radial solution, not the upgoing one (see Eqs.(\ref{eq:DeltaG in terms of Deltag}) and (\ref{eq:DeltaG in terms of Deltag,Teuk})). Therefore, the radial-dependence
at late times of the GF itself also only comes in through the ingoing radial solution.
For this reason, we apply the mentioned Barnes trick to give explicitly the first three leading orders of the ingoing radial solution, whereas we only give the leading order (for which the Barnes
trick is not necessary) of the upgoing solution.
We show, however, how the Barnes trick can be used to obtain the behaviour of the ingoing solution up to arbitrary order in the frequency within the perturbative regime and we note that it could be similarly applied to the upgoing solution\fixme{Is that so?}.

The technique we use for obtaining the small-frequency expansion of the radial solutions can be applied just the same to the RW or to the BPT solutions.
Besides, from the expansion for the RW solutions one can readily obtain the expansion for the BPT solutions via the Chandrasekhar transformation of Sec.\ref{sec:transf RW-BPT}.
For this reason, we only give the expansions explicitly for one type of solutions: the BPT solutions for the upgoing modes and the RW solutions for the ingoing modes.


\subsection{Upgoing radial solution}

The results in this subsection correspond to the particular normalization of the BPT solutions that is chosen in ST.
Taking the small-$\ob$ asymptotics of Eq.4.9 MSTa\fiximp{Write this expression?} we obtain
\begin{equation}
\RTupN
= 
\frac{\RAMmLead\Gamma(2\ell+1)\Gamma(2\ell+2)}{2^{\ell}\Gamma(\ell+1)\Gamma^2(\ell+s+1)\ob^{\ell+s-1}}R_0^{\RAM}
+o\left(\ob^{-\ell-s+2}\right),
\end{equation}
where we have used Eqs.(\ref{eq:Kmu}) and (\ref{eq:Kmmu}).
From
Eq.(\ref{eq:R0})
\fixme{Make sure we use the same normalization as in Eq.138 ST} for $R_0^{\RAM}$ we then obtain the leading-order asymptotics:
\fixme{Same attention as below Eq.(\ref{eq:Page R}): 
this assumess that $n=0$ is the dominant term for $\ob \to 0$ in the $n$-sum in Eq.138ST, which should be checked}

\begin{equation}
\label{eq:Rup lead order}
\RTupN=
\frac{\RAMmLead\Gamma^2(2\ell+1)\Gamma(2\ell+2)}{ 2^{\ell}\Gamma^2(\ell+1)\Gamma^2(\ell+s+1)\Gamma(\ell-s+1)}
\frac{(\rb-1)^{-s}{}_2F_1\left(-\ell,-\ell-s;-2\ell;\frac{1}{\rb}\right)}{\ob^{\ell+s+1}}
a_0^\text{T} 
+o\left(\ob^{-\ell-s}\right).
\end{equation}


\subsection{Ingoing radial solution}

\fixme{Do we have the small freq. exp. of just $\f$ or of $\g$ as well? do we have it for BPT as well as RW?}
In this subsection we 
consider the ingoing radial solutions of the RW equation.
We start with 
Eqs.
(\ref{eq:rln X and f,g}), (\ref{eq:coeff tra norm MST}), 
(\ref{eq:RW Xin}) and (\ref{eq:MSTRWin})\footnote{We note that there is a typographical error on the right hand side of Eq.(6)~\cite{PhysRevLett.109.111101}: there is a factor $\rb^{s+1}$ missing and the $j$-sum in the denominator is missing $\frac{\Gamma(a)\Gamma(b)}{\Gamma(c)}$ as in Eq.(\ref{eq:gamma's args}) but with $k\to j$.\fixme{Check. This is the 2nd typo that we mention in this paper that we made in~\cite{PhysRevLett.109.111101}, should we instead write an Errata for ~\cite{PhysRevLett.109.111101}?}},
\begin{align}  \label{eq:f small-nu}
   &
\f(r,\omega) =
\frac{X_s^\text{in}}
{
N_s^\text{in}
 \sum_{k=-\infty}^{\infty}a_k \frac{\Gamma(a)\Gamma(b)}{\Gamma(c)}}=
\\&
\rb^{s+1}
\frac{e^{-i\ob\rb}(\rb-1)^{-i\ob}}{ \sum_{k=-\infty}^{\infty}a_k \frac{\Gamma(a)\Gamma(b)}{\Gamma(c)}}
  \sum_{k=-\infty}^{\infty}a_k \frac{\Gamma(a)\Gamma(b)}{\Gamma(c)}
{}_2F_1(a,b;c;1-\rb),
\nonumber
\end{align}
where 
\begin{align} \label{eq:gamma's args}
a\equiv k+\RAM+s+1-i\ob, \quad
b\equiv -k-\RAM+s-i\ob,\quad
c\equiv 1-2i\ob.
\end{align}

Let us here carry out a basic comparison of MST's leading order behaviour with other results in the literature.
We can use the small-frequency asymptotics for the `in' solution of the Teukolsky equation given by Eqs.3.6.13 and 3.6.15~\cite{th:CasalsPhD} \footnote{Note that there is a typographical error in Eq.3.6.15~\cite{th:CasalsPhD}: 
a factor $\rb^{-s+i\ob}$ is missing on its right hand side.} 
(see also~\cite{ar:PageI'76,ar:J&McL&Ott'95}). Setting $a=0$, we obtain:

\begin{align} \label{eq:Page R}
&
\Rin{s}(r,\omega)\sim
r_h^{-2s} 
e^{-\ob i} \Rintra{s} (\rb-1)^{-s-\ob i}\rb^{-s+\ob i}{}_2F_1(-\ell-s,\ell-s+1,1-s-2\ob i;1-\rb)=
\\ &
=r_h^{-2s}  e^{-\ob i} \Rintra{s} (\rb-1)^{-s-\ob i}\rb^{-\ob i}{}_2F_1(-\ell-2i\ob,\ell+1-2i\ob,1-s-2\ob i;1-\rb),
\qquad (\rb-1)\ob \ll \ell+1.
\nonumber
\end{align}
\fixme{
To leading order for small $\nb$, Eqs.137, 138, 167 ST differ (assuming that $n=0$ is the dominant term for $\ob \to 0$ in the $n$-sum in Eq.138ST, which should be checked) from Page's Eq.(\ref{eq:Page R}) by a factor `2' (not even $r_h$)
in the case $s\leq 0$?
}
In order to obtain the RW ingoing solution from BPT's one, we need to apply the Chandrasekhar transformation as per Eq.(\ref{eq:Chandr BPT to RW}).
Specifically setting $s=0$ in Eq.(\ref{eq:Page R}),  and taking into account of the normalizations as per Eqs.(\ref{eq: bc f}) and (\ref{eq: bc Rin}), we obtain
\begin{equation} \label{eq:Page f}
\f(r,\omega)\sim
e^{-\ob i} (\rb-1)^{-\ob i}\rb^{1+\ob i}{}_2F_1(-\ell,\ell+1,1-2\ob i;1-\rb),
\qquad (\rb-1)\ob \ll \ell+1,\quad s=0.
\end{equation}
The leading order ($\ob=0$) behaviour of Eq.(\ref{eq:Page f}) clearly agrees with that from Eq.(\ref{eq:f small-nu}).
Note that the asymptotics of Eq.(\ref{eq:Page R}) are not amenable to carrying out Fourier frequency-integrations.
We now proceed to give a prescription for expanding the ingoing RW solution to arbitrary orders in the frequency.

Now that we know from Sec.\ref{sec:Tail an and nu} the behaviour of the coefficients $a_k$ 
 as $\ob\to0$ our challenge is obtain a
suitable expansion for the hypergeometric functions in Eq.(\ref{eq:f small-nu}).
To this end we employ the Barnes integral representation (e.g., Eq.15.6.6~\cite{bk:onlineAS}) which gives
\begin{align} 
\label{eq:barnes}
\frac{\Gamma(a)\Gamma(b)}{\Gamma(c)} {}_2F_1(a,b,c,z) = \frac{1}{2\pi i}
\int\limits_{-i\infty}^{i\infty}\frac{\Gamma(a+t)\Gamma(b+t)\Gamma(-t)}{\Gamma(c+t)}
(-z)^t dt,\quad a,b\neq 0,-1,-2,\dots
\end{align}
where the path of integration must be chosen to separate the poles of $\Gamma(-t)$ at $t=0,
1,2,\dots$ from those of  $\Gamma(a+t)$ and $\Gamma(b+t)$, respectively at $t=-a, -a-1,-a-2,\dots$ and
$t=-b, -b-1,-b-2,\dots$ 
 
For small $\ob$ we know from Eqs.(\ref{eq:small-omega nu}) and (\ref{eq:2nd order nu}) that
\begin{align*}
\RAM \equiv \ell+\delta\RAM= \ell - \RAMmLead \ob^2 + O(\ob^3),
\end{align*}
where $\RAMmLead$ is positive so that the corresponding poles lie at unit intervals left from
\begin{align*}
   -&k-\ell -s-1 + i \ob + \RAMmLead \ob^2 +O(\ob^3),\\
    & k+\ell-s+ i \ob - \RAMmLead \ob^2 +O(\ob^3).\\
\end{align*}
Note that when $\omega=0$ we have double poles and it is not possible to find a splitting contour and the representation breaks down.

\begin{figure}[h!]
\begin{center}
                \includegraphics[width=12.2cm]{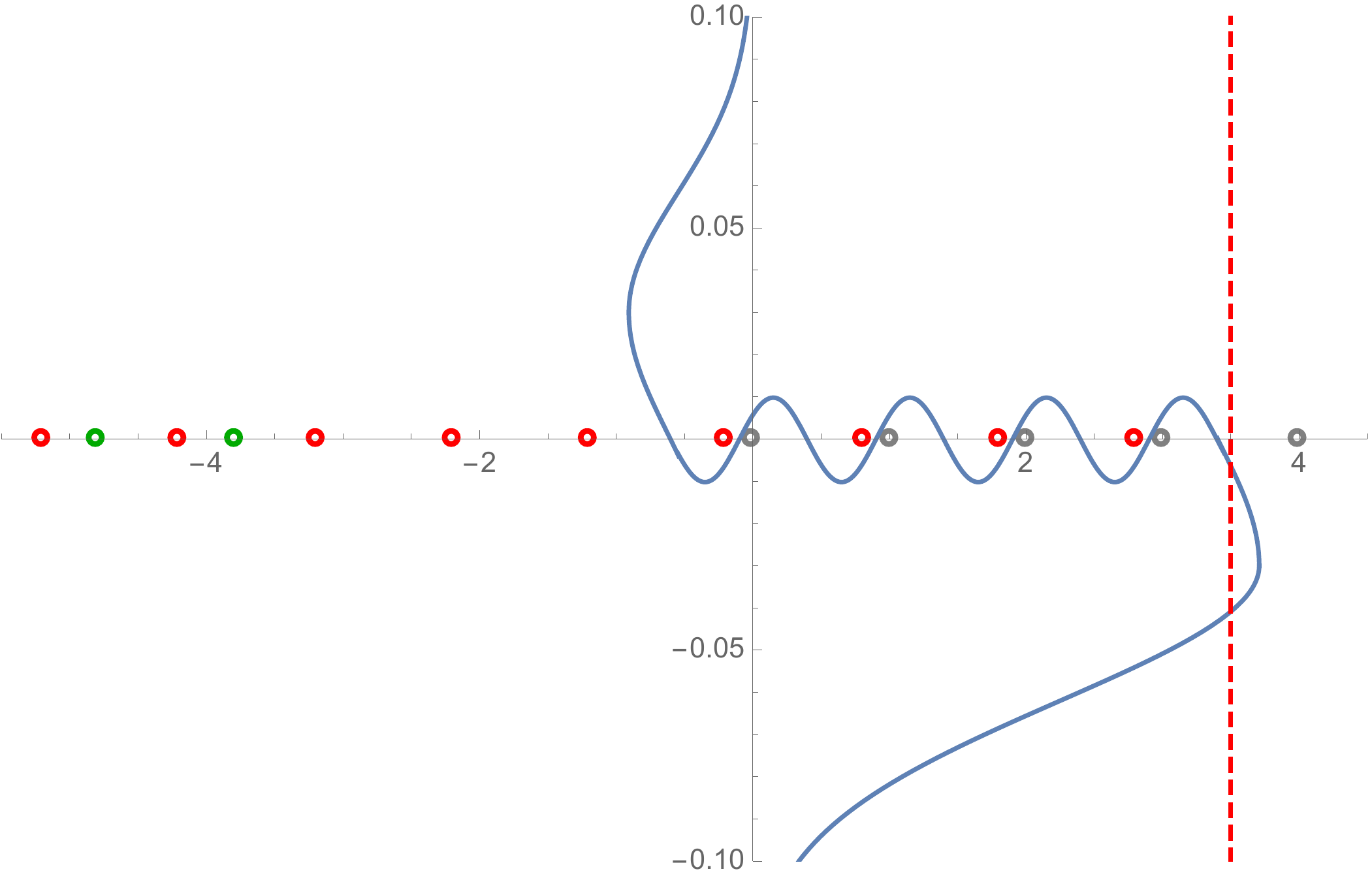}
\end{center}
\caption{
\fiximp{Is it ok that the red poles are slightly shifted towards the left of the integers for $t>0$? }
(Color online).
Plot of the contour deformation in the complex $t$-plane for the Barnes integral representation Eq.(\ref{eq:barnes}) for the  hypergeometric function.
The straight blue curve is the contour ( $C_1$ ) in  Eq.(\ref{eq:barnes}).
The dashed red vertical line is the contour  ( $C_2$ ) in  Eq.(\ref{eq:C2}).
The circled crosses correspond to the poles of the various $\Gamma$ functions in the integrand of Eq.(\ref{eq:barnes}):
the blue ones (at $t=0,1,2,\dots$) are those of $\Gamma(-t)$; the green ones (at $t=-a,-a-1,-a-2,\dots$) are those of $\Gamma(a+t)$;
the red ones  (at $t=-b,-b-1,-b-2,\dots$) are those of $\Gamma(b+t)$.
This plot is for the case $k=1$, $\ell=2$, $s=0$, $\ob=-0.1 i$, which yields 
$a=3.9+O(\ob^2)$, $b=-3.1+O(\ob^2)$
 and $N=3$. 
}
\label{fig:contour}
\end{figure}

In Fig.~\ref{fig:contour} we illustrate the poles and contour $C_1$ of the Barnes integral representation for the term of the MST series Eq.~(\ref{eq:f small-nu}) for $k=1$, $\ell=2$, $s=0$, $\ob=-0.1 i$. The proximity of the poles for small $\ob$ leads to threading contour $C_1$ complicates evaluation of the integral however we may overcome this simply by deforming the contour to the contour $C_2$ lying to the right of all the poles of $\Gamma  (a+t) $ and $\Gamma(b+t)$ collecting the residues from the poles of $\Gamma(-t)$ as we do\fixme{Fix sentence}.  The contour $C_2$ can now be taken well away from any poles
for example along the line $7/2+i t$ in Fig.~\ref{fig:contour} and is exponentially convergent.

Letting $N\equiv \max (k+\ell-s,-k-\ell-s-1)$ denote the rightmost pole of $\Gamma  (a+t) $ and $\Gamma(b+t)$, the sum over residues gives
\begin{align}\label{eq:Barnes poles}
F_1(k)\equiv\sum\limits_{n=0}^{ N} \frac{(-1)^n}{n!} \frac{\Gamma(n+k+\ell+s+1-i\ob+\delta\RAM)\Gamma(n-k-\ell+s-i\ob-\delta\RAM)}
{\Gamma(n+1-2i\ob)}(\rb-1)^n,
\end{align}
where the coefficients of the polynomial may readily be expanded about $\ob=0$. The corresponding contribution from $C_2$ is given by
\begin{align} 
\label{eq:C2}
&F_2(k)\equiv \\ &
 \frac{1}{2\pi }
\int\limits_{-\infty}^{\infty}\frac{\Gamma(N+k+\ell+s+3/2-i\ob +\delta\RAM + i y)\Gamma(N -k-\ell+s+1/2-i\ob -\delta\RAM+ i y)\Gamma(-N-1/2 - i y)}{\Gamma(N+3/2-2i\ob + i y)}
(r-1)^{N+1/2 + i y} dy.
\nonumber
\end{align}

From the above, we obtain
\begin{align}  \label{eq:f small-nu Barnes}
   &
\f (r,\omega)=
\rb^{s+1}\frac{e^{-i\ob\rb}(\rb-1)^{-i\ob}}{ \sum_{k=-\infty}^{\infty}a_k \frac{\Gamma(a)\Gamma(b)}{\Gamma(c)}}
\sum_{k=-\infty}^{\infty}a_k\cdot (F_1(k)+F_2(k)).
\end{align}
We note that the integral in $F_2$ is exponentially convergent.
All that one has to do is completely trivial small-$\ob$ expansions of the various coefficients appearing in Eqs.(\ref{eq:Barnes poles}) and (\ref{eq:C2}).
For that, the  small-$\ob$ expansions for $\RAM$ and for $a_k$ of the previous section are required.

\subsection{Cases $\ell=-s=0,1,2$}

The results of the previous subsection are valid for {\it any} values of the spin $s$ (except for, obviously, Eq.(\ref{eq:Page f})) and  multipole number $\ell\ge |s|$.
In this subsection we use
those results
 to give the small-$\ob$ expansions for the ingoing RW solutions specifically for the modes $\ell=-s=0$, $1$ and $2$, as these
are the modes that we will use in the next section in order to calculate the Schwarzschild black hole response to a specific perturbation.

\fixme{The following expressions for $\f$ for $\ell=-s=0$, $1$ and $2$ for general $r$ should be checked (particularly for overall factors)}
Specifically, for $\ell=s=0$  we have, for general $r$,
\begin{align} \label{eq:Barnes ell=s=0}
&\f(r,-i\fNIA)=
\rb
+\rb\left\{
-\log \left(\rb-1\right)-1
+\int_{-\infty}^{\infty}d\iv
\frac{(\rb-1)^{\frac{1}{2}+i \iv} \text{sech}(\pi  \iv)}{1+2 i \iv}
\right\}\nb+
\nonumber\\ &
\frac{\rb}{6}\left\{
\rb^2+8 \rb-6+3 \log \left(\rb-1\right) \left(\log \left(\rb-1\right)+2\right)
-
\right.\nonumber \\ & \left.
\int_{-\infty}^{\infty}d\iv\frac{ (\rb-1)^{\frac{1}{2}+i \iv} \text{sech}(\pi  \iv)}{(4 \iv^2-8 i\iv-3)} \left(21-3\rb -6i\rb \iv+10 i \iv+\frac{24}{1+2  i \iv}-6 (3+2 i \iv) \log \left(\rb-1\right)\right)
\right\}\nb^2
+o\left(\nb^2\right).
\end{align}

For $\ell=-s=1$ we have,  for general $r$,
\begin{align} \label{eq:Barnes ell=-s=1}
&\f(r,-i\fNIA)=
\rb^2
-\left\{
\frac{ \rb }{2^4}\left(3-2 \rb+\rb \log \left(\rb-1\right)\right)
+\frac{1}{\pi }\int_{-\infty}^{\infty}d\iv
\left(\rb-1\right)^{\tfrac{5}{2}+i \iv} \Gamma \left(-i \iv-\tfrac{5}{2}\right) \Gamma \left(i \iv+\tfrac{1}{2}\right)
\right\}\nb
+
\nonumber\\ &
\bigg\{
 20 - 168 \rb + 144 \rb^2 + 16 \rb^3 + 3 \rb^4
 +15 \rb \log \left(\rb-1\right) \left(6-4 \rb+\rb \log \left(\rb-1\right)\right)
+
\nonumber \\ & \left.
\frac{1}{4}\int_{-\infty}^{\infty}d\iv
\frac{ (\rb-1)^{\frac{5}{2}+i \iv} \text{sech}(\pi  \iv) \Gamma \left(i \iv+\tfrac{1}{2}\right) }{ \Gamma \left(i \iv+\tfrac{9}{2}\right)}
\left(809-15\rb (1+2 i \iv)-60 (7+2 i \iv) \log
   \left(\rb-1\right)+
   \right. \right.\nonumber \\ & \left. \left.
   154 i \iv+\frac{720}{2 i\iv+1}+\frac{480}{2 i\iv+3}+\frac{240}{2 i\iv+5 }\right)
\right\}\frac{\nb^2}{30}
+o\left(\nb^2\right).
\end{align}

For $\ell=-s=2$ we have, for general $r$,
\begin{align} \label{eq:Barnes ell=-s=2}
&\f(r,-i\fNIA)=
\rb^3
\nonumber\\ &
- \left\{
\frac{1}{6}  \left(6 \rb^3 \log \left(\rb-1\right)-19 \rb^3+30 \rb^2-15 \rb+10\right)
+\frac{12}{\pi  \rb}
\int_{-\infty}^{\infty}d\iv
  \left(\rb-1\right)^{\frac{9}{2}+i \iv} \Gamma \left(-\tfrac{9}{2}-i \iv\right) \Gamma \left(\tfrac{1}{2}+i \iv\right)
\right\}\nb+
\nonumber\\ &
 \left\{
\frac{1}{5040 \rb}\left[
840 \rb \log \left(\rb-1\right) \left(3 \rb^3 \log \left(\rb-1\right)-19 \rb^3+30 \rb^2-15 \rb+10\right)+
\right. \right. \nonumber \\  & \left.\left.
360 \rb^6+1728 \rb^5+40385 \rb^4-67020 \rb^3+49470 \rb^2-397760 \rb+24860
   \right]
\right.\nonumber \\ & \left.
+\frac{1}{35 \pi  \rb}
\int_{-\infty}^{\infty}d\iv 
  \frac{(\rb-1)^{\frac{9}{2}+i \iv} \Gamma \left(i \iv+\frac{1}{2}\right)}{ \Gamma \left(i \iv+\frac{13}{2}\right)}\left[
  175 \pi  (\rb-1) (13+2 i \iv) \text{sech}(\pi  \iv)+
   \right.
    \right.\nonumber \\ & \left. 
 \Gamma \left(-\tfrac{9}{2}-i   \iv\right) \Gamma \left(\tfrac{13}{2}+i \iv\right) \left(420 \rb+420 \log \left(\rb-1\right)+420 \psi\left(i \iv+\tfrac{1}{2}\right)-420 \psi \left(i \iv+\tfrac{11}{2}\right)-1019\right)
   \right]
\Bigg\}
\nb^2
+o\left(\nb^2\right),
\end{align}
where $\psi$ is the digamma function.


\section{Late-time Tail} \label{sec:plot perturbation}

In this section we illustrate how one can apply the small-frequency expansions of the previous sections to obtain physically-relevant results:
the late-time behaviour to high order of the black hole response, at an arbitrary point, to a field perturbation of arbitrary integer spin.
Before we do that in Sec.\ref{sec:pert}, we first derive the late-time tail of the Green function in the next subsection.


\subsection{Late-time Tail of the Green Function} \label{sec:late-time GF}

The late-time behaviour of the RW Green function is, via Eq.(\ref{eq:Green}), given by that of its $\ell$-modes ${}_sG^\text{ret}_{\ell}$.
In its turn, the late-time behaviour of ${}_sG^\text{ret}_{\ell}$ is dominated by the BC $\ell$-modes $\GlBC$ in Eq.(\ref{eq: G^BC integral}).
Finally, it follows from the latter equation that  $\GlBC$ at late times is given by the small-frequency behaviour of $\DGw$.
We now give more detailed expressions for these expansions.

The radius-independent part of the BC integrand in the BPT case is given in general (except for $\ell=s=0$) in Eq.(\ref{q/W^2 gral s,l}),
and specifically in 
Eqs.(\ref{q/W^2 s=0=l})--(\ref{q/W^2 -s=2=l})   for the  lower modes. 
Let us generically write its small-frequency expansion as:
\begin{equation} \label{eq:q/W2 BPT gral}
\frac{\qT}{\left|\WT\right|^2}=v_0\nb^{2\ell+1}+v_1\nb^{2\ell+2}+\left(v_{2a}+v_{2b} \ln \nb\right)\nb^{2\ell+3}+o(\nb^{2\ell+3}),
\end{equation}
where the $\nb$-independent constant coefficients $v_0$, $v_1$ and $v_{2a/b}$ are readily readable from the mentioned equations.
In order to obtain  the corresponding RW quantity, we can trivially use Eq.(\ref{eq:q/W^2 vs qT/WT^2}). We write the small-frequency expansion of the 
proportionality constant in Eq.(\ref{eq:q/W^2 vs qT/WT^2}) as
\begin{equation}\label{eq:C small-freq gral}
\mathcal{C}=\mathcal{C}_0+\mathcal{C}_1\nb+\mathcal{C}_2\nb^2+O(\nb^3),
\end{equation}
where the $\nb$-independent constant coefficients $\mathcal{C}_{0/1/2}$ can be read off from
from Eqs.(\ref{eq:C}) and (\ref{eq:C spin-2}).

The radius-dependent part of the BC integrand in the RW case is given in general in Eq.(\ref{eq:f small-nu Barnes}); all that one has to do is a trivial small-$\ob$ expansion of the summand in Eq.(\ref{eq:Barnes poles})
and the integrand in Eq.(\ref{eq:C2}) (with the use of Eqs.(\ref{eq:small-omega nu}) and (\ref{eq:gen}) in general, and specifically Eq.(\ref{eq:specific an's}) for the  lower modes).
Again, let us write the small-frequency expansion of the ingoing RW solution as
\begin{equation} \label{eq:f small-freq gral}
\f(r,-i\sigma)=f_{\ell 0}(r)+f_{\ell 1}(r)\nb+f_{\ell 2}(r)\nb^2+O(\nb^3),
\end{equation}
where the $\nb$-independent (but radius-dependent) coefficients $f_{\ell,0/1/2}$ can be readily obtained in the manner just indicated.

The analytic small-frequency expansions of the radius-independent (Eq.(\ref{eq:q/W2 BPT gral}) times  Eq.(\ref{eq:C small-freq gral})) and radius-dependent  (Eq.(\ref{eq:f small-freq gral}) evaluated at $r$ times the same expression but evaluated at $r'$) parts
of the BC integrand in the RW case are then to be  put together in Eq.(\ref{eq:DeltaG in terms of Deltag}) and
integrated as per Eq.(\ref{eq: G^BC integral}).
The result, for late times is, straight-forwardly,
\begin{align} \label{eq:late-time GF}
&
\GlBC=
-\frac{ (2 \ell+2)!}{4 \pi  M^2}
\Bigg\{
\mathcal{C}_0v_0f_{00}(r,r')+
(2 \ell+3)
\bigg[
\mathcal{C}_0 v_0 f_{01}(r,r')+\left( \mathcal{C}_0v_1+\mathcal{C}_1v_0\right)f_{00}(r,r')
\bigg]
   {\bar t}^{-1}
   +
 \\ &
2(\ell+2) (2 \ell+3)
\bigg[
\mathcal{C}_0 v_0 f_{012}(r,r') +\left( v_0 \mathcal{C}_1+\mathcal{C}_0v_1 \right)f_{01}(r,r')   
+   
 \nonumber   \\ &
\left(  \mathcal{C}_0 v_{2b} \left(\psi (2   \ell+5)-  \ln \bar t   \right)
   +\mathcal{C}_0v_{2a}+v_0\mathcal{C}_2+\mathcal{C}_1v_1\right)f_{00}(r,r')
  \bigg]
    {\bar t}^{-2}
  \Bigg\}
  {\bar t}^{-2\ell-3}+
o\left(  {\bar t}^{-2\ell-5}\right),
  \nonumber
\end{align}
where
\begin{align}
&
f_{00}(r,r')\equiv f_{\ell 0}(r)f_{\ell 0}(r'),\quad
f_{01}(r,r')\equiv f_{\ell 0}(r)f_{\ell 1}(r')+f_{\ell 0}(r')f_{\ell 1}(r),
\\ &
f_{012}(r,r')\equiv f_{\ell 0}(r)f_{\ell 2}(r')+f_{\ell 0}(r')f_{\ell 2}(r)+f_{\ell 1}(r)f_{\ell 1}(r').
\nonumber
\end{align}
Here we are using the dimensionless time $\bar t\equiv t/(2M)$.
The leading-order behaviour at late times is, therefore, 
$\GlBC\sim \dt^{-2\ell-3}$
for all integer spins. 
We also note that the logarithmic behaviour in Eq.(\ref{q/W^2 gral s,l}) for $\qT/ |\WT|^2$ for small frequencies led to the appearance of a  logarithmic behaviour in $\GlBC$ as
$\dt^{-2\ell-5}\ln \dt$.
The late-time behaviour of the GF of the $4$-dimensional RW equation
 is then obtained by replacing ${}_sG^\text{ret}_{\ell}$ in Eq.(\ref{eq:Green}) by the  expansion in Eq.(\ref{eq:late-time GF}).

A similar analysis can be done for the BPT GF, but using the corresponding  equations instead: (\ref{eq:DeltaG in terms of Deltag,Teuk}) and (\ref{eq:GF Teuk s=2}).
The radius-independent part of the integrand,  $\qT/ |\WT|^2$, we have already given for the BPT case. The radius-dependent part can be obtained from the 
expansions  Eq.(\ref{eq:f small-freq gral}) of the ingoing RW  solutions via the Chandrasekhar transformation of Eq.(\ref{eq:Chandr RW to BPT}).
The late-time behaviour of the $\ell$-modes of the BPT GF is of the same leading order as that in the RW case (i.e., $\dt^{-2\ell-3}$)
and the logarithmic term also generally appears at the same order as in the RW case (i.e., $\dt^{-2\ell-5}\ln \dt$).



\subsection{Late-time Tail of an Initial Perturbation} \label{sec:pert}

We shall now give a particular application of our late-time results for the GF.
Let us here consider an initial field perturbation  given by $u_{\ell}^{ic}(r'_*)\equiv  u_{\ell}(r'_*, t'=0)$ for the $\ell$-multipole of the field and 
 $\dot u_{\ell}^{ic}(r'_*)\equiv  \partial_t u_{\ell}(r'_*, t'=0)$ for the $\ell$-multipole  of the time derivative of the field.
Then
the  response of 
a Schwarzschild black hole
 is given by
\begin{align} \label{eq:perturbation}
u_{\ell}(r_*, t)=\int_{-\infty}^{\infty}dr'_*\left[ \Glret(r,r'; t)\dot u_{\ell}^{ic}(r'_*)+u_{\ell}^{ic}(r'_*)\frac{\partial}{\partial t}\Glret(r,r'; t)\right],
\end{align}
where $G^{ret}_{\ell}(r,r'; t)$ is the $\ell$-multipole of the retarded GF of the wave equation satisfied by the field.
 In this section we will take the RW equation as the wave equation.
The late-time asymptotics of the perturbation $u_{\ell}$ are given by 
replacing $\Glret$ by $\GlBC$ in Eq.(\ref{eq:perturbation}) and approximating  $\GlBC$ by performing a small-frequency expansion
of $\DGw$ in Eq.(\ref{eq: G^BC integral}).

As in~\cite{PhysRevLett.109.111101},
 let us consider the following initial perturbation  for general spin\footnote{As an alternative to the initial data of Eq.(\ref{eq:ic}) we could consider the following initial data (as used in~\cite{CDGOWZ}): zero for the field and  a Gaussian distribution of small width  for the time-derivative of the field.  In this case, the perturbation response is an approximation to the Green function. However, we found that 
such approximation is not as good as (\ref{eq:ic}) for assessing the validity of the late-time asymptotics to the higher-than-leading orders as  intended in this paper.}
\begin{equation} \label{eq:ic}
u_{\ell}^{ic}(r_*)=
\frac{1}{M}\text{exp}\left(\frac{-\left(r_*-x_0\right)^2}{2M^2}\right),
\quad
\dot u_{\ell}^{ic}(r_*)=-\frac{\left(r_*-x_0\right)}{M^2}u_{\ell}^{ic}(r_*),
\end{equation}
 with $x_0\equiv r_*(10M)$.

In Figs.\ref{fig:perturbation,l=0,s=0,r=10}--\ref{fig:perturbation,l=2,s=2,r=10}
 we plot the 
time evolution via the RW equation of
 a spin-$s$ field for various multipole numbers $\ell$ using the initial data Eq.(\ref{eq:ic}), similarly to~\cite{PhysRevLett.109.111101} where we only presented the case $s=0$, $\ell=1$. 
We calculate the time evolution numerically  up to $T=3000M$,  where $T\equiv t-2x_0$, and we compare it with late-time asymptotics.
The numerical solution is obtained using the code in~\cite{Wardell-scalarwave1d} 
for the $(1+1)$-dimensional differential equation which results from the 4-d RW equation after factorizing out  the angle dependence of the solution via scalar spherical harmonics.
We obtain the late-time asymptotics, up to  four leading orders, from  the results of the previous section.
In order to obtain the small-frequency expansion of $\DGw$ via Eq.(\ref{eq:DeltaG in terms of Deltag}), we need the small-frequency expansion of $q/|W|^2$ and of the `in' radial solution $\f$.
We now give the specific expansions of the perturbation response for different spins $s$ and $\ell$ modes.

We found the late-time asymptotics for the case $s=0$ 
and  $\ell=0$ using Eq.(\ref{eq:Barnes ell=s=0}) for the  radial solution and Eq.(\ref{q/W^2 s=0=l}) 
for $q/|W|^2$ ;
for $\ell=1,2$ we merely inserted the corresponding value of $\ell$ into Eq.(\ref{q/W^2 s=0 gral l}).
The results
are the following (the approximation sign is due to the fact
that we have rounded up the coefficients to seven significant figures):
\begin{equation}\label{eq:late-time s=l=0}
\ulBC\approx
\frac{20.02467}{\bar T^3}-\frac{269.1100}{\bar T^4}+\frac{-881.0855 \ln \left(\bar T\right)+7561.913}{\bar T^5}+o\left(\bar T^{-5}\right),\quad s=\ell=0,
\end{equation}
\begin{equation}\label{eq:late-time s=0,l=1}
\ulBC\approx
-\frac{1144.318}{\bar T^5}+\frac{52385.24}{\bar T^6}+\frac{43484.10\ln \left(\bar T\right)-1746947}{\bar T^7}
+o\left(\bar T^{-7}\right),\quad s=0,\ \ell=1,
\end{equation}
\begin{equation}\label{eq:late-time s=0,l=2}
\ulBC\approx
\frac{42634.22}{\bar T^7}-\frac{3114525}{\bar T^8}+\frac{- 1796322 \ln \left(\bar T\right)+146090980}{\bar T^9}+o\left(\bar T^{-9}\right),
\quad s=0,\ \ell=2.
\end{equation}

Figs.\ref{fig:perturbation,l=0,s=0,r=10}--\ref{fig:perturbation,l=2,s=0,r=10}
 show excellent agreement at late times between the numerical solution of the RW equation for $s=0$ and $\ell=0,1,2$ and the late-time asymptotics of Eqs.(\ref{eq:late-time s=l=0})--(\ref{eq:late-time s=0,l=2}).

\begin{figure}[h!]
      \includegraphics[width=12cm]{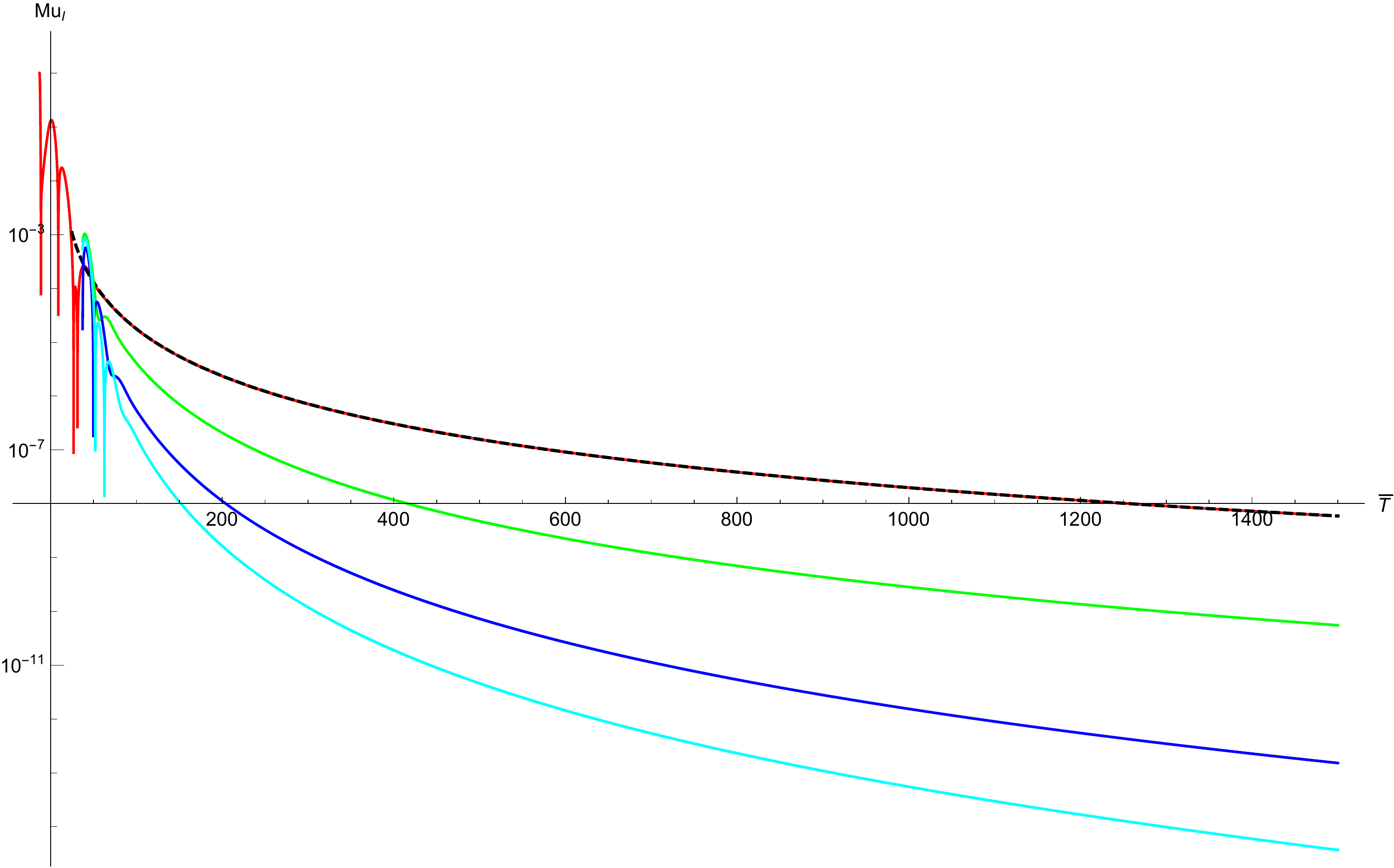}
\caption{
Scalar field $Mu_{\ell}$ at $r_*=x_0$ as a function of time $T/M$ for the initial data Eq.(\ref{eq:ic}) with $x_0=r_*(10M)$ and $s=\ell=0$
(i.e., similar to Fig.1 in~\cite{PhysRevLett.109.111101}  and with the same initial data but here it is  for $s=\ell=0$).
Red curve: numerical solution.
Dashed black curve (overlapping with red curve): late-time expression given by Eq.(\ref{eq:late-time s=l=0})
(which comes from Eqs.(\ref{q/W^2 s=0=l}) together with the small-$\nb$ results in Sec.\ref{sec:radial} for the radial function).
Green curve: numerical solution minus the leading order (i.e., $O(\bar T^{-3})$) in the late-time expression.
Solid blue curve: numerical solution minus the two leading orders (i.e., $O(\bar T^{-3})$ and $O(\bar T^{-4})$) in the late-time expression.
Solid cyan curve: numerical solution minus the four leading orders (i.e., $O(\bar T^{-3})$, $O(\bar T^{-4})$, $O(\bar T^{-5}\ln(\bar T))$ and $O((\bar T)^{-5})$) in the late-time expression.
}
\label{fig:perturbation,l=0,s=0,r=10}
\end{figure}

\begin{figure}[h!]
            \includegraphics[width=12cm]{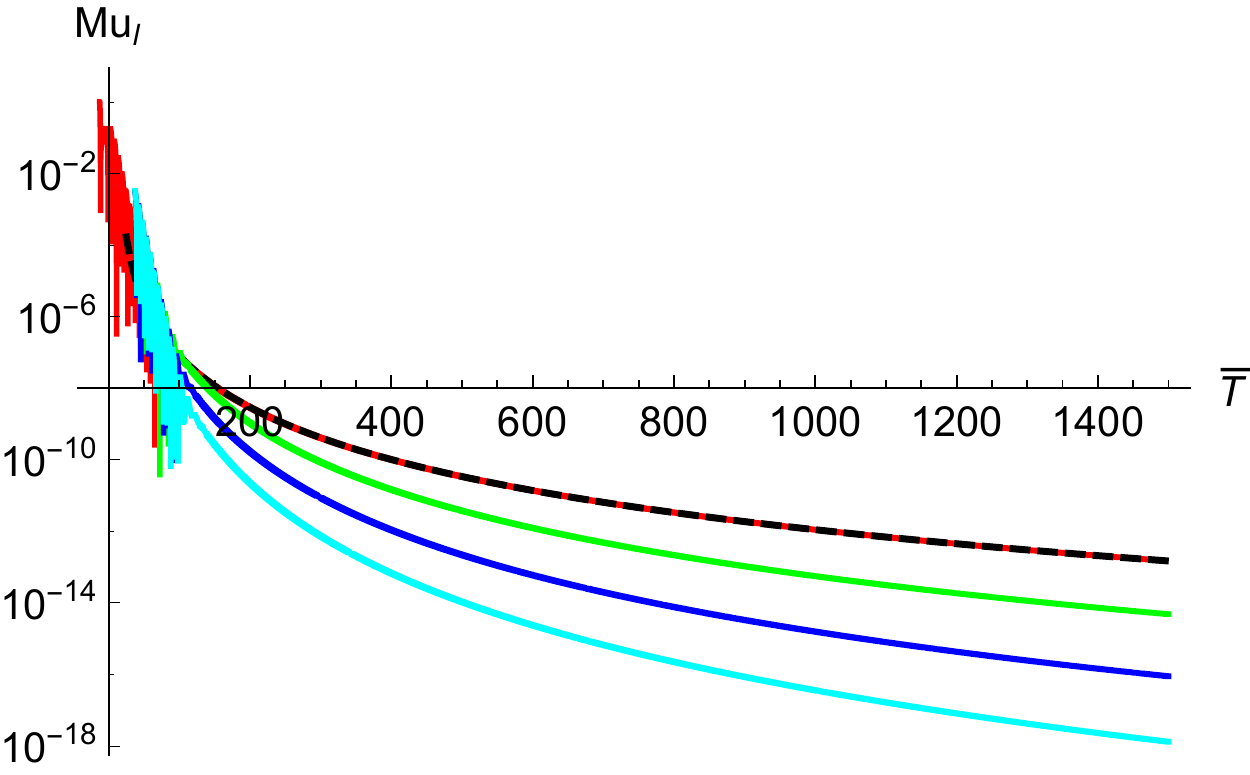}
\caption{
Similar to Fig.\ref{fig:perturbation,l=0,s=0,r=10} but here it is for $s=0$, $\ell=1$
(the orders subtracted from the numerical solution are the corresponding leading order, next-to-leading order and four leading orders for this case).
The late-time expression is given by Eq.(\ref{eq:late-time s=0,l=1}).
}
\label{fig:perturbation,l=1,s=0,r=10}
\end{figure}

\begin{figure}[h!]
            \includegraphics[width=12cm]{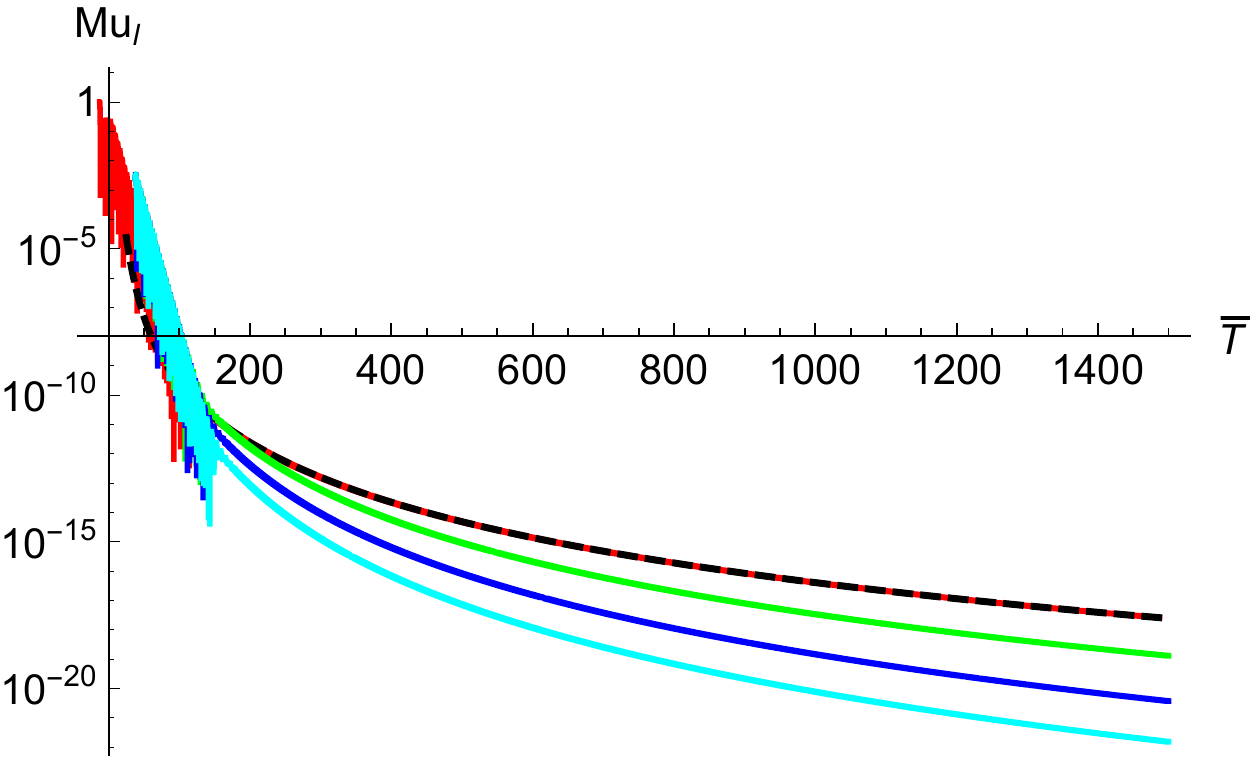}
\caption{
Similar to Fig.\ref{fig:perturbation,l=0,s=0,r=10} but here it is for $s=0$, $\ell=2$
(the orders subtracted from the numerical solution are the corresponding leading order, next-to-leading order  and four leading orders for this case).
}
\label{fig:perturbation,l=2,s=0,r=10}
\end{figure}



The late-time asymptotics for the RW cases $|s|=\ell=1$ and $2$ are obtained similarly to those above for $s=0$:
we used Eqs.(\ref{eq:Barnes ell=-s=1}) and (\ref{eq:Barnes ell=-s=2}) for the small-frequency expansions of the radial solution
and Eqs.(\ref{q/W^2 -s=1=l}) and (\ref{q/W^2 -s=2=l}) for that of $\qT/|\WT|^2$, which is then simply converted to RW version with the use of Eqs.(\ref{eq:qT vs q final}) and
(\ref{eq:Wronskian RW to BPT}).
The final results for the late-time perturbation are the following:
\begin{equation}\label{eq:late-time s=l=1}
\ulBC\approx
- \frac{1338.214}{\bar T^5}+\frac{59157.01}{\bar T^6} - \frac{1970216-62896.05\ln\left(\bar T\right)}{\bar T^7} +o\left(\bar T^{-7}\right)
,\quad  |s|=\ell=1,
\end{equation}
and
\begin{equation}\label{eq:late-time s=l=2}
\ulBC\approx
\frac{60748.13}{\bar T^7} - \frac{4318418}{\bar T^8} + \frac{ (198106997- 3466693 \ln\left(\bar T\right)}{\bar T^9} +o\left(\bar T^{-9}\right),
\quad  |s|=\ell=2.
\end{equation}

Figs.\ref{fig:perturbation,l=1,s=1,r=10}--\ref{fig:perturbation,l=2,s=2,r=10}
 show excellent agreement at late times between the numerical solution of the RW equation for $|s|=\ell=1,2$ and the late-time asymptotics of Eqs.(\ref{eq:late-time s=l=1})--(\ref{eq:late-time s=l=2}).

The above examples illustrate the appearance of a higher-order logarithmic behaviour of a perturbation response, appearing after Price's well-known power-law tail~\cite{Price:1971fb,Price:1972pw}.
We note that, as already pointed out in Sec.\ref{sec:q/W^2},  this logarithmic behaviour 
 appears at {\it third} leading order at late times. It appears at third leading order, not at second leading order as one
might have naively expected from the fact that the logarithmic behaviour at small frequency appears at the second leading order both in the BC strength $\qT$ and in the Wronskian.
This is due to there being delicate cancellations between certain terms of $\qT$ and the Wronskian, which we have been able to derive using the precise values of the coefficients in the expansions.

\begin{figure}[h!]
      \includegraphics[width=12cm]{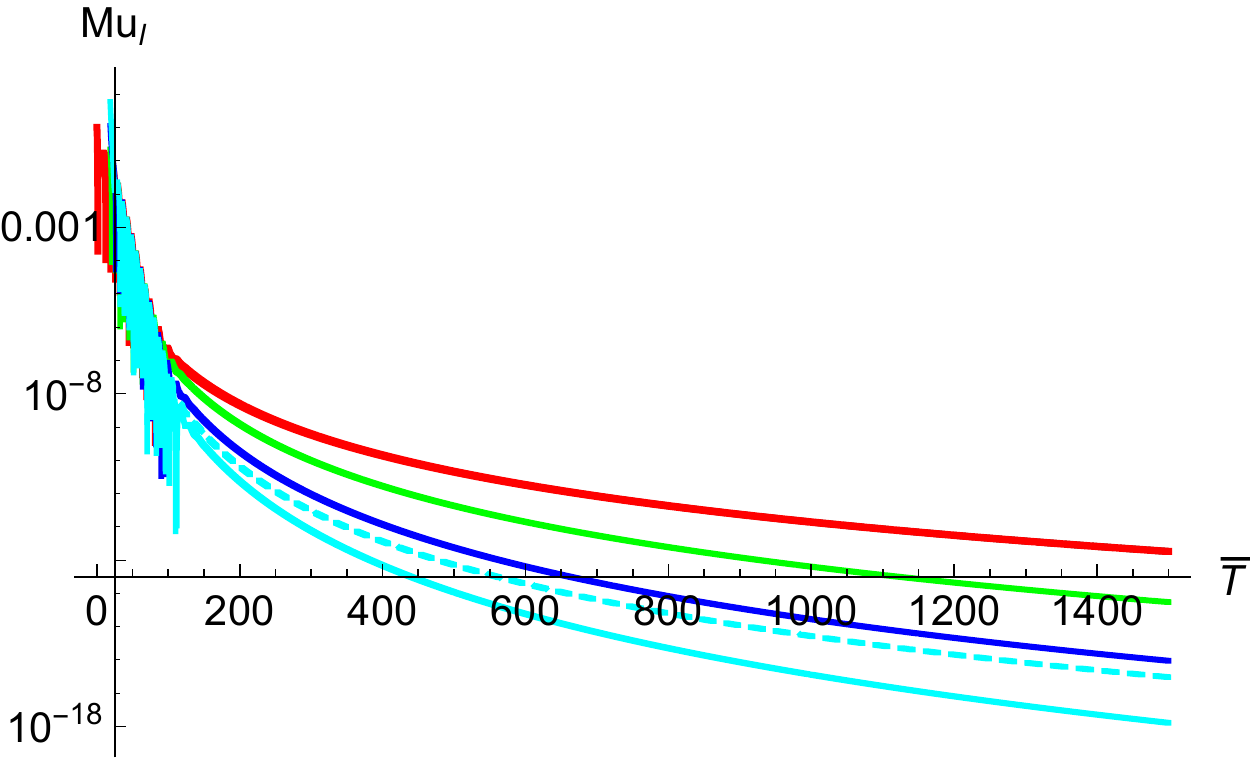}
\caption{
Similar to Fig.\ref{fig:perturbation,l=0,s=0,r=10} but here it is the RW solution for $|s|=\ell=1$
(the orders subtracted from the numerical solution are the corresponding leading order, next-to-leading order  and four leading orders for this case).
}
\label{fig:perturbation,l=1,s=1,r=10}
\end{figure} 

\begin{figure}[h!]
      \includegraphics[width=12cm]{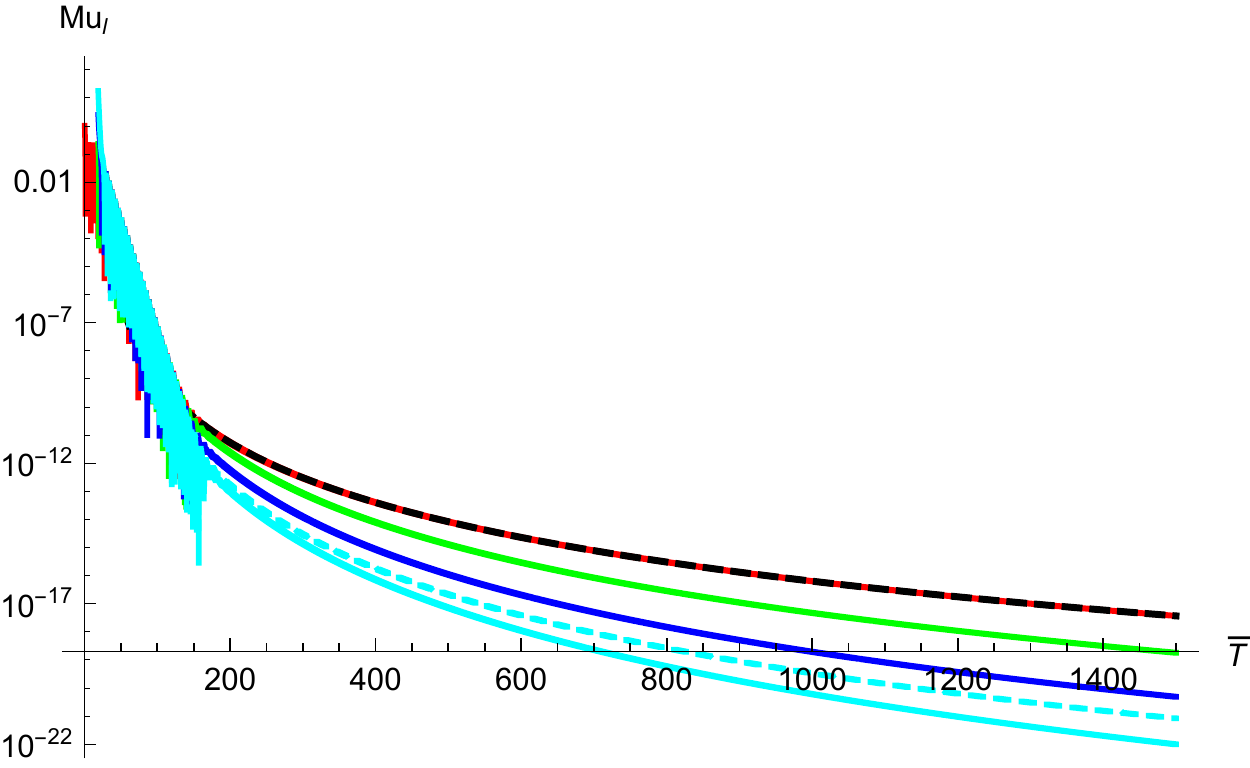}
\caption{
Similar to Fig.\ref{fig:perturbation,l=0,s=0,r=10} but here it is the RW solution for $|s|=\ell=2$
(the orders subtracted from the numerical solution are the corresponding leading order, next-to-leading order  and four leading orders for this case).
}
\label{fig:perturbation,l=2,s=2,r=10}
\end{figure}

\section{Conclusions}

In this paper we have developed the MST formalism for the solutions of the radial Regge-Wheeler and Bardeen-Press-Teukolsky equations, which are obeyed
by linear field perturbations of a Schwarzschild black hole space-time.
We have derived, for the first time, the MST formalism for the solutions of the RW equation for spin-1  as well as 
the MST expressions for the branch-cut-relevant quantities for general spin.
We have given explicit expansions for small frequency up to the first four leading orders for the various MST quantities for general spin.
In principle, the MST series could be expanded to arbitrarily large order in the frequency.
The main difficulty  in achieving that is the fact that 
the  small-frequency expansion of the renormalized angular momentum parameter $\RAM$ that we currently use
does not reproduce the numerically-observed behaviour 
of a {\it sudden} appearance
of an imaginary part in $\RAM$ as the frequency is increased from 0 to larger real values. 

We have used our small-frequency expansions in order to obtain the late-time behaviour  for arbitrary radius of spin-field perturbations of a Schwarzschild black hole  up
to the first four leading orders (for a specific multipole-$\ell$). Our results explicitly reveal a new logarithmic behaviour  at  third order for late times as $\dt^{-2\ell-5}\ln \dt$.
We note that the appearance of a logarithmic behaviour was already predicted by other works (see, e.g.,~\cite{Ching:1994bd,Ching:1995tj,Hod:1999cu,Hod:2009my}).
However, to the best of our knowledge, the order at which the logarithmic behaviour appears was not correctly predicted anywhere (nor was the calculation of the coefficients for general radius carried out).
As we noted at the end of Sec.\ref{sec:q/W^2}, there is a delicate cancellation between different logarithmic terms which averts the appearance of a logarithmic behaviour at a lower order;
this delicate cancellation is probably hard to predict unless an exact and detailed analysis is carried out such as the one in this paper (see also~\cite{Smith:2005sd}, where they find numerical
evidence that the logarithmic behaviour does not appear at first nor second leading orders).

We  succinctly presented in the Letter~\cite{PhysRevLett.109.111101}  the final results that we have derived in this paper. 
We already used some of  these results in the calculation of the scalar self-force in Schwarzschild space-time carried out in~\cite{CDOW13}.
The natural extension of our results 
 is the explicit calculation of the  small-frequency expansions of the MST quantities for the Teukolsky equation in Kerr space-time, with the corresponding late-time analysis
 of perturbations of a Kerr black hole.
 We expect to present the Kerr analysis in the near future, together with its application to a self-force calculation.
 
 \fixme{Unsure whether to include the following}
Finally, while the calculation of the quasinormal modes has been applied to the modelling of radiation after the inspiral of two black holes
via a matching to a numerical relativity solution,  to the best of our knowledge, the branch cut has never been taken into account for such purposes.
We would expect that the inclusion  in the modelling  of the branch cut, together with the quasinormal modes, would help match the analytical solution from perturbation theory
with the numerical relativity one.


\begin{acknowledgments}
We are thankful to Chris Kavanagh and Barry Wardell for useful discussions.
A.C.O. acknowledges support from Science Foundation Ireland under Grant No. 10/RFP/PHY2847.
\end{acknowledgments}

\appendix


\section{Validation of BC results} \label{sec:plots BC}

In this section we compare the small-$\nb$ results presented in this paper with an independent method which
we presented in~\cite{Casals:2012ng}.
The latter method is naturally-adapted to a `mid-frequency' regime (which, although valid for all frequencies, is not practical to use in the asymptotic small- or large-frequency regimes).
In this appendix we illustrate with plots that there is a region of overlap between the results obtained with the  small-$\nb$ method presented here and those obtained with the ÔÔmid-frequencyÕÕ method of~\cite{Casals:2012ng}.
In~\cite{Casals:2012ng}\footnote{We note that there are two typos in Eq.3.6~\cite{Casals:2012ng}: there should be an overall factor `$-1$' in the expression for
$\Delta\tilde h_{\ell}(r,\nu)$, using the notation of that paper, and, in the expression for $\Delta\tilde h_{n,\ell}$, it should be $\tilde a_{n +}$ instead of  $\tilde a_{n -}$. 
The minus sign typo carried over to Fig.10~\cite{Casals:2012ng}, which is therefore a plot of `$+ie^{\nu r_*}\Delta \tilde{g}_{\ell}$',
instead of `$-ie^{\nu r_*}\Delta \tilde{g}_{\ell}$' as stated there.} we already presented plots of quantities focused mainly on the $s=\ell=2$ case. Therefore, for variety, here we will focus on the $s=0$ and $\ell=0,1$ cases.

As explained in~\cite{Casals:2012ng}, the ingoing radial function $\f(r,\omega)$ possesses simple poles along the NIA (which are ultimately irrelevant, as they are cancelled out in the Fourier modes of the GF by the corresponding poles in the Wronskian). It is therefore useful to define the following radial function:
$\Chi(r,\omega)\equiv \sin(-2\pi i\ob)\f(r,\omega)$.

In Figs.\ref{fig:f wrt small nu,l=1,s=0,r=10}, \ref{fig:q,s=0}, \ref{fig:wronskian,s=0,l=1,r=2.8} 
and \ref{fig:DeltaG s=0,l=1}
we respectively plot the radial function $\f$, the BC `strength' $q(\fNIA)$, the Wronskian $W$ and the BC mode $\DGw\left(r,r',-i\fNIA\right)$, all as functions of the frequency $\nb$ along the NIA.

\begin{figure}[h!]
      \includegraphics[width=8cm]{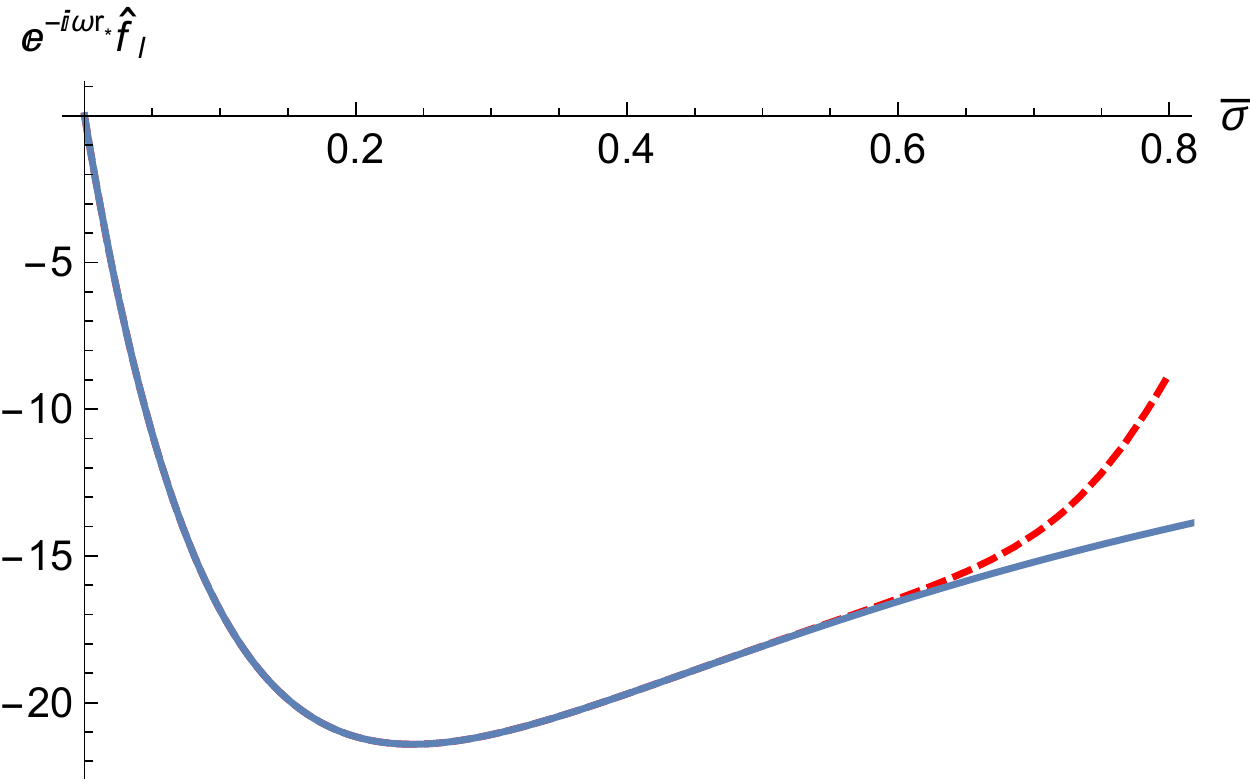}
\caption{
The radial function
$e^{-i\omega r_*}\fb(r,\omega)$ at $r=10M$ as a function of $\nb$ for $s=0$, $\ell=1$.
We calculated the straight blue curve  using the so-called Jaff\'e series (see, e.g.,~\cite{Leaver:1986a,Casals:2012ng}).
and the dashed red curve using the
small-frequency asymptotics presented in this paper
 up to $O(\nb^{15})$ (specifically, using Eq.137~\cite{Sasaki:2003xr} as in this case we did not need to obtain the coefficients of the expansion analytically via the use of the Barnes integral method).
}
\label{fig:f wrt small nu,l=1,s=0,r=10}
\end{figure}

\begin{figure}[h!]
\begin{center}
\includegraphics[width=8cm]{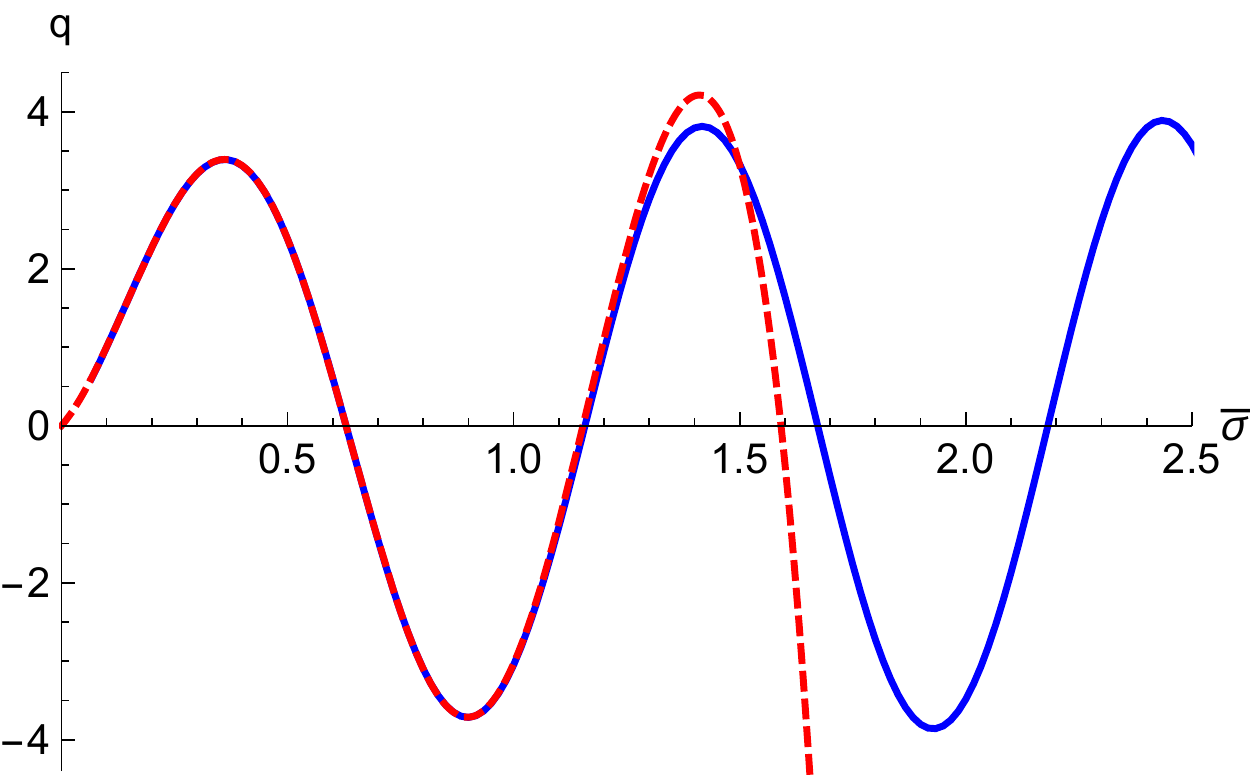}
\includegraphics[width=8cm]{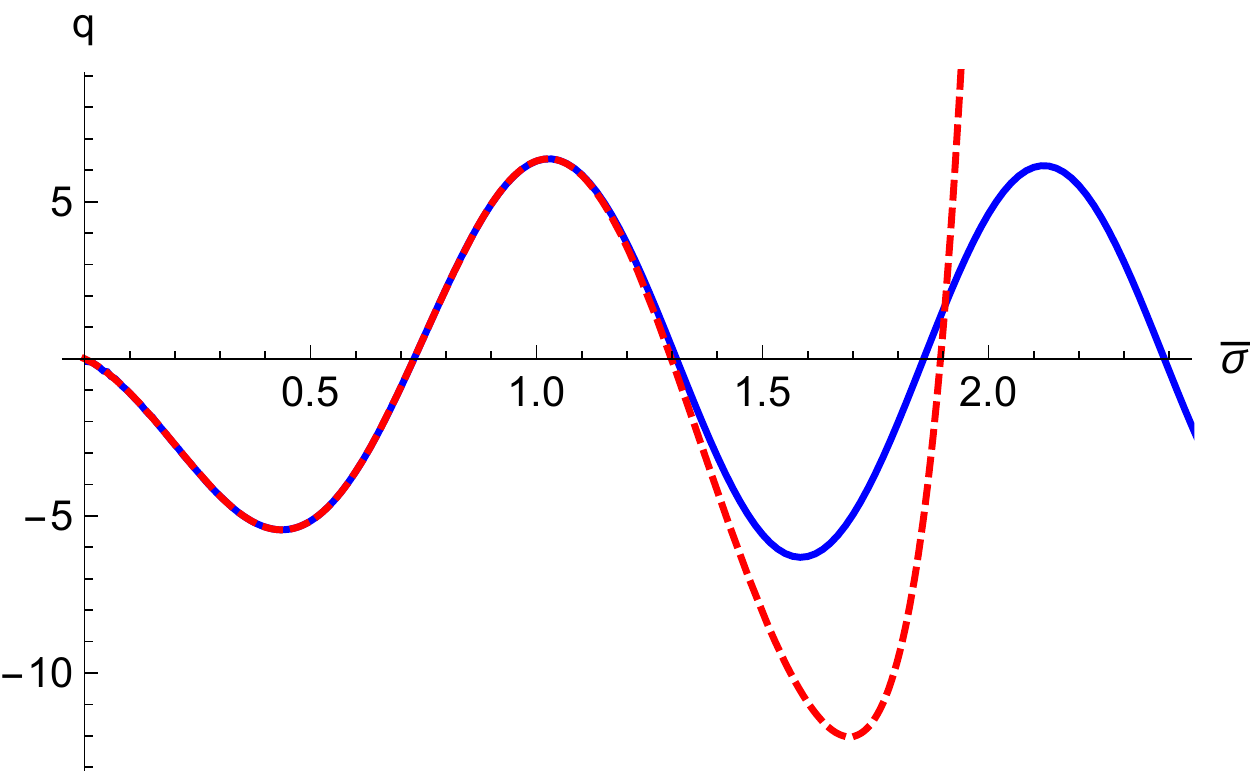}
 \end{center}
\caption{
BC `strength' $q(\fNIA)$ of Eq.(\ref{eq:def q}) (or, equivalently for $s=0$, $\qT(\fNIA)$ of Eq.(\ref{eq:def qT})) as a function of $\nb$.
In the straight blue curves we calculated $q(\fNIA)$ using 
 the mid-frequencyÕ method of~\cite{Casals:2012ng} (with the radial functions calculated
at  $r=5M$).
In the dashed red curves we calculated $q(\fNIA)$ using 
the small-frequency asymptotics presented in this paper
 up to $O(\nb^{15})$.
(a) For $s=0$, $\ell=0$.
(b) For $s=0$, $\ell=1$.
}
\label{fig:q,s=0}
\end{figure}

\begin{figure}[h!]
\begin{center}
 \includegraphics[width=8cm]{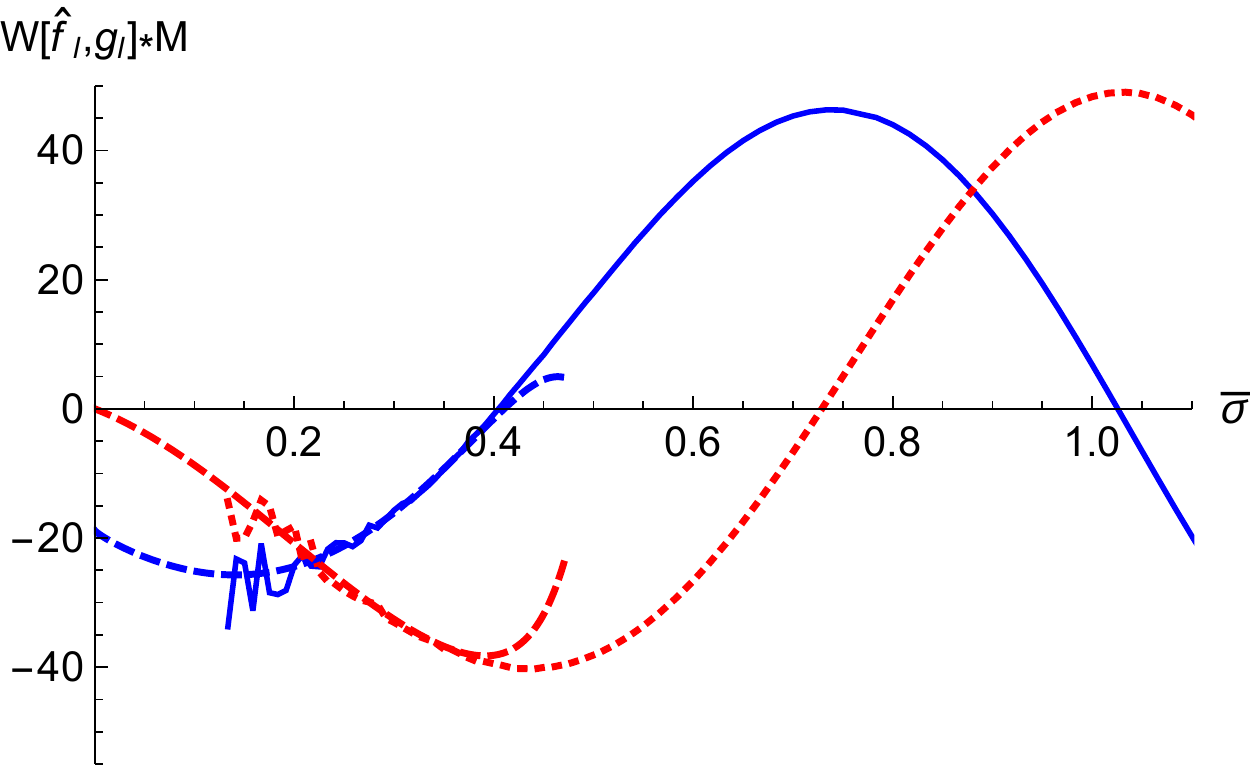}
        \includegraphics[width=8cm]{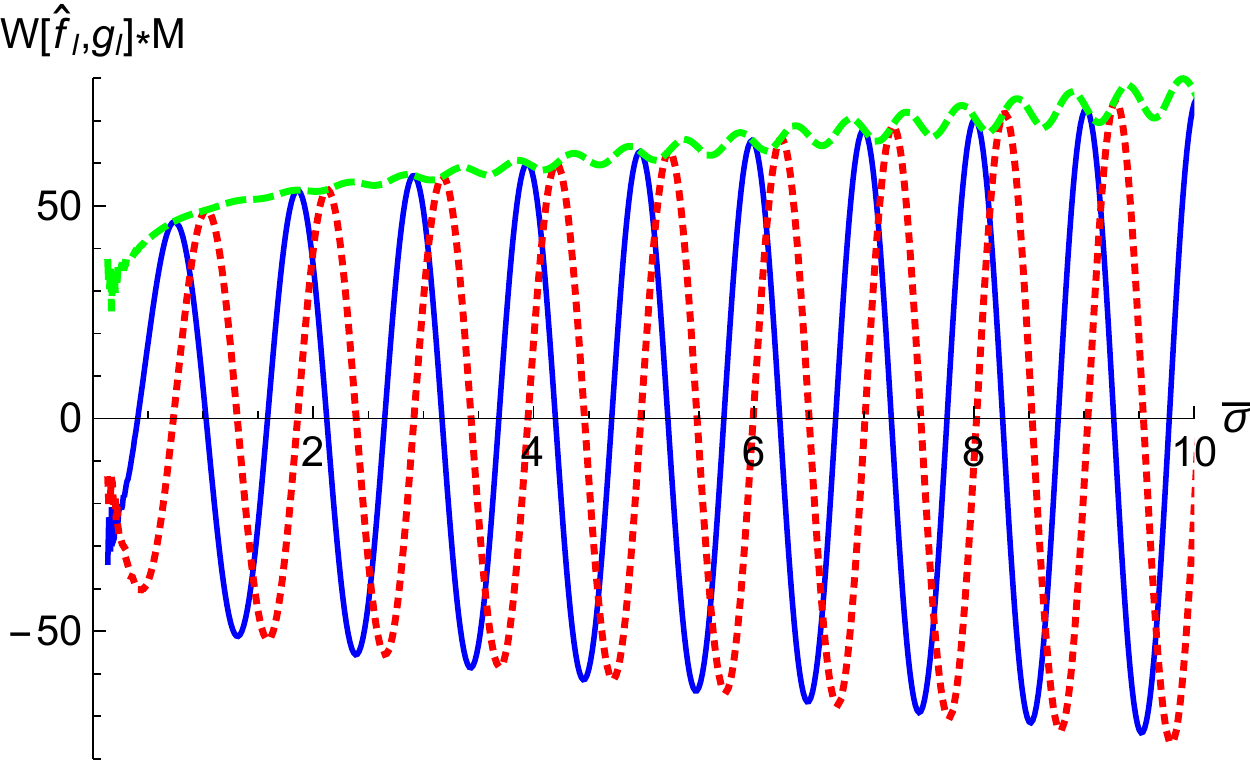}
            \end{center}
\caption{
The Wronskian $W\left[\Chi,\g\right]$ 
as a function of $\nb$ for $s=0$, $\ell=1$. 
The straight-blue and dotted-red curves respectively correspond to the real and imaginary parts of the Wronskian
evaluated using  the mid-frequencyÕ method of~\cite{Casals:2012ng} with the radial functions calculated at
at $r=2.8M$; the convergence of the series used in   the mid-frequencyÕ method becomes slower as the frequency becomes smaller.
The dashed blue and red curves 
respectively correspond to the real and imaginary parts of the Wronskian
evaluated using the small-frequency asymptotics presented in this paper
 up to $O(\nb^{14})$.
In figure (b) we plot the mid-frequency  to larger values of the frequency so as to show its general form; the green-dashed curve here corresponds to the absolute value of the Wronskian. Cf. Fig.13~\cite{Casals:2012ng} for $s=\ell=2$.
}
\label{fig:wronskian,s=0,l=1,r=2.8}
\end{figure} 


\begin{figure}[h!]
\begin{center}
\includegraphics[width=8cm]{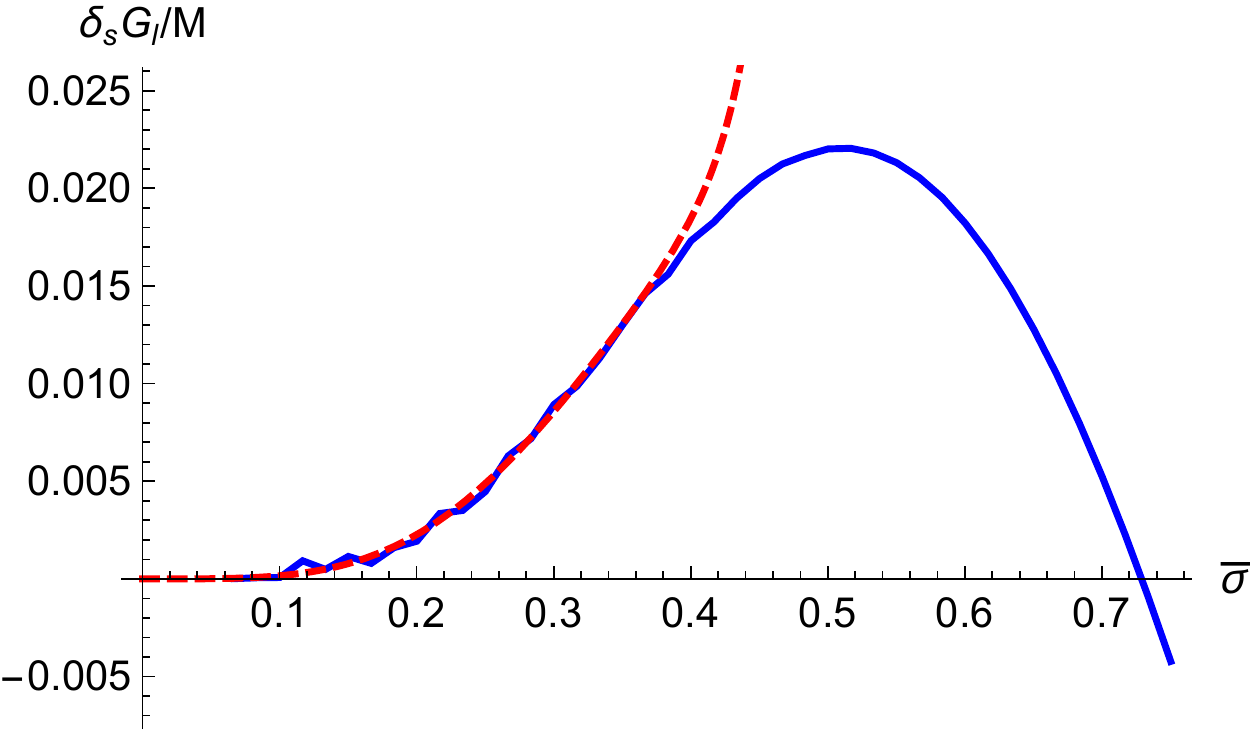}
\end{center}
\caption{
BC mode 
$\DGw\left(r,r',-i\fNIA\right)$ 
of Eq.(\ref{eq:DeltaG in terms of Deltag}) as a function of $\nb$ at $r=r(r_*=0.4M)$, $r'=r'(r_*=0.2M)$ for  $s=0$ and $\ell=1$.
In the straight blue curve we calculated tbe BC mode using  the mid-frequencyÕ method of~\cite{Casals:2012ng} and 
in the dashed red curve we calculated it using the small-frequency asymptotics presented in this paper up to $O(\nb^{15})$.
}
\label{fig:DeltaG s=0,l=1}
\end{figure}



\section{Radial Coefficients} \label{sec:App radial coeffs}

In this appendix we use the Chandrasekhar transformation Eq.(\ref{eq:RW toTeuk}) in order to relate the BPT and RW radial coefficients and Wronskians.
Let us first add higher orders to the asymptotics in Eq.(\ref{eq: bc f}) of the RW solution:
\begin{align}
\label{eq: bc Rin higher}
\f\sim 
\begin{cases}
\left(1+\alpha_+\Delta +\beta_+\Delta^2\right) e^{-i\omega r_*},& r_*\to -\infty, \\ 
\Ain \left(1+\frac{\alpha_{\infty}}{r}+\frac{\beta_{\infty}}{r^2}\right) e^{-i\omega r_*}+\Aout e^{+i\omega r_*},& r_*\to +\infty,
\end{cases} 
\end{align}
where $\alpha_+$, $\beta_+$, $\alpha_{\infty}$  and $\beta_{\infty}$ are coefficients to be determined.
By imposing that these asymptotics satisfy  the RW equation, we find
\begin{align} \label{eq:RW asympt coeffs}
&
\alpha_{\infty}=-\frac{i\lambda}{2\omega},\quad
\beta_{\infty}=\frac{\lambda(2-\lambda)+4Mi(s^2-1)\omega}{8\omega^2}, \\
&
\alpha_+= \frac{i \left(\lambda+1 -s^2\right)}{4 M^2 (4 M \omega +i)} ,\quad
\beta_+=\frac{4 -s^4-\lambda ^2+\lambda  (2-16 i M \omega )+s^2 (2 \lambda +20 i M \omega -3)-20 i M \omega}{64 M^4 (2 M \omega +i) (4 M \omega +i)}.
\nonumber
\end{align}
We  note that $\alpha_{\infty},\alpha_+,\beta_{\infty},\beta_+$ are all real-valued when $\omega$ is purely imaginary.

Using the transformation Eq.(\ref{eq:RW toTeuk}) it is straight-forward to find the following coefficients of BPT solutions:
\begin{equation} \label{eq:Rininc RW to BPT}
\RinincNfg{s}=
\begin{cases}
\Ain, & s=0,\\
\displaystyle-\alpha_{\infty} \Ain & s=-1, \\
\displaystyle 2\beta_{\infty}\Ain,  & s=-2.
\end{cases}
\quad
\RintraNfg{s}=
\begin{cases}
\frac{1}{2M}, & s=0,\\
\displaystyle \alpha_+ & s=-1, \\
\displaystyle \frac{2(\alpha_++2M^2\beta_+)}{M},  & s=-2.
\end{cases}
\quad
\RuptraNfg{s}=
\begin{cases}
1, & s=0,\\
\displaystyle 2i\omega & s=-1, \\
\displaystyle -4\omega^2,  & s=-2.
\end{cases}
\end{equation}
It then immediately follows, using Eqs.(\ref{eq:Wronsk}) and (\ref{eq:WT}), that
\begin{equation} \label{eq:Wronskian RW to BPT}
\WT=W\cdot
\begin{cases}
2M, & s=0,\\
\displaystyle\frac{-\alpha_{\infty}}{\alpha_+}, & s=-1, \\
\displaystyle\frac{M\beta_{\infty}}{\alpha_++2M^2\beta_+},  & s=-2.
\end{cases} 
\end{equation}
Note that $\WT$ is, by definition, independent of the chosen normalization for $\Rin{s}$ and $\Rup{s}$.

\end{document}